\documentclass[12pt,letterpaper]{article}    
\pdfoutput=1 
\usepackage{jheppub}    

\usepackage{amsthm}
\usepackage{amsmath,soul}        
\usepackage{enumerate}  
\usepackage{graphicx}       
\usepackage{color}
\usepackage{float}
\usepackage{amsfonts}
\usepackage{caption}
\usepackage{cancel}
\usepackage[dvipsnames]{xcolor} 
\usepackage{subcaption} 

\newcommand{\reef}[1]{(\ref{#1})}
\newcommand{\beq}{\begin{equation}} 
\newcommand{\eeq}{\end{equation}}
\newcommand{\beqa}{\begin{eqnarray}}
\newcommand{\eeqa}{\end{eqnarray}} 
\newcommand{\beqar}{\begin{eqnarray*}}
\newcommand{\eeqar}{\end{eqnarray*}}

\newcommand{\hyperf}[4]{{}_2\text{F}_1\!\left( #1 , #2 \,\text{;}\, #3 \,\text{;}\, #4 \right)}
\newcommand{\bspliteq}{\begin{equation}\begin{split}}
\newcommand{\espliteq}{\end{split}\end{equation}}

\newcommand{\labell}[1]{\label{#1}}

\newcommand{\mcO}{\mathcal{O}}

\newcommand{\bthesis}{\iffalse}

\newcommand{\up}[1]{^{\scriptscriptstyle(#1)}}

\newcommand{\Do}{\Delta_0} 
\newcommand{\Oo}{\mathcal{O}_0} 

\makeatletter
\def\setc{\edef\getc{\pdfliteral{\current@color}}}
\makeatother

\newif\ifthesis
\thesisfalse
\newif\ifblue
\bluetrue

\title{Quantum critical response: from conformal perturbation theory to holography}{\tiny }

\author[a]{Andrew Lucas,}
\author[b,c]{Todd Sierens}
\author[d]{and William Witczak-Krempa}

\affiliation[a]{Department of Physics, Stanford University, Stanford, CA 94305, USA}
\affiliation[b]{Perimeter Institute for Theoretical Physics,
Waterloo, Ontario N2L 2Y5, Canada}
\affiliation[c]{Department of Physics \& Astronomy and
Guelph-Waterloo Physics Institute,  
University of Waterloo, Waterloo, Ontario N2L 3G1, Canada}
\affiliation[d]{Department of Physics and Regroupement qu\'eb\'equois sur les mat\'eriaux de pointe, Universit\'e de Montr\'eal, Montr\'eal, Qu\'ebec, H3C 3J7, Canada} 

\emailAdd{ajlucas@stanford.edu}
\emailAdd{tsierens@perimeterinstitute.ca}
\emailAdd{w.witczak-krempa@umontreal.ca}

\abstract{We discuss dynamical response functions near quantum critical points, allowing for both a finite temperature 
and detuning by a relevant operator. When the quantum critical point is described by a conformal field theory (CFT), 
conformal perturbation theory and the operator product expansion can be used to fix the first few leading terms at high frequencies. 
Knowledge of the high frequency response allows us then to derive non-perturbative sum rules. We show, via explicit computations, 
how holography recovers the general results of conformal field theory, and the associated sum rules, for any holographic field theory 
with a conformal UV completion -- regardless of any possible new ordering and/or scaling physics in the IR. We numerically obtain holographic response functions at all frequencies, allowing us to probe the breakdown of the asymptotic high-frequency regime. 
Finally, we show that high frequency response functions in holographic Lifshitz theories are quite similar to their conformal counterparts, even though they are not strongly constrained by symmetry.} 
\begin{document}

\maketitle

\section{Introduction}
\begin{figure}   
\centering
\includegraphics[width=3in]{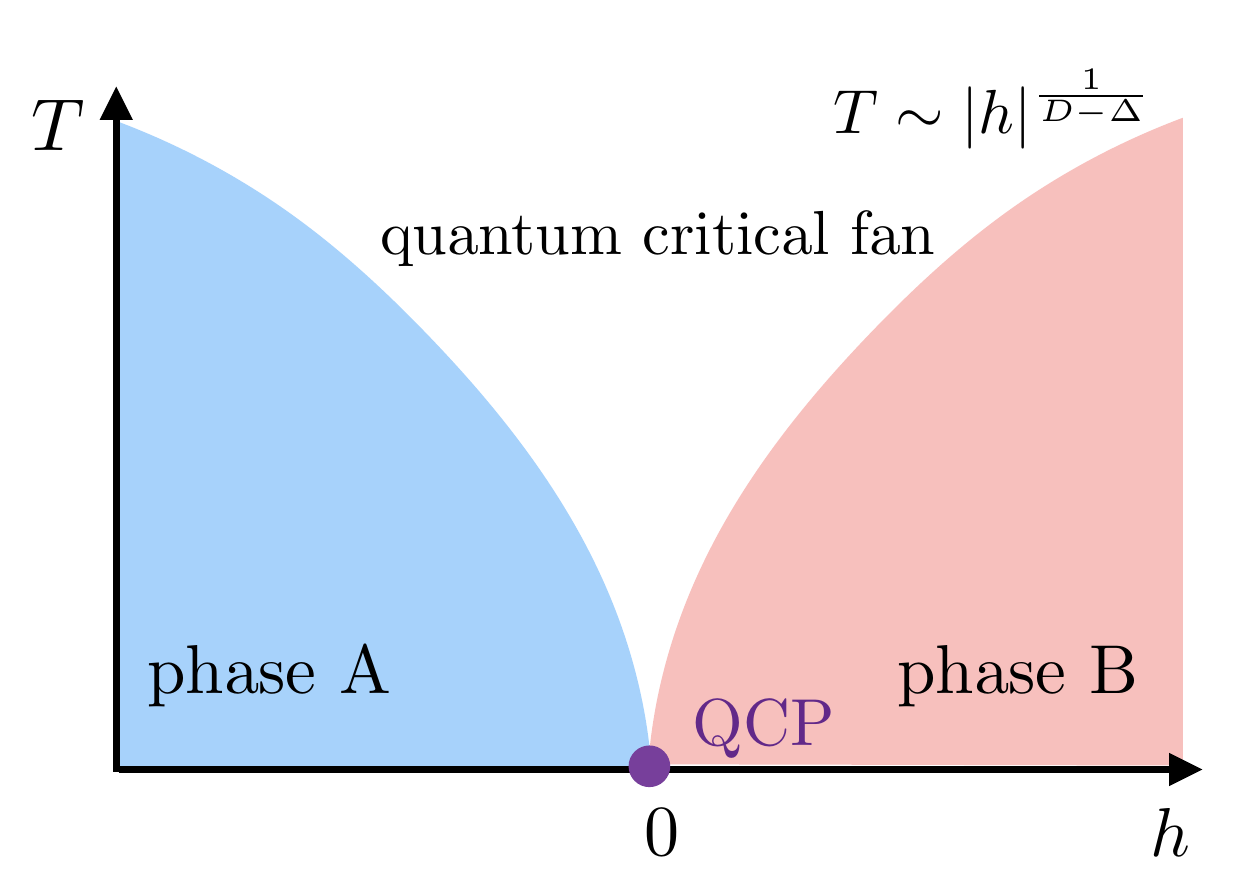} 
\caption{Schematic phase diagram of a theory with a (conformal) quantum critical point separating two phases at zero temperature.   The non-thermal detuning parameter $h$ must be set to zero at the critical point.  At finite temperature $T$,  when $T\gtrsim |h|^{\frac{1}{D-\Delta}}=|h|^\nu$,  there is a window in between the two phases where the system behaves qualitatively like the quantum critical point $h=0$, detuned only by temperature. In this paper, 
we demonstrate that certain fingerprints in the dynamical response functions of 
the quantum critical point can extend throughout this entire phase diagram.} 
\label{fig:QCP} 
\end{figure}

A quantum critical point (QCP) is a zero-temperature \emph{continuous} phase transition which arises 
as a coupling parameter is tuned through a critical value. Typically, one finds that at zero temperature (away from the QCP) there are two distinct `conventional' gapped or gapless phases of matter,   whereas at finite temperature there is a finite region of the phase diagram dominated strongly by critical fluctuations:  see Figure \ref{fig:QCP}. A canonical example is the transverse field quantum Ising model in $D=1+1,2+1$ and $3+1$ spacetime dimensions, where the QCP separates a paramagnet from a broken symmetry phase, the ferromagnet. The QCP is described at low
energies by the $D$-dimensional Ising CFT \cite{book}. Other examples involve  
fermions such as the Gross-Neveu model \cite{ZinnJustin,book}, and/or gauge fields (e.g.\ Banks-Zaks fixed points \cite{Banks-Zaks}).    

  Proximity to a QCP is believed to be the origin of exotic response functions in many experimentally interesting condensed matter systems \cite{Sachdev:2011cs}.    Even when the ground state is conventional, so long as it lies `near' a QCP,  both thermodynamic and dynamical response functions are significantly modified by quantum critical 
fluctuations. Because interesting quantum critical theories are strongly coupled and lack long-lived quasiparticle excitations,   obtaining  quantitative predictions for  critical dynamical response functions, where both quantum and thermal fluctuations must be accounted for, has proven to be challenging. In this respect, quantum Monte Carlo 
studies have proven useful to obtain the response functions 
in imaginary time \cite{natphys,chen,Gazit14,katz,Swanson2013}. The precise analytic continuation to real frequencies
of such data constitutes a difficult open problem. The results of this paper  
impose constraints on the response functions that will constrain the continuation procedure.

In the case of conformal field theories (CFTs), one can derive rigorous and general constraints on the dynamics at zero and finite temperature by using the operator product expansion (OPE). In addition, one can study the response functions in the neighborhood of the QCP by using conformal perturbation theory, where the expansion parameter is the detuning of the coupling parameter from its critical value.   Another technique is the gauge-gravity duality \cite{lucasreview},  which relates certain large-$N$ matrix quantum field theories to classical gravity theories in higher dimensional spacetimes.      The power of this approach is that we are very easily able  to access finite temperature and finite frequency response functions, directly in real time.    Unfortunately, experimentally realized condensed matter systems do not have weakly coupled gravity duals,  and some care is required in identifying which features of these holographic theories are relevant for broader field theories.

In this paper, we will discuss the asymptotics of a variety of response functions at high frequency \cite{Son09,caron09,sum-rules,natphys,David11,David12,will-hd, katz, willprl,will-susy, Myers:2016wsu, Lucas:2016fju}. We will show that in holographic field theories, this asymptotics is universal and does not depend in any detailed fashion on the nature of the ground state. As a consequence,  we will then derive non-perturbative sum rules which may be tested directly in experiments.    In the special case when the QCP is a conformal field theory (CFT), we will also be able to compare our holographic calculation of the response functions with an independent computation using techniques of conformal field theory, and will find perfect agreement.     
We will also discuss the generalization of these calculations to a richer class of Lifshitz field theories, which are less sharply constrained by symmetry.    

\subsection{Main Results}  
We study a quantum critical system in $D$ spacetime dimensions deformed by a relevant operator $\mathcal{O}$:   
\begin{equation}
S = S_{\mathrm{QCP}} -  h \int \mathrm{d}^D x \; \mathcal{O},   \label{eq:hdef}
\end{equation}
where for simplicity we have chosen the coupling $h$ to be $x$-independent. We briefly discuss the case of an inhomogeneous coupling in the
conclusion \ref{sec:conclusion}.  
This formula is presented in real time;  in Euclidean time, there is a relative minus sign for the second term. Now, let $\Delta$ be the scaling dimension of $\mathcal{O}$:    
\begin{equation}
\langle \mathcal{O}(\mathbf{x},t) \mathcal{O}(\mathbf{y},t)\rangle_0 = C_{\mathcal{OO}} |\mathbf{x}-\mathbf{y}|^{-2\Delta},
\end{equation}
where the $0$ subscript means that the correlation function is evaluated at the QCP $T=h=0$.  
The parameter $h$ is commonly called the (non-thermal) detuning parameter; it drives the quantum phase transition. We also will study this detuned theory at finite temperature $T$.   The dynamical critical exponent $z$ of this QCP is defined as the `dimension' of energy:
namely, a dilation of space $\mathbf x\to \lambda\mathbf x$ is accompanied by a temporal dilation $t\to \lambda^z t$ in order to be a symmetry. 
This implies that the excitations of this theory (generally there are no well-defined quasi-particles at the QCP) disperse as $\omega(q) \sim q^z$.   For simplicity, in much of this paper we will focus on the case $z=1$, specializing to the case where the QCP is a conformal field theory, returning to the more general case $z\ne 1$ at the end of the paper.     

 Let $\mathcal{X}$ be a local operator of scaling dimension $\Delta_{\mathcal{X}}$ in this critical theory, 
and assume for now that $\Delta_{\mathcal{X}}\neq (D/2)+n$, where $n$ is an integer, which is the generic case
and ensures that both powers in the brackets do not differ by an integer (otherwise, additional logarithms can appear).  
We will show that, when $z=1$, the generic asymptotic expansion of the 
dynamical response function $\langle \mathcal{XX}\rangle$ at high frequency is:   
\begin{equation}
\langle \mathcal{X}(\omega) \mathcal{X} (-\omega)\rangle = (\mathrm{i}\omega)^{2\Delta_{\mathcal{X}} - D} \left[\mathcal{C}_{\mathcal{XX}} + \mathcal{A}\frac{h}{(\mathrm{i}\omega)^{D-\Delta}} + \mathcal{B}\frac{\langle \mathcal{O}\rangle}{(\mathrm{i}\omega)^\Delta} + \cdots \right],
\label{eq:mainres}
\end{equation}
where $\langle \cdots\rangle$ denotes an expectation value evaluated 
at non-zero values of $h$ and $T$, in contrast to $\langle \cdots\rangle_0$, which is evaluated directly in the quantum critical theory, at $h=T=0$. 
We emphasize that the first correction is analytic in the coupling $h$ and independent of temperature. 
The expansion (\ref{eq:mainres}) holds when $\omega\gg T, |h|^{1/(D-\Delta)}$ ($\hbar=k_{\mathrm{B}}=1$) 
and when the detuned system has a finite correlation length. It can fail in regions separated from the quantum critical ``fan'' by a phase transition, 
where potentially new gapless modes can arise. An example of this failure was shown in the broken symmetry (Goldstone) phase in the vicinity of the
Wilson-Fisher $O(N)$ QCP at large $N$ \cite{Lucas:2016fju}. 
The leading order coefficient $\mathcal{C}_{\mathcal{XX}}$ is an operator normalization in the vacuum of the CFT.  
We compute the coefficients $\mathcal{A}$ and $\mathcal{B}$ exactly for arbitrary spacetime dimension $D$ and dimension $\Delta$ of the (relevant) operator, in three practical examples: $\langle \Oo \Oo \rangle$, $\langle J^x J^x\rangle$,  and $\langle T^{xy}T^{xy}\rangle$.    Here $\Oo$ is a scalar operator,  $J^x$ is a spatial component of a conserved current, and $T^{xy}$ is an off-diagonal spatial component of the conserved stress-energy tensor.    Remarkably, $\mathcal{A}$ and $\mathcal{B}$ will not depend on either $h$ or $T$, and are properties of the pure CFT, and the choice of operator $\mathcal{X}$.    Furthermore, the ratio $\mathcal{A}/\mathcal{B}$ is universal, depending only on $\Delta$ and $D$.  

\begin{table}
  \begin{center}  
    \begin{tabular}{|c|c|}\hline
       CFT operator $\mathcal{X}$ &\ $\mathcal{A}/\mathcal{B}$ \\
      \hline
      $T_{xy}$ (stress tensor) &\ (\ref{eq:ABratioshear})  \\
      $J_x$ (conserved current) &\  (\ref{eq:ABratiocurrent}) \\
      $\Oo$ (scalar operator) &\  (\ref{eq:ABratioscalar}) \\
      \hline
          \end{tabular}
  \end{center}
  \caption{Where to find our main results for the ratio of coefficients $\mathcal{A}/\mathcal{B}$ in the asymptotic expansion of $\langle \mathcal X\mathcal X\rangle$ in CFTs,
(\ref{eq:mainres}).}    
\label{table1}
\end{table}

When $\Delta_{\mathcal{X}}$ and/or $\Delta_{\mathcal{X}}-\Delta$ is an integer or half-integer, depending on the dimension $D$, logarithmic corrections to (\ref{eq:mainres}) appear.   These special cases are important because conserved currents have such dimensions, and we will discuss them later in this paper.

The reason that $\mathcal{O}$ be a relevant operator is to ensure that the asymptotic expansion (\ref{eq:mainres}) is well-behaved;  if $\mathcal{O}$ is irrelevant, the second term in (\ref{eq:mainres}) dominates.   This is not surprising, as we are probing the UV physics (which is sensitive to the dynamics of irrelevant operators).

One of the main purposes of the asymptotic expansion (\ref{eq:mainres}) is to verify when the following kind of sum rule holds: 
\begin{equation}
\int\limits_0^\infty \mathrm{d}\omega \; \mathrm{Re}\left(\frac{\langle \mathcal{X}(-\omega)\mathcal{X}(\omega)\rangle}{\mathrm{i}\omega} - \mathcal{C}_{\mathcal{XX}} \cdot(\mathrm{i}\omega)^{2\Delta_{\mathcal{X}}-D}\right) =0.  \label{eq:mainsum}
\end{equation} 
Clearly, this sum rule is automatically satisfied at the critical point,  but it can also be satisfied off-criticality.   When $h=0$, we will see that such a sum rule will only hold if $2\Delta_{\mathcal{X}} - D < \Delta$.   When $h\ne 0$, an additional inequality $\Delta < 2(D-\Delta_{\mathcal{X}})$ is also required.

 Eq.~(\ref{eq:mainres}) is presented in real time.   Our derivation of this result will occur in Euclidean time.    It is usually straightforward to analytically continue back to real time.  There can be subtleties for certain values of $D$ and $\Delta$, for certain operators $\mathcal{X}$, and we will clarify the restrictions later.   
 
We will present two derivations of (\ref{eq:mainres}): 
one using conformal perturbation theory applied to general CFTs (such an approached was succinctly presented for the conductivity in \cite{Lucas:2016fju}), 
one using holography.  
We will show that these approaches \emph{exactly} agree.   A priori, as the conformal perturbation theory approach does not rely on the large-$N$ matrix limit of holography, it may seem as though the holographic derivation is superfluous.  However, we will argue that this is not the case:   that conformal perturbation theory is completely controlled at finite $h$ and finite $T$ is not always clear,  especially if the ground state of the detuned theory is not a conventional gapped phase. 
  Our holographic computation reveals that, independent of the details of the ground state,  the asymptotics of the response functions are given by (\ref{eq:mainres}), with all coefficients in the expansion determined completely by the UV CFT.   We expect that -- with some important exceptions in broken symmetry phases \cite{Lucas:2016fju} -- this is a robust result of our analysis, valid beyond the large-$N$ limit.    

For holographic models, we will also study the case $z\ne 1$, and find a simple generalization of (\ref{eq:mainres}).  While we will prove the universality of (\ref{eq:mainres}) even beyond holography when $z=1$,   we cannot prove this universality for $z\ne 1$.

We also present the full frequency dependence of these response functions in multiple holographic models.    In addition to verifying that the asymptotics (\ref{eq:mainres}) and the associated sum rules such as (\ref{eq:mainsum}) hold,  we will be able to probe frequency scales past which conformal perturbation theory fails, both at a finite temperature critical point,  and in systems strongly detuned off-criticality. 

\subsection{Outline}
The remainder of the paper is organized as follows.  In Section \ref{sec2}, we will describe the asymptotics of response functions using conformal perturbation theory.    In Section \ref{sec3}, we will describe a very general class of holographic theories with conformal UV completions, and compute all necessary CFT data required to compare with Section \ref{sec2}.   Section \ref{sec4} contains a direct computation of high frequency holographic response functions,  in which exact agreement with conformal perturbation theory is found.   We numerically compute the full frequency response functions in Section \ref{sec5}.    We generalize our approach in Section \ref{sec6} to holographic theories with `Lifshitz' field theory duals,  where there are, as of yet, no other approaches like conformal field theory which can be used.    We discuss sum rules in Section \ref{sec7}.  We conclude the paper by discussing the possible breakdown of (\ref{eq:mainres})  in broken symmetry phases, and the generalization of (\ref{eq:mainres}) to QCPs deformed by disorder.

\section{Conformal Perturbation Theory}\label{sec2}
In this section, we present a computation of the high-frequency asymptotic behavior of the three advertised two-point functions, $\langle \Oo \Oo \rangle$, $\langle J^x J^x\rangle$,  $\langle T^{xy}T^{xy}\rangle$, using conformal perturbation theory for generic spacetime dimension $D$.   The computation of $\langle J^x J^x\rangle$ was recently presented by two of us \cite{Lucas:2016fju}.   We will present the computation of $\langle \Oo \Oo \rangle$ in detail -- the computation of the latter two is extremely similar, and so we simply summarize the results. 
\subsection{Scalar Two-Point Functions}
We begin with the correlator $\langle \Oo\Oo\rangle$ of a scalar operator $\Oo$ with dimension $\Do$.   In the vacuum of a unitary CFT,  one finds that in position space \begin{equation}
\langle \Oo(x)\Oo(y)\rangle_0 = \frac{C_{\Oo\Oo}}{|x-y|^{2\Do}}.
\end{equation}
In this work, we are interested in momentum space correlation functions, which are readily computed (up to an overall $\delta$ function): \begin{align}
\langle \Oo(k)\Oo(-k)\rangle_0 &= \int \mathrm{d}^Dx \,\mathrm{e}^{\mathrm{i}k\cdot x} \langle \Oo(x)\Oo(0)\rangle_0  = \int\limits_0^\infty \frac{\mathrm{d}s}{s} \int \mathrm{d}^Dx  \mathrm{e}^{\mathrm{i}k\cdot x - sx^2} \frac{s^{\Do}}{\Gamma(\Do)}C_{\Oo\Oo} \notag \\
&= \frac{2^{D-2\Delta}\pi^{D/2} \Gamma(\frac{D}{2}-\Do)}{\Gamma(\Do)} C_{\Oo\Oo} k^{2\Do-D} \equiv \mathcal{C}_{\Oo\Oo} k^{2\Do-D}.   \label{eq:OOkk}
\end{align}
If $\Do-\frac{D}{2}=0,1,2,\ldots $ is an integer, then we find an extra factor of $\log k$ in (\ref{eq:OOkk}); henceforth in the main text, we will assume that $\Do-\frac{D}{2}$ is not an integer.  We also note that $C_{\Oo\Oo}>0$, but depending on $\Delta$ and $D$, $\mathcal{C}_{\Oo\Oo}$ can take either sign.   

We can now deform away from the exact critical point by turning on a finite temperature $T$ or detuning $h$.   (\ref{eq:OOkk}) will now receive corrections due to $h$ and $T$ -- the leading order corrections are written in (\ref{eq:mainres}).   We first outline the argument for how the structure (\ref{eq:mainres}) arises,  after which it will be relatively straightforward to do a direct computation to fix $\mathcal{A}$ and $\mathcal{B}$.   Suppose that we set $T=0$, for simplicity, but detune our CFT by $h$, (\ref{eq:hdef}).
We express the expectation value using a Euclidean path integral: \begin{align} 
\langle \Oo(\Omega)\Oo(-\Omega)\rangle_h &= \frac{1}{\mathcal Z[h]} \int \mathrm{D}\Phi_{\mathrm{CFT}} \mathrm{e}^{-S_{\mathrm{CFT}} + h\mathcal{O}(0)} \Oo(\Omega)\Oo(-\Omega) \notag \\
&= \frac{\mathcal Z[0]}{\mathcal Z[h]} \left\langle \Oo(\Omega)\Oo(-\Omega) \mathrm{e}^{h\mathcal{O}(0)}\right\rangle_{0} \notag \\ 
&= \left\langle \Oo(\Omega)\Oo(-\Omega) (1+h\mathcal{O}(0)+\cdots)\right\rangle_{0} \notag \\ 
&= \mathcal{C}_{\Oo \Oo} \Omega^{2\Do - D} + \frac{\mathcal{A}h}{\Omega^{D-\Delta}} \Omega^{2\Do - D} + \cdots  \label{eq:21}
\end{align}
Here $\mathcal{O}(0)$ is the zero momentum mode of $\mathcal{O}$.   The first term of (\ref{eq:21}) follows from (\ref{eq:OOkk}). 
The power of $\Omega$ in the second term has been fixed by dimensional analysis; below we explicitly evaluate this term by evaluating the 3-point function $\langle\Oo\Oo \mathcal O\rangle_0$. 
This 3-point function is fixed by the conformal Ward identities in momentum space up to an overall coefficient;
alternatively, one can perform an analytic continuation from position to momentum space, although one must be
careful with regularization \cite{Bzowski:2013sza}. Holography naturally produces the correct answer as we show below. 
There are further terms in (\ref{eq:21}) containing subleading integer powers of $h/\Omega^{D-\Delta}$.   

However, the $\Oo$ 2-point correlation function will also contain terms non-analytic in $h/\Omega^{D-\Delta}$. It will often be the case, in the pure CFT, 
that the  operator product expansion (OPE) will take the form, here expressed in momentum-space:
\begin{equation}
\Oo(\Omega)\Oo(-\Omega) = \mathcal{C}_{\Oo\Oo}\Omega^{2\Do-D} +  \mathcal{B} \Omega^{2\Do-D} \frac{\mathcal{O}(0)}{\Omega^{\Delta}} + \cdots   \label{eq:OPE21}
\end{equation} 
We have $\langle \mathcal{O}\rangle_{0}=0$, since the pure CFT has no dimensionful parameter.  Hence, upon averaging this equation in a CFT, the $\mathcal{B}$ contribution to the OPE vanishes. However, once we detune with $h$ at $T=0$, 
we expect:  
\begin{equation}
\langle \mathcal{O}\rangle_{h} = A_\pm |h|^{\frac{\Delta}{D-\Delta}},  \label{eq:Ohscale}
\end{equation}
with a coefficient that generally depends on the sign of $h$. 
The OPE (\ref{eq:OPE21}) is a high energy property of the theory, and should be valid even at finite $h$, so long the frequency $\Omega$ is large enough: $h \ll \Omega^{D-\Delta}$.   Hence, we expect that we may apply the expectation value to the local OPE and find an additional contribution to the two-point function, linear in $\langle \mathcal{O}\rangle$: \begin{equation} \label{eq:OOexp}
\langle \Oo(\Omega)\Oo(-\Omega)\rangle_h = \mathcal{C}_{\Oo \Oo} \Omega^{2\Do - D} + \frac{\mathcal{A}h}{\Omega^{D-\Delta}} \Omega^{2\Do - D} + \cdots + \mathcal{B}\frac{\langle \mathcal{O}\rangle}{\Omega^\Delta}\Omega^{2\Do - D} + \cdots.
\end{equation}
In the context of classical critical phenomena, an analogous expansion for the short-distance spatial correlators of the order 
parameter have been found for thermal Wilson-Fisher fixed points in $D=3$ \cite{fisherlanger}, and subsequently 
more quantitatively in other theories using the OPE \cite{Guida95,Guida96,Caselle16}.  
These classical results   
are a special case of (\ref{eq:OOexp}) analytically continued to imaginary times at $T=0$.   
In the context of the classical 3D Ising model, such an expansion was recently used together with conformal bootstrap results to make 
predictions for the correlators of various scalars near the critical temperature $T_c$ \cite{Caselle16}.     
We also see that the coefficient $\mathcal{B}$ is nothing more than an OPE coefficient in the CFT, as found in \cite{katz, Lucas:2016fju}. 

Obviously, for generic $\Delta$, the power of $h$ in (\ref{eq:Ohscale}) is not an integer.  
Hence, the $\mathcal{B}$ term in this expansion cannot be captured at any finite order in the conformal perturbation theory expansion (\ref{eq:21}).   Still, we will see that it is possible to compute the ratio $\mathcal{A}/\mathcal{B}$ in terms of CFT data.

We claim that upon generalizing to finite temperature $T$, the essential arguments above follow through, 
despite the fact that when $T>0$, $\langle \mathcal{O} \rangle$ will be an analytic function of $h$ near $h=0$. Our holographic computation will confirm this insight in a broad variety of models. 

The argument above contains essentially all of the physics of conformal perturbation theory.  It remains to explicitly fix the coefficient $\mathcal{A}$ in terms of $\mathcal{B}$.  To do so, we carefully study 
the 3-point function $\langle \Oo (p_1) \Oo(p_2)\mathcal{O}(p_3)\rangle_0$, upon choosing the momenta to be \begin{subequations}\label{eq:p1p2p3}\begin{align}  
p_1 &= (\Omega,0), \\
p_2 &= (-\Omega-p,0), \\
p_3 &=  (p,0).
\end{align}\end{subequations}
where $p\ll\Omega$.
In a conformal field theory, this correlation function is completely fixed up to an overall prefactor  $A_{\Oo \Oo \mathcal{O}} $ \cite{Bzowski:2013sza, Barnes:2010jp}:\begin{equation}
\langle \Oo (p_1) \Oo(p_2)\mathcal{O}(p_3)\rangle_0 = A_{\Oo \Oo \mathcal{O}} 
I\!\left(\tfrac{D}{2}-1, \Do - \tfrac{D}{2}, \Do - \tfrac{D}{2}, \Delta - \tfrac{D}{2}\right)  \label{eq:sec2scalar}
\end{equation}
where we have defined (for later convenience):\begin{equation}
I(a,b,c,d) \equiv \int\limits_0^\infty \mathrm{d}x \; x^a p_1^b p_2^cp_3^d \, \mathrm{K}_b(p_1x) \mathrm{K}_c(p_2x)\mathrm{K}_d(p_3x). \label{eq:Idef}
\end{equation}
Here $\mathrm{K}_\nu(x)$ is the modified Bessel function which exponentially decays as $x\rightarrow \infty$.  From (\ref{eq:p1p2p3}), we see that $p_{1,2}\gg p_3$.  Hence, a good approximation to (\ref{eq:Idef}) should come from an asymptotic expansion of $\mathrm{K}_d(p_3x)$ in (\ref{eq:Idef}).   Defining \begin{equation}
\mathcal{Z}(b) \equiv 2^{b-1}\Gamma(b)\labell{eq:Z}
\end{equation} and using the asymptotic expansion \begin{equation}
\mathrm{K}_b(x) = \mathcal{Z}(b) x^{-b}\left[1 + \mathrm{O}\left(x^2\right)\right] + \mathcal{Z}(-b) x^b\left[1 + \mathrm{O}\left(x^2\right)\right] ,  \label{eq:bessel}
\end{equation}
we take the limit of (\ref{eq:p1p2p3}) and simplify (\ref{eq:Idef}) to 
\begin{align}
&\langle \Oo (p_1) \Oo(p_2)\mathcal{O}(p_3)\rangle_0 \approx A_{\Oo \Oo \mathcal{O}}\mathcal{Z}
\left(\Delta-\tfrac{D}{2}\right)\Psi\left(D-\Delta;\Delta_0-\tfrac{D}{2}\right) \Omega^{2\Do+\Delta-2D} \notag \\
&\;\;\;\;\;\;\;+ A_{\Oo \Oo \mathcal{O}} \mathcal{Z}\left(\tfrac{D}{2}-\Delta\right)  \Psi\left(\Delta;\Delta_0-\tfrac{D}{2}\right) p^{2\Delta-D}\Omega^{2\Do-\Delta-D},  \label{eq:29}
\end{align}
where we have defined \begin{equation}
\Psi(a;b) \equiv \int\limits_0^\infty \mathrm{d}x \; x^{a-1} \mathrm{K}_b(x)^2 =  \frac{\sqrt{\pi} \Gamma(\frac{a}{2}) \Gamma(\frac{a}{2}+b) \Gamma(\frac{a}{2}-b)}{4\Gamma(\frac{1+a}{2})} .  \label{eq:Psidef}
\end{equation}

The first term of (\ref{eq:29}) is completely regular as $p\rightarrow 0$.  The correlation function $\langle \Oo (\Omega)\Oo(-\Omega)\mathcal{O}(0)\rangle_{0}$ which appeared in conformal perturbation theory in (\ref{eq:21}) should be interpreted as this first contribution.  Thus we obtain \begin{equation}
\mathcal{A} = A_{\Oo \Oo \mathcal{O}} \mathcal{Z}\!\left(\Delta-\tfrac{D}{2}\right) \Psi\!\left(D-\Delta;\Delta_0-\tfrac{D}{2}\right).  \label{eq:21A}
\end{equation} 
The second term of (\ref{eq:29}) can be interpreted as follows.   At a small momentum $p$, the OPE (\ref{eq:OPE21}) generalizes to \begin{equation}
\Oo(\Omega)\Oo(-\Omega-p) = \mathcal{B} \Omega^{2\Do-D} \frac{\mathcal{O}(-p)}{\Omega^{\Delta}} + \cdots .  \label{eq:OPE212}
\end{equation}
Upon contracting both sides of (\ref{eq:OPE212}) with $\langle \cdots \mathcal{O}(p)\rangle_{\mathrm{CFT}}$, we obtain a contribution which is non-analytic in $p$: 
\begin{align}
\langle \Oo (p_1) \Oo(p_2)\mathcal{O}(p_3)\rangle_{0} &= \cdots + \mathcal{B} \Omega^{2\Do-\Delta -D} \langle \mathcal{O}(-p)\mathcal{O}(p)\rangle_{0}  \notag \\
&= \cdots + \mathcal{B}\mathcal{C}_{0} p^{2\Delta-D}\Omega^{2\Do-\Delta-D}.  \label{eq:213}
\end{align}  
Comparing (\ref{eq:213}) with (\ref{eq:29}) we conclude \begin{equation}
\mathcal{BC}_{\mathcal{OO}} = A_{\Oo \Oo \mathcal{O}} \mathcal{Z}\left(\tfrac{D}{2}-\Delta\right) \Psi\left(\Delta;\Delta_0-\tfrac{D}{2}\right).  \label{eq:21B}
\end{equation}
The ratio $\mathcal{A}/\mathcal{B}$ is independent of $A_{\Oo\Oo\mathcal{O}}$: \begin{equation}
\frac{\mathcal{A}}{\mathcal{B}} = \mathcal{C}_{\mathcal{OO}} \frac{\mathcal{Z}\left(\Delta-\frac{D}{2}\right) \Psi\left(D-\Delta;\Delta_0-\frac{D}{2}\right)}{\mathcal{Z}\left(\frac{D}{2}-\Delta\right) \Psi\left(\Delta;\Delta_0-\frac{D}{2}\right)}.  \label{eq:ABratioscalar}
\end{equation}
As promised, we have fixed $\mathcal{A}$ and $\mathcal{B}$ completely in terms of the CFT data $\Delta_0$, $\Delta$, $\mathcal{C}_{\mathcal{OO}}$ and $A_{\Oo\Oo\mathcal{O}}$.

The assumption that $\Delta-D/2$ is not an integer is relaxed in Appendix \ref{app:integer}.   For the most part, (\ref{eq:mainres}) is qualitatively unchanged, with an interesting exception when $\Delta=D/2$.

\subsection{Conductivity}
The conductivity at finite (Euclidean) frequency is defined as \begin{equation}
\sigma(\mathrm{i}\Omega) \equiv -\frac{1}{\Omega} \langle J_{x}(-\Omega)J_{x}(\Omega)\rangle,
\end{equation}
where the spatial momentum has been set to zero. 
Note that the operator dimension $\Delta_J=D-1$ is fixed by symmetry.   In a pure CFT, at zero temperature, we find (in position space) \cite{Osborn:1993cr}
\begin{equation}
\langle J_\mu(x)J_\nu(0)\rangle_0 = \frac{C_{JJ}}{x^{2D-2}}\mathcal{I}_{\mu\nu}(x)
\end{equation} 
with $C_{JJ}>0$ and 
\begin{equation} 
\mathcal{I}_{\mu\nu} \equiv \delta_{\mu\nu} -2 \frac{x_\mu x_\nu}{x^2}.
\end{equation}
   This can be Fourier transformed, analogous to (\ref{eq:OOkk}), to give 
\begin{equation} 
\langle J_\mu(k)J_\nu(-k)\rangle_0 = -\mathcal{C}_{JJ} k^{D-2} \pi_{\mu\nu}(k)
\end{equation}with \begin{equation}
\mathcal{C}_{JJ}  \equiv -\frac{2^{2-D}\pi^{D/2}(D-2)\Gamma(1-\frac{D}{2})}{\Gamma(D)}C_{JJ}   \label{eq:CJJ}
\end{equation} 
and the transverse (Euclidean) projector \begin{equation}
\pi_{\mu\nu}(k) \equiv \delta_{\mu\nu} - \frac{k_\mu k_\nu}{k^2}.
\end{equation}
In this paper, we will generically take $\mu=\nu=x$ and $k_0 = \Omega$, $\mathbf k=0$.   Hence we find \begin{equation}
\sigma(\mathrm{ i}\Omega) = \mathcal{C}_{JJ} \Omega^{D-3}
\end{equation}
in a pure CFT.   

Actually, this argument was a bit fast.  In even dimensions $D=4,6,8,\ldots$, the coefficient $\mathcal{C}_{JJ}$ diverges due to a pole in the $\Gamma$ function.   Upon regularization, this implies a logarithm in the momentum space correlator, and so the conductivity: \begin{equation}
\sigma(\mathrm{i}\Omega) =  (-1)^{\frac{D}{2}+1}\tilde{\mathcal{C}}_{JJ} \Omega^{D-3} \log \frac{\Omega}{\Lambda},   \label{eq:sigmalog}
\end{equation}
with $\Lambda$ a UV cutoff scale and \begin{equation}
\tilde{\mathcal{C}}_{JJ} \equiv  C_{JJ} \frac{2^{3-D}\pi^{\frac{D}{2}}(D-2)}{\Gamma(D)\Gamma(\frac{D}{2})}.
\end{equation}
Note that $\tilde{\mathcal{C}}_{JJ} >0$, and so in real time  
(setting $\mathrm{i}\Omega=\omega$) we find that the dissipative real part of the conductivity is finite and independent of $\Lambda$:\begin{equation}
\mathrm{Re}\left(\sigma(\omega)\right) = \frac{\pi}{2}\tilde{\mathcal{C}}_{JJ} \omega^{D-3} > 0.  \label{eq:2resigma}
\end{equation}
The dimension $D=2$ is special.  Indeed, one finds that the limit $D\rightarrow 2$ is regular in (\ref{eq:CJJ}):  and hence in Euclidean time (at zero temperature) 
\begin{equation} 
\sigma(\mathrm i \Omega) = \frac{2\pi C_{JJ}}{\Omega}.
\end{equation}

Now, we are ready to perturb the pure CFT by a finite $T$ and $h$. 
Upon doing so, the asymptotics of $\sigma(\mathrm i \Omega)$ at large $\Omega$ was recently computed in \cite{Lucas:2016fju} using conformal perturbation theory.  Conformal invariance demands \cite{Bzowski:2013sza}: \begin{align}
\left\langle J_x (p_1) J_x (p_2)\mathcal{O}(p_3)\right\rangle_0  &= A_{JJ\mathcal{O}} \left[I\left(\tfrac{D}{2}, \tfrac{D}{2}-1, \tfrac{D}{2}-1, \Delta - \tfrac{D}{2}+1\right) \right. \notag \\
& \left.+ \frac{\Delta}{2}(D-2-\Delta) I\left(\tfrac{D}{2} - 1, \tfrac{D}{2}-1, \tfrac{D}{2}-1, \Delta - \tfrac{D}{2} \right) \right]  \label{eq:JJOformal} 
\end{align}
so long as the $x$-component of all momenta vanishes, as in (\ref{eq:p1p2p3}).   
Following the exact same procedure as the previous section, we find that (\ref{eq:mainres}) becomes \begin{equation}
\sigma(\mathrm i \Omega) = \Omega^{D-3}\left[\mathcal{C}_{JJ} + \mathcal{A}\frac{h}{\Omega^{D-\Delta}} + \mathcal{B} \frac{\langle \mathcal{O}\rangle}{\Omega^\Delta}+\cdots\right]
\end{equation}
with \begin{subequations}\label{eq:ABsigma}\begin{align}
\mathcal{A} &= -A_{JJ\mathcal{O}} (D-\Delta) \left(1-\tfrac{\Delta}{2}\right) \mathcal{Z}\left(\Delta-\tfrac{D}{2}\right) 
\Psi\left(D-\Delta; \tfrac{D}{2}-1\right),\\
\mathcal{B} &= -\frac{A_{JJ\mathcal{O}}}{\mathcal{C}_{\mathcal{OO}}} \Delta \left(1-\tfrac{D-\Delta}{2}\right) \mathcal{Z}\left(\tfrac{D}{2}-\Delta\right) \Psi\left(\Delta; \tfrac{D}{2}-1\right).
\end{align}\end{subequations}
The ratio \begin{equation}
\frac{\mathcal{A}}{\mathcal{B}} = \mathcal{C}_{\mathcal{OO}}\frac{(D-\Delta)(2-\Delta)}{\Delta(2+\Delta-D)} \frac{\mathcal{Z}(\Delta-\frac{D}{2}) \Psi(D-\Delta;\frac{D}{2}-1)}{\mathcal{Z}(\frac{D}{2}-\Delta) \Psi(\Delta;\frac{D}{2}-1)} . \label{eq:ABratiocurrent}
\end{equation}

\subsection{Viscosity}
The viscosity at finite (Euclidean) frequency is defined as \begin{equation}
\eta(\mathrm i \Omega) \equiv -\frac{1}{\Omega} \langle T_{xy}(-\Omega)T_{xy}(\Omega)\rangle.
\end{equation}
As with $J_\mu$, the stress tensor $T_{\mu\nu}$ has a protected operator dimension $\Delta_T=D$. 
In a pure CFT, one finds in position space \cite{Osborn:1993cr} 
\begin{equation} 
\langle T_{\mu\nu}(x)T_{\rho\sigma}(0)\rangle_0 = \frac{C_{TT}}{x^{2D}} \left(\frac{\mathcal{I}_{\mu\rho}(x)\mathcal{I}_{\nu\sigma}(x) + \mathcal{I}_{\mu\sigma}(x)\mathcal{I}_{\nu\rho}(x)}{2} - \frac{\delta_{\mu\nu}\delta_{\rho\sigma}}{D}\right).
\end{equation} 
After a Fourier transform to momentum space: \begin{equation}
\langle T_{\mu\nu}(k)T_{\rho\sigma}(-k)\rangle_0 = -\mathcal{C}_{TT} k^D \left(\frac{\pi_{\mu\rho}(k)\pi_{\nu\sigma}(k) +\pi_{\nu\rho}(k)\pi_{\mu\sigma}(k) }{2} - \frac{\pi_{\mu\nu}(k)\pi_{\rho\sigma}(k)}{D-1}\right)  
\end{equation}
with \begin{equation}
\mathcal{C}_{TT} = -\frac{D-1}{D+1} \frac{\pi^{\frac{D}{2}}\Gamma(-\frac{D}{2})}{2^D \Gamma(D)} C_{TT}.
\end{equation}
We now find that in all even dimensions $D$, there are logarithmic corrections to this correlation function. 

At finite $T$ and $h$, we follow the same arguments as above to determine $\mathcal{A}$ and $\mathcal{B}$.   Conformal invariance demands \cite{Bzowski:2013sza}: 
\begin{align}
&\left\langle T_{xy} (p_1) T_{xy} (p_2)\mathcal{O}(p_3)\right\rangle_0  = A_{TT\mathcal{O}} \left[I\!\left(\tfrac{D}{2}+1, \tfrac{D}{2}-1, \tfrac{D}{2}-1, \Delta - \frac{D}{2}+2\right) \right. \notag \\
&\;\;\;\;\;\;\;\; + \tfrac{\Delta+2}{2} (D-\Delta-2) I\!\left(\tfrac{D}{2}, \tfrac{D}{2}-1, \tfrac{D}{2}-1, \Delta - \tfrac{D}{2}+1\right) \notag \\
&\;\;\;\;\;\;\;\; \left.+ \frac{\Delta}{8}(D-\Delta)(\Delta+2)(D-2-\Delta) I\!\left(\tfrac{D}{2} - 1, \tfrac{D}{2}, \tfrac{D}{2}, 
\Delta - \tfrac{D}{2} \right) \right]  \label{eq:TTOformal} 
\end{align}
so long as the momenta are of the form (\ref{eq:p1p2p3}).   Following an identical procedure to before we find the asymptotic
expansion at large frequencies
\begin{equation} 
\eta(\mathrm i \Omega) = \Omega^{D-1}\left[\mathcal{C}_{TT} + \mathcal{A}\frac{h}{\Omega^{D-\Delta}} + \mathcal{B} \frac{\langle \mathcal{O}\rangle}{\Omega^\Delta}+\cdots\right]
\end{equation}with \begin{subequations}\label{eq:ABvisc}\begin{align}
\mathcal{A} &= A_{TT\mathcal{O}} \frac{\Delta(D-\Delta)(D-\Delta+2)(\Delta-2)}{8}  \mathcal{Z}\!\left(\Delta-\tfrac{D}{2}\right) \Psi\!\left(D-\Delta; \tfrac{D}{2}\right),\\
\mathcal{B} &= \frac{A_{TT\mathcal{O}}}{\mathcal{C}_{\mathcal{OO}}} \frac{\Delta(D-\Delta)(\Delta+2)(D-\Delta-2)}{8}  \mathcal{Z}\!\left(\tfrac{D}{2}-\Delta\right) \Psi\!\left(\Delta; \tfrac{D}{2}\right). 
\end{align}\end{subequations} The ratio \begin{equation}
\frac{\mathcal{A}}{\mathcal{B}} = \mathcal{C}_{\mathcal{OO}}\frac{(D-\Delta+2)(\Delta-2)}{(D-\Delta-2)(2+\Delta)} \frac{\mathcal{Z}(\Delta-\frac{D}{2}) \Psi(D-\Delta;\frac{D}{2})}{\mathcal{Z}(\frac{D}{2}-\Delta) \Psi(\Delta;\frac{D}{2})} .  \label{eq:ABratioshear}
\end{equation}

\section{Holography: Bulk Action}\labell{sec3}
Now, we begin our holographic derivation of these results. The ``minimal" holographic model which we will study has the 
following action (in Euclidean time): \begin{align} 
S &= \int \mathrm{d}^{D+1}x\sqrt{g}\left[\frac{R}{2\kappa^2} + \frac{1}{2}(\partial \Phi)^2 + V(\Phi) +  \frac{Y(\Phi)}{\kappa^2}C_{abcd}C^{abcd} +\frac{Z(\Phi)}{4e^2}F_{ab}F^{ab} \right. \notag \\
 &\left. + \frac{1}{2}(\partial \psi)^2 + \frac{W(\Phi)}{2}\psi^2  + \cdots \right].  \labell{eq:holomodel}
\end{align}
As usual, $R$ is the Ricci scalar, $F_{ab}$ is the field strength of a bulk gauge field $A_\mu$,  and $C_{abcd}$ is the Weyl curvature tensor (traceless part of the Riemann tensor). Bulk dimensions are denoted with $ab\cdots$.   We additionally have two scalar fields $\Phi$ and $\psi$, dual to $\mathcal{O}$ and $\mathcal{O}_0$ respectively.   The purpose of each bulk field is to source correlation functions of different operators in the boundary theory, as described in Table \ref{table1}.   The functions $V$, $Y$, $Z$ and $W$ are all functions of the bulk scalar field $\Phi$, dual to the operator $\mathcal{O}$ whose influence on the asymptotic behavior of correlation functions is the focus of this paper.  We impose the following requirements: 
\begin{subequations}\labell{eq:holointeractions}\begin{align}    
V(\Phi\rightarrow 0) &= - \frac{D(D-1)}{2L^2\kappa^2} +  \frac{\Delta(\Delta-D)}{2L^2} \Phi^2 + \cdots, \\
Z(\Phi\rightarrow 0) &= 1 + \alpha_Z L^{\frac{D-1}{2}} \Phi + \cdots, \label{eq:ZPhi} \\
Y(\Phi \rightarrow 0) &=  \alpha_Y L^{\frac{D+3}{2}} \Phi + \cdots, \\
W(\Phi\rightarrow 0) &= \frac{\Do(\Do-D)}{L^2} + \alpha_W L^{\frac{D-5}{2}}  \Phi + \cdots   \label{eq:WPhi}
\end{align}\end{subequations}
Table \ref{table2} explains how the details of this model will be related to the CFT data of the theory dual to  (\ref{eq:holomodel}).   We will make these connections precise later in this section.

\begin{table}
  \begin{center}  
    \begin{tabular}{|c|c|}\hline
      bulk field & CFT operator \\
      \hline
      $g_{\mu\nu}$ (metric) &\ $T_{\mu\nu}$ (stress tensor) \\
      $A_\mu$ (gauge field) &\ $J_\mu$ (conserved U(1) current) \\
      $\Phi$ &\  $\mathcal{O}$ (relevant scalar)\\
      $\psi$ &\ $\Oo$ (probe scalar) \\
      \hline
          \end{tabular}
  \end{center}
  \caption{The four bulk fields of (\ref{eq:holomodel}), and their dual operators in the CFT of interest.  Our interest will be in the influence of $\mathcal{O}$ (and hence $\Phi$) on the two-point correlators of $T_{\mu\nu}$, $J_\mu$ and $\Oo$.}
\label{table1}
\end{table}

\begin{table}
  \begin{center}  
    \begin{tabular}{|c|c|}\hline
      bulk coupling & CFT data \\
      \hline
      $L^{D-1}/\kappa^2$ &\ $\mathcal{C}_{TT}$ \\
      $L^{D-3}/e^2$ &\ $\mathcal{C}_{JJ}$ \\ \hline
      $\alpha_Z$ &\  $\mathcal{C}_{JJ\mathcal{O}}$ \\
      $\alpha_Y$ &\  $\mathcal{C}_{TT\mathcal{O}}$ \\
      $\alpha_W$ &\  $\mathcal{C}_{\Oo\Oo\mathcal{O}}$ \\
      \hline
          \end{tabular}
  \end{center}
  \caption{The couplings in the holographic model (\ref{eq:holomodel}) are related to the CFT data of its dual theory.  Here we have listed the qualitative connection between certain dimensionless numbers associated with (\ref{eq:holomodel}) and CFT data of interest for our computation.   One thing missing from this table is $\mathcal{C}_{\mathcal{OO}}$ -- this is related to the normalization of the kinetic term for $\Phi$ in (\ref{eq:holomodel}), which we have set to $\frac{1}{2}$ as per usual.}
\label{table2}
\end{table}

Unless  otherwise stated, in this paper we will assume that the only non-trivial background fields are the metric $g_{\mu\nu}$ and $\Phi$.    The field theory directions will be denoted with $x^\mu$, and the emergent bulk radial direction with $r$.   We take $r\rightarrow 0$ to be the asymptotically AdS region of the bulk spacetime, where an approximate solution to the equations of motion from (\ref{eq:holomodel}) is \begin{equation}
\mathrm{d}s^2(r\rightarrow 0) = \frac{L^2}{r^2} \left(\mathrm{d}r^2 + \mathrm{d}t_{\textsc{e}}^2 + \mathrm{d}\mathbf{x}^2\right),  \label{eq:adsmetric}
\end{equation}
and 
\begin{equation}
\Phi(r\rightarrow0) = \frac{1}{L^{\frac{D-1}{2}}}\left[ hr^{D-\Delta} + \cdots + \frac{\langle \mathcal{O}\rangle}{2\Delta-D}r^\Delta + \cdots \right].  \labell{eq:phinormalization}
\end{equation}
With the normalizations in (\ref{eq:phinormalization}), $h$ is exactly equal to the detuning parameter $h$ as defined in (\ref{eq:hdef}),  and $\langle \mathcal{O}\rangle$ is identically the expectation value of the dual operator \cite{lucasreview}.    Exceptions arise when $\Delta-D/2$ is an integer, as the holographic renormalization procedure becomes more subtle \cite{Skenderis:2002wp};  these cases are discussed further in Appendix \ref{app:integer}. 

Let us briefly comment on the logic behind the construction (\ref{eq:holomodel}).  The $W$, $Z$ and $Y$ terms are chosen to be the simplest diffeomorphism/gauge invariant couplings of $\Phi$ to $\psi$, $A_a$ and $g_{ab}$ respectively.    In the last case, we further demand that $\Phi$ is not sourced by the AdS vacuum -- this forbids couplings such as $R^2$, $R_{abcd}R^{abcd}$, etc.   The reason for this is that in a pure CFT, any relevant scalar operator $\mathcal{O}$ satisfies $\langle \mathcal{O}\rangle=0$.  Hence, we cannot linearly couple $\Phi$ to any geometric invariant which is non-vanishing on the background (\ref{eq:adsmetric}).   Remarkably, we will see that the $Y(\Phi)$ coupling, which was introduced in \cite{Myers:2016wsu} so that $\langle \mathcal{O}\rangle \ne 0$ at finite temperature, also is precisely the term in the bulk action which leads to $A_{TT\mathcal{O}}\ne 0$.   

We will be computing two point functions holographically.  Hence, it is important to understand the asymptotic behavior of $\psi$, $A_x$ and $g_{xy}$ -- related to $\langle \Oo\Oo\rangle$, $\sigma$ and $\eta$ respectively.   One finds \begin{subequations}\label{eq:Ooasym}\begin{align}
\psi &=  \frac{1}{L^{\frac{D-1}{2}}}\left[h_0 r^{D-\Delta} + \cdots + \frac{\langle \Oo\rangle}{2\Do-D} r^{\Do} + \cdots\right],   \\
A_x &= A_x^0 + \frac{e^2}{L^{D-3}} \langle J_x\rangle \frac{r^{D-2}}{D-2}+ \cdots, \\
g_{xy} &= \frac{L^2}{r^2}\left[g^0_{xy} + \cdots + \frac{2\kappa^2}{L^{D-1}} \langle T_{xy}\rangle \frac{r^D}{D} + \cdots\right]
\end{align}\end{subequations}
where $h_0$, $A_x^0$ and $g_{xy}^0$ are the source operators for $\Oo$, $J_x$ and $T_{xy}$ respectively.   $A_x^0$ can  be thought of as an external gauge field, and $g_{xy}^0$ an external metric deformation in the boundary theory.   In practice, we will compute two-point functions by solving appropriate linearized equations and looking at the ratio of response to source:  for example,  \begin{equation}
\langle \Oo(-\Omega)\Oo(\Omega)\rangle = \left.\frac{\langle \Oo\rangle}{h_0}\right|_{\Omega}.  \label{eq:sec32pt}
\end{equation}

Let us briefly note that for $D=2$, the model we have introduced describes a CFT with a marginal $J_\mu J^\mu$ deformation \cite{Faulkner:2012gt}. In this dimension, the expansion of the gauge field in (\ref{eq:Ooasym}) becomes $\tilde A_x^0 + Le^2 \langle J_x\rangle \log r$;  the coefficient $\tilde A_x^0$ is sensitive to the `cutoff' in the logarithm, and this ambiguity is related to logarithmically running couplings associated with the marginal deformation.    The asymptotic corrections to the conductivity which we will compute in this paper are regular in the $D\rightarrow 2$ limit.  

\subsection{Fixing CFT Data}
Our ultimate goal is to compare a holographic computation of two-point functions to the general expansion obtained using  
conformal field theory. First, however, we will fix four CFT coefficients in the holographic model presented above:  $\mathcal{C}_{\mathcal{OO}}$, $A_{\Oo\Oo\mathcal{O}}$,  $A_{JJ\mathcal{O}}$ and $A_{TT\mathcal{O}}$.   As we are computing CFT data, we may assume in this section that the background metric is given exactly by (\ref{eq:adsmetric}).   Though we will compute two-point correlators analogously to (\ref{eq:sec32pt}), we will compute three-point correlators by directly evaluating the on-shell bulk action. 

Although these are rather technical exercises, we will see a beautiful physical correspondence between holographic Witten diagrams and the $I(a,b,c,d)$ integrals over Bessel functions, defined in (\ref{eq:Idef}) and found in \cite{Bzowski:2013sza} to be a consequence of conformal invariance alone.  

\subsubsection{$\mathcal{C}_{\mathcal{OO}}$} \label{app:COO}
In $D$ spacetime dimensions in a Euclidean AdS background (\ref{eq:adsmetric}), the equation of motion for the field $\Phi(r)\mathrm{e}^{\mathrm{i}\Omega t}$, dual to operator $\mathcal{O}(\Omega)$, is: \begin{equation}
0 = \frac{1}{\sqrt{g}}\partial_r \left(\sqrt{g}\partial_r \Phi\right) - \Omega^2 g^{tt} \Phi = m^2 \Phi + \cdots,
\end{equation}which becomes \begin{equation}
r^2 \partial_r \Phi - (D-1)r\partial_r \Phi - \Omega^2 r^2 \Phi  = \Delta(\Delta-D)\Phi + \cdots. \label{eq:freescalar}
\end{equation}
Further nonlinear terms in this equation of motion are not relevant for the present computation. The regular solution in the 
bulk (which vanishes as $r\to \infty$) is given by \begin{align}
\Phi &= r^{D/2}\mathrm{K}_{\Delta-D/2}(\Omega r) = Ar^{D-\Delta}+Br^\Delta + \cdots.  \label{eq:sec4phi}
\end{align}
The two-point function $\langle \mathcal{O}(\Omega)\mathcal{O}(-\Omega)\rangle \equiv \mathcal{C}_{\mathcal{OO}}\Omega^{2\Delta-D}$ is proportional to the ratio $B/A$.  More precisely, using (\ref{eq:bessel}) and the definition of $\mathcal{Z}$ in  \reef{eq:Z}, we find 
\begin{equation}
\langle \mathcal{O}(-\Omega)\mathcal{O}(\Omega)\rangle_0  = (2\Delta-D)\frac{B}{A}=  (2\Delta-D) \Omega^{2\Delta-D} \frac{\mathcal{Z}(\frac{D}{2}-\Delta)}{\mathcal{Z}(\Delta-\frac{D}{2})} \labell{eq:OOdef} 
\end{equation}
which fixes \begin{equation}
\mathcal{C}_{\mathcal{OO}} = (2\Delta-D)\frac{\mathcal{Z}(\frac{D}{2}-\Delta)}{\mathcal{Z}(\Delta-\frac{D}{2})}.  \label{eq:COOholography}
\end{equation}
This derivation assumes that (as noted previously)  $\Delta-\frac{D}{2}$ is not an integer;  see Appendix \ref{app:integer} for the changes to this derivation for this special case. 

\subsubsection{$A_{\Oo\Oo\mathcal{O}}$}\label{sec:AOOO}
It is easiest to compute three-point correlators in holography by directly evaluating the classical bulk action on-shell (see e.g. \cite{Freedman:1998tz, Chowdhury:2012km}):
\begin{equation}
\langle \Oo (p_1)\Oo(p_2)\mathcal{O}(p_3)\rangle_0 = -\frac{\delta^3 S_{\mathrm{bulk}}}{\delta j_{\Oo}(p_1) \delta j_{\Oo}(p_2)\delta j_{\mathcal{O}}(p_3)  }
\end{equation}
We are denoting here $j_{\Oo}$ and $j_{\mathcal{O}}$ as coefficients in the asymptotic expansion \begin{subequations}\label{eq:phip}\begin{align}
\Phi(p) &= \frac{j_{\mathcal{O}}(p)r^{D-\Delta}}{L^{\frac{D-1}{2}}} + \cdots, \\
\psi(p) &= \frac{j_{\Oo}(p)r^{D-\Do}}{L^{\frac{D-1}{2}}} + \cdots.
\end{align}\end{subequations}
To compute $A_{\Oo\Oo\mathcal{O}}$ (to leading order in $N$), we know that the boundary-bulk propagators for $\Phi$ and $\psi$ are nothing more than the solutions to their free (in the bulk) equations of motion.  Hence, from (\ref{eq:bessel}) and (\ref{eq:sec4phi}): \begin{subequations}\begin{align}
\Phi(p,r) &= \frac{r^{\frac{D}{2}} p^{\Delta-\frac{D}{2}} \mathrm{K}_{\Delta-\frac{D}{2}}(pr)}{L^{\frac{D-1}{2}}\mathcal{Z}(\Delta-\frac{D}{2})} j_{\mathcal{O}}(p), \\
\psi(p,r) &= \frac{r^{\frac{D}{2}} p^{\Do-\frac{D}{2}} \mathrm{K}_{\Do-\frac{D}{2}}(pr)}{L^{\frac{D-1}{2}}\mathcal{Z}(\Do-\frac{D}{2})} j_{\Oo}(p).
\end{align}\end{subequations}
Hence, \begin{align}
-&\langle \Oo (p_1)\Oo(p_2)\mathcal{O}(p_3)\rangle_0  = \frac{\delta^3 }{\delta j_{\Oo}(p_1) \delta j_{\Oo}(p_2)\delta j_{\mathcal{O}}(p_3)  } \int \mathrm{d}^{D+1}x \sqrt{g} \frac{\alpha_W}{2} L^{\frac{D-5}{2}} \times \notag \\
&\;\;\;\;\;\;\; \left[\int \mathrm{d}^D p_1 \mathrm{e}^{\mathrm{i}p_1\cdot x} \psi(p_1,r) \right]\left[\int \mathrm{d}^D p_2 \mathrm{e}^{\mathrm{i}p_2\cdot x} \psi(p_2,r) \right]\left[\int \mathrm{d}^D p_3 \mathrm{e}^{\mathrm{i}p_3\cdot x} \Phi(p_3,r) \right] \notag \\
&\;\;\;\;\;\;\;\;\;\;\;= \int \mathrm{d}r \mathrm{d}^Dx  \left(\frac{L}{r}\right)^{D+1} \alpha_W L^{\frac{D-5}{2}}   \mathrm{e}^{\mathrm{i}(p_1+p_2+p_3)\cdot x}  \times \notag \\
&\;\;\;\;\;\;\;\;\;\;\;\;\;\;\;\;\;\;\;   \frac{r^{\frac{D}{2}} p_1^{\Do-\frac{D}{2}} \mathrm{K}_{\Do-\frac{D}{2}}(p_1r)}{L^{\frac{D-1}{2}}\mathcal{Z}(\Do-\frac{D}{2})}\frac{r^{\frac{D}{2}} p_2^{\Do-\frac{D}{2}} \mathrm{K}_{\Do-\frac{D}{2}}(p_2r)}{L^{\frac{D-1}{2}}\mathcal{Z}(\Do-\frac{D}{2})}\frac{r^{\frac{D}{2}} p_3^{\Delta-\frac{D}{2}} \mathrm{K}_{\Delta-\frac{D}{2}}(p_3r)}{L^{\frac{D-1}{2}}\mathcal{Z}(\Delta-\frac{D}{2})} \notag \\ 
&\;\;\;\;\;\;\;\;\;\;\;= \frac{\alpha_W \delta^D(p_1+p_2+p_3) }{\mathcal{Z}(\Do - \frac{D}{2})^2 \mathcal{Z}(\Delta-\frac{D}{2})} I\left(\tfrac{D}{2}-1, \Do - \tfrac{D}{2}, \Do - \tfrac{D}{2}, \Delta-\tfrac{D}{2}\right).
\end{align}
Comparing with (\ref{eq:sec2scalar}), we conclude that \begin{equation}
A_{\Oo\Oo\mathcal{O}} = -\frac{\alpha_W}{\mathcal{Z}(\Do - \frac{D}{2})^2 \mathcal{Z}(\Delta-\frac{D}{2})}. \label{eq:AOOOholography}
\end{equation}

\subsubsection{$A_{JJ\mathcal{O}}$}\label{sec:AJJO}
This computation proceeds similarly to before, but is a bit more technically involved.   In this subsection and in the next, we assume that all momenta are in the $t$ direction, which simplifies many tensorial manipulations.    The $A_x$ propagator in the bulk is \begin{equation}
A_x(p,r) =  \frac{\hat{\mathrm{K}}_{\frac{D}{2}-1}(pr)}{\mathcal{Z}(\frac{D}{2}-1)} j_{A_x}(p),  \label{eq:Ax43}
\end{equation}
where $j_{A_x}$ is the boundary source for the dual current operator $J_x$, and we have defined, for later convenience, \begin{equation}
\hat{\mathrm{K}}_b(x) = x^b \mathrm{K}_b(x).
\end{equation}
(\ref{eq:Ax43}) follows from the bulk Maxwell equations in pure AdS:  \begin{equation}
0 = \frac{1}{\sqrt{g}}\partial_a \left(\sqrt{g}F^{ax}\right) = \left(\frac{L}{r}\right)^{3-D} \partial_r \left(\left(\frac{L}{r}\right)^{D-3}\partial_r A_x \right) - p^2 A_x.  \label{eq:maxwellads}
\end{equation}

Now, we need to evaluate the $\Phi A_x A_x$ contribution to the bulk action, just as before.   This is \begin{equation}
S_{\mathrm{bulk}} = \int \mathrm{d}^{D+1}x \sqrt{g} \frac{\alpha_Z L^{\frac{D-1}{2}}}{e^2} \Phi(p_3)\left[g^{ab}g^{xx} \partial_a A_x(p_1) \partial_b A_x(p_2)\right].
\end{equation}
Now, suppose we integrate by parts to remove derivatives from $A_x(p_1)$: \begin{align}
S_{\mathrm{bulk}} &= - \frac{\alpha_Z L^{\frac{D-1}{2}}}{e^2}  \int \mathrm{d}^{D+1}x \sqrt{g} A_x(p_1) g^{xx}g^{ab} \partial_b A_x(p_2) \partial_a \Phi(p_3) \notag \\
&\;\;\;\; - \frac{\alpha_Z L^{\frac{D-1}{2}}}{e^2}  \int \mathrm{d}^{D+1}x A_x(p_1) \Phi(p_3) \partial_a \left[\sqrt{g}g^{ab}g^{xx}\partial_b A_x(p_2)\right]. \label{eq:416}
\end{align}
The second line of the above equation vanishes by (\ref{eq:maxwellads}).   Now, we combine (\ref{eq:416}) with an equivalent equation where we integrate by parts to remove derivatives from $A_x(p_2)$, leading to: \begin{align}
S_{\mathrm{bulk}} &= - \frac{\alpha_Z L^{\frac{D-1}{2}}}{2e^2}  \int \mathrm{d}^{D+1}x \sqrt{g} g^{xx}g^{ab}\partial_a \Phi(p_3)\partial_b (A_x(p_1)A_x(p_2)).
\end{align}
We now integrate by parts to remove all derivatives from $A_x$: \begin{align}
S_{\mathrm{bulk}} &= \frac{\alpha_Z L^{\frac{D-1}{2}}}{2e^2}  \int \mathrm{d}^{D+1}x A_x(p_1)A_x(p_2) g^{xx} \left[ \partial_a \left(\sqrt{g}g^{ab}\partial_b \Phi(p_3)\right)  + \frac{2}{r}\sqrt{g} g^{rr}\partial_r \Phi(p_3) \right] \notag \\
&=  \frac{\alpha_Z L^{\frac{D-1}{2}}}{2e^2}  \int \mathrm{d}^{D+1}x \sqrt{g}A_x(p_1)A_x(p_2) g^{xx} \left[ \frac{\Delta(\Delta-D)}{L^2}\Phi(p_3)+ \frac{2}{r} g^{rr}\partial_r \Phi(p_3)\right],
\end{align}
where we have employed (\ref{eq:freescalar}) in the second step.  Using the derivative identities \begin{subequations}\label{eq:besselders}\begin{align}
\partial_x \left(x^b\mathrm{K}_b(x)\right) &= -x^b \mathrm{K}_{b-1}(x) = 2bx^{b-1}\mathrm{K}_b(x) - x^b \mathrm{K}_{b+1}(x), \\
\partial_x \left(x^b\mathrm{I}_b(x)\right) &= x^b \mathrm{I}_{b-1}(x),
\end{align}\end{subequations}
and the explicit expression (\ref{eq:phip}) for $\Phi(p_3)$,  we obtain \begin{align}
\partial_r \Phi(p_3) = \frac{r^{D-\Delta-1}}{L^{\frac{D-1}{2}}\mathcal{Z}(\Delta-\frac{D}{2})} \left[\Delta  \hat{\mathrm{K}}_{\Delta-\frac{D}{2}}(p_3r) - \hat{\mathrm{K}}_{\Delta-\frac{D}{2}+1}(p_3r)\right]
\end{align}
Hence, we find \begin{align}
S_{\mathrm{bulk}} = -&\delta^D(p_1+p_2+p_3)  \frac{\alpha_Z L^{D-3}}{e^2} \int \mathrm{d}r  \left\lbrace\frac{\hat{\mathrm{K}}_{\frac{D}{2}-1}(p_1r)\hat{\mathrm{K}}_{\frac{D}{2}-1}(p_2r)\hat{\mathrm{K}}_{\Delta -\frac{D}{2} +1 }(p_3r)}{\mathcal{Z}(\frac{D}{2}-1)^2 \mathcal{Z}(\Delta -\frac{D}{2}) r^{\Delta-1} } \right. \notag \\
& \left.+ \frac{\Delta(D-\Delta-2)}{2}\frac{\hat{\mathrm{K}}_{\frac{D}{2}-1}(p_1r)\hat{\mathrm{K}}_{\frac{D}{2}-1}(p_2r)\hat{\mathrm{K}}_{\Delta -\frac{D}{2} }(p_3r)}{\mathcal{Z}(\frac{D}{2}-1)^2 \mathcal{Z}(\Delta -\frac{D}{2}) r^{\Delta-1} } \right\rbrace,
\end{align}
and therefore holography gives (for transverse momenta):
\begin{align}
\langle J^x(p_1)&J^x(p_2)\mathcal{O}(p_3)\rangle_0 =  \frac{\alpha_Z L^{D-3}}{e^2 \mathcal{Z}(\frac{D}{2}-1)^2 \mathcal{Z}(\Delta -\frac{D}{2})} \bigg\lbrace I\left(\tfrac{D}{2},\tfrac{D}{2}-1,\tfrac{D}{2}-1,\Delta-\tfrac{D}{2}+1\right) \notag \\
&+ \frac{\Delta(D-\Delta-2)}{2} I\left(\tfrac{D}{2}-1,\tfrac{D}{2}-1,\tfrac{D}{2}-1,\Delta-\tfrac{D}{2}\right)\bigg\rbrace \delta^D(p_1+p_2+p_3).   \label{eq:42Axfin} 
\end{align}
Comparing with (\ref{eq:JJOformal}), we find complete agreement in the functional form, and fix the coefficient \begin{equation}
A_{JJ\mathcal{O}} =  \frac{\alpha_Z L^{D-3}}{e^2 \mathcal{Z}(\frac{D}{2}-1)^2 \mathcal{Z}(\Delta -\frac{D}{2})}. \label{eq:AJJOholography}
\end{equation}

\subsubsection{$A_{TT\mathcal{O}}$}
The computation of $A_{TT\mathcal{O}}$ proceeds using similar tricks to the computation of $A_{JJ\mathcal{O}}$.   However, due to the fact that the relevant correction to the bulk action is much higher derivative, this computation is much more technically challenging.    We leave the details of this computation to Appendices \ref{app:weyl} and \ref{app:ATTO}, and here only quote the main result: \begin{equation}
A_{TT\mathcal{O}} = -\frac{L^{D-1}}{2\kappa^2} \times \frac{16 \alpha_Y}{\mathcal{Z}(\frac{D}{2})^2 \mathcal{Z}(\Delta-\frac{D}{2})}.  \label{eq:ATTOhol}
\end{equation}

\section{Holography:  High Frequency Asymptotics}\labell{sec4}
In this section, we reproduce (\ref{eq:mainres}) directly from holography.   The derivation presented in Section \ref{sec2} does not rely on any ``matrix large $N$" limit, so it is natural to question what we can learn from the holographic derivation.   The key advantage of our holographic formulation is that we will be able to derive the full asymptotic behavior of (\ref{eq:mainres}) non-perturbatively in $h$.   Even if the true ground state of the theory at finite $h$ and $T$ is far from a CFT,  the existence of a UV CFT is sufficient to impose (\ref{eq:mainres}).    Holography geometrizes this intuition in a very natural way -- we will see that (\ref{eq:mainres}) is universally independent of the low-energy details of the theory in these holographic models.

\subsection{Scalar Two-Point Functions}\labell{sec:41}
As in Section \ref{sec2},  we begin by studying $\langle \Oo(\Omega)\Oo(-\Omega)\rangle$.   To do so, we must perturb the field $\psi$ by a source at frequency $\Omega$ and compute the response, using (\ref{eq:Ooasym}) and (\ref{eq:sec32pt}).   We will begin by assuming that $\psi=0$ identically on the background geometry.  In this case, $\psi$ is the only field which will be perturbed in linear response, as there is no linear in $\psi$ term in (\ref{eq:holomodel}).   We will relax this assumption at the end of the derivation.   We also assume that $\Delta - D/2$ is not an integer:  this assumption is relaxed in Appendix \ref{app:integer}. For now, we must solve the differential equation \begin{equation}
\nabla_a \nabla^a \psi = W(\Phi)\psi,
\end{equation}
which leads to \begin{equation}
\frac{1}{\sqrt{g}} \partial_r\left(\sqrt{g}g^{rr} \partial_r\psi\right) - \Omega^2 g^{tt} \psi  = \left(\frac{\Do(\Do-D)}{L^2}  +  \alpha_W L^{\frac{D-5}{2}}  \Phi(r) \cdots\right)\psi  \label{eq:scalar41}
\end{equation}
In general, the solution of this equation for all $\Omega$ requires knowledge of the full bulk geometry.  

However, if we restrict ourselves to studying asymptotic behavior at large $\Omega$, a great simplification occurs.   To see this, it is helpful to rescale the radial coordinate to \begin{equation}
R \equiv \Omega r.  \label{eq:r2R}
\end{equation}
We claim that the function $\psi(R)$ is essentially non-vanishing when $R\sim 1$.   To see why, note that (\ref{eq:scalar41}) becomes 
\begin{align}
&\frac{\Omega^2}{\sqrt{g}} \partial_R\! \left(\sqrt{g}\, g^{rr}\, \partial_R\psi \right) - \Omega^2 g^{tt}\psi = \left(\frac{\Do(\Do-D)}{L^2}  +  \alpha_W L^{\frac{D-5}{2}}  \Phi \cdots\right)\psi.  \label{eq:scalar412}
\end{align}
If this equation is only non-trivial when $R\sim 1$, then perturbation theory about $\Omega=\infty$ should be well-defined. 
Making use of the expansion for $\Phi$ (\ref{eq:phinormalization}), the right hand side of (\ref{eq:scalar412}) becomes 
\begin{equation} 
\left(\frac{\Do(\Do-D)}{L^2}  +  \frac{\alpha_W}{L^2}  \left(\frac{h}{\Omega^{D-\Delta}}R^{D-\Delta}+\frac{\langle \mathcal{O}\rangle}{(2\Delta-D)\Omega^\Delta}R^\Delta\right)+ \cdots\right)\psi\label{eq:scalar41RHS}
\end{equation}
and so as $\Omega \rightarrow \infty$, we may neglect the scalar corrections to the equations of motion.   The $\Omega \rightarrow \infty$ limit is given by the $r\rightarrow 0$ limit (keeping $R$ fixed), and so we may approximate the metric at leading order by (\ref{eq:adsmetric}).  The left hand side of (\ref{eq:scalar412}) is hence \begin{equation}
\Omega^2 \left(\frac{R}{\Omega L}\right)^{D+1} \partial_R \left(\left(\frac{R}{\Omega L}\right)^{1-D}\partial_R\psi\right) - \Omega^2 \left(\frac{R}{\Omega L}\right)^2\psi  = \frac{R^2 \partial_R^2 \psi + (1-D)R\partial_R\psi - R^2 \psi }{L^2}  \label{eq:scalar41LHS}
\end{equation}
Corrections due to the deviation of the IR geometry from pure AdS (both from finite $T$ and finite $h$) are subleading for the same reason that $W(\Phi)$ corrections are subleading. Indeed, comparing (\ref{eq:scalar41RHS})  and (\ref{eq:scalar41LHS}), we see that perturbation theory about $\Omega = \infty$ is well-controlled.   Changing variables to \begin{equation}
\chi(R) \equiv R^{\Do-D} \psi(R),
\end{equation}
and expanding $\chi$ as a perturbative expansion in small $\alpha_W$
\begin{equation}
\chi = \chi\up0 + \alpha_W \chi\up1 + \cdots\,.
\end{equation}
equation \reef{eq:scalar412} reduces to \begin{equation}
R^{D+2 - \Do} \left[\partial_R^2 \chi\up0 + \frac{D+1-2\Do}{R} \partial_R \chi\up0 - \chi\up0 \right] = 0.  \label{eq:scalar41lead}
\end{equation}

The regular solution to equation, conveniently normalized so that $\chi(0)=1$, is \begin{equation}
\chi\up0(R) = \frac{R^{b} \mathrm{K}_{b}(R)}{\mathcal{Z}(b)},  \label{eq:scalar41leadS}
\end{equation}
where \begin{equation}
b \equiv \Delta_0 - \frac{D}{2}.
\end{equation}
We begin by assuming that $b>0$, which means that the $\mathrm{O}(R^0)$ term in $\chi$ describes the source, and the $\mathrm{O}(R^{2b})$ term in $\chi$ describes the response.  We will discuss the case of $b<0$ in Appendix \ref{app:analcont}.

At large but finite $\Omega$, we must correct (\ref{eq:scalar41lead}).   The dominant contribution comes from the scalar field corrections in (\ref{eq:scalar41RHS}), and so (\ref{eq:scalar41lead}) becomes \begin{equation}
\partial_R^2 \chi\up1 + \frac{1-2b}{R} \partial_R \chi\up1 - \chi\up1 \approx \frac{ \alpha_W }{R^2} \left(\frac{h}{\Omega^{D-\Delta}}R^{D-\Delta}+\frac{\langle \mathcal{O}\rangle}{(2\Delta-D)\Omega^\Delta}R^\Delta\right) \chi\up0.  \label{eq:scalar413}
\end{equation}
Since $\Omega$ is large, the right hand side is perturbatively small and we may correct the leading order solution (\ref{eq:scalar41leadS}) perturbatively.   Keeping the boundary conditions $\chi(0)=1$ and $\chi(\infty)=0$ fixed, the perturbative correction to (\ref{eq:scalar41leadS}) is unique, and it is straightforward to extract $\langle \Oo\Oo\rangle$.   The perturbative computation of the finite $\Omega$ corrections to this correlator proceed in a few simple mathematical steps.

First, we construct the Green's function $G(R;R_0)$ for this differential equation: \begin{equation}
\partial_R^2 G(R;R_0) + \frac{1-2b}{R} \partial_R G(R;R_0) - G(R;R_0) = -\delta(R-R_0).
\end{equation}
Employing Bessel function identities including (\ref{eq:besselders}) and \cite{abramowitz}\begin{equation}
\mathrm{I}_b(x)\mathrm{K}_{b-1}(x) + \mathrm{K}_b(x)\mathrm{I}_{b-1}(x) = \frac{1}{x},
\end{equation} and our boundary conditions that $G(0;R_0) = G(\infty;R_0)=0$, we find: \begin{equation}
G(R;R_0) = \left\lbrace\begin{array}{ll} R_0 \mathrm{K}_b(R_0)\mathrm{I}_b(R) (R/R_0)^b &\ R<R_0 \\   R_0 \mathrm{I}_b(R_0)\mathrm{K}_b(R) (R/R_0)^b &\ R>R_0 \end{array}\right..   \label{eq:green41}
\end{equation}
Using this Green's function, we can readily construct the solution to the differential equation \begin{equation}
\partial_R^2 \chi + \frac{1-2b}{R} \partial_R \chi - \chi = -\frac{c}{\mathcal{Z}(b)}R^{a+b}\mathrm{K}_b(R),
\end{equation}
which is: \begin{equation}
\chi = \frac{c}{\mathcal{Z}(b)}\int\limits_0^\infty \mathrm{d}R_0 \; G(R;R_0) R_0^{a+b}\mathrm{K}_b(R_0).
\end{equation}  
Keeping in mind our holographic application, we need only extract the $\mathrm{O}(R^{\Do})$ contribution to $\psi$, or equivalently the $\mathrm{O}(R^{2b})$ contribution to $\chi$.    We now assume that $a+2>2b$ for simplicity -- the opposite case is studied in Appendix \ref{app:analcont}.  In this case, the leading order term as $R\rightarrow 0$ of the perturbation to $\chi$ is $\mathrm{O}(R^{2b})$:  \begin{equation}
\chi\up1(R\rightarrow 0) \approx \frac{c}{\mathcal{Z}(b)} R^b \mathrm{I}_b(R) \int\limits_0^\infty \mathrm{d}R_0 \; R^{1+a} \mathrm{K}_b(R)^2 = \frac{c}{\mathcal{Z}(b)} R^b \mathrm{I}_b(R) \Psi(a+2;b).  \label{eq:scalarint}
\end{equation}

We now employ (\ref{eq:scalarint}), using \begin{equation}
(a,c) = \left(D-\Delta-2,-\alpha_Wh\right)\text{  or  }   \left(\Delta-2,-\frac{\alpha_W\langle \mathcal{O}\rangle }{2\Delta-D}\right),
\end{equation}
to write down the leading order corrections to (\ref{eq:scalar413}):  \begin{align}
\chi(R) = 1 + &R^{2b}\left[\frac{\mathcal{C}_{\Oo\Oo}}{2\Do-D}- \frac{\alpha_W h}{2^{2b-1}\Gamma(b)\Gamma(1+b)\Omega^{D-\Delta}}\Psi(D-\Delta; b) \right. \notag \\ 
&\left.- \frac{\alpha_W\langle \mathcal{O}\rangle }{2^{2b-1}\Gamma(b)\Gamma(1+b)(2\Delta-D)\Omega^\Delta}\Psi(\Delta;b)\right] + \cdots
\end{align}
where $\mathcal{C}_{\Oo\Oo}$ is the holographic normalization of the operator $\Oo$, defined analogously to $\mathcal{C}_{\mathcal{OO}}$ in (\ref{eq:COOholography}).     Hence, our explicit non-perturbative calculation gives  
\begin{align}
\langle \Oo(-\Omega)\Oo(\Omega)\rangle &= \Omega^{2\Do-D} \left[\mathcal{C}_{\Oo\Oo} - \frac{\alpha_W \Psi(D-\Delta;\Do-\frac{D}{2})}{2^{2\Do - D-2}\Gamma(\Do-\frac{D}{2})^2  }  \,\frac{h}{\Omega^{D-\Delta}} \right. \notag \\ 
&\left. - \frac{\alpha_W \Psi(\Delta;\Do-\frac{D}{2})}{2^{2\Do - D-2}\Gamma(\Do-\frac{D}{2})^2 (2\Delta-D)} \,\frac{\langle \mathcal{O}\rangle}{\Omega^\Delta} \right] + \cdots .\label{eq:51final1}
\end{align} 
%
It is simple to compare to the predictions of conformal perturbation theory.   Combining (\ref{eq:21A}), (\ref{eq:21B}), (\ref{eq:COOholography}) and (\ref{eq:AOOOholography}) and simplifying, we indeed obtain (\ref{eq:51final1}). 

\subsubsection{Subleading Orders in the Expansion}
In the above derivation, we assumed that the geometry was AdS.   In fact, the metric is not quite AdS in the presence of a background $\psi$ field:  away from $r=0$,  there will be deformations of the form \begin{equation}
\mathrm{d}s^2 = \frac{L^2}{r^2}\left(\mathrm{d}r^2 + \mathrm{d}t_{\textsc{e}}^2 + \mathrm{d}\mathbf{x}^2\right) + \mathrm{O}\left(r^{2\Delta-2}, r^{2(D-\Delta)-2}, r^{D-2}\right).  \label{eq:adsmetriccor}
\end{equation}
Because Einstein's equations are sourced by $g^{ab}\psi^2$ and $(\partial \psi)^2$, the leading order terms in the asymptotic expansion of (\ref{eq:phinormalization}) imply the first subleading corrections in (\ref{eq:adsmetriccor}).   These subleading corrections are small as $r\rightarrow0$ (or for $R\sim 1$ as $\Omega \rightarrow \infty$), and so they will lead to perturbative corrections to $\langle \Oo\Oo\rangle$ of the form \begin{equation}
\langle \Oo(-\Omega)\Oo(\Omega)\rangle = \Omega^{2\Do-D}\left[\mathcal{C}_{\Oo\Oo} + \cdots +  \# \frac{ h^2}{\Omega^{2(D-\Delta)}}+  \# \frac{ \langle \mathcal{O}\rangle^2}{\Omega^{2\Delta}}+  \# \frac{ h\langle \mathcal{O}\rangle}{\Omega^{D}}+\cdots\right].   \label{eq:sec411}
\end{equation}
The last most term in (\ref{eq:sec411}), which leads to $\Omega^{-D}$ corrections to the correlation function, can likely be interpreted as corrections to $\langle T_{\mu\nu}\rangle$.   Schematically, one finds  \cite{katz}\begin{equation}
\langle \Oo(-\Omega)\Oo(\Omega)\rangle = \Omega^{2\Do-D}\left[\mathcal{C}_{\Oo\Oo} + \cdots 
+  \mathcal{C}_{\Oo\Oo T}^{\mu\nu}\frac{ \langle T_{\mu\nu}\rangle }{\Omega^D}+\cdots\right].  \label{eq:TmunuOPE}
\end{equation}
The logic for such terms follows analogously to the appearance of the $\mathcal{B}$-term in (\ref{eq:mainres}) -- the stress tensor will appear in the $\Oo\Oo$ OPE, and so terms proportional to pressure and energy density will appear in the asymptotic expansion of generic correlation functions. For instance, this was noted for the conductivity \cite{sum-rules, katz} and shear viscosity \cite{Son09,caron09,sum-rules,willprl}.  

There are other types of corrections that can arise,  which are rather straightforward.   In all cases, the asymptotic expansion of the correlation function in powers of $1/\Omega$ is cleanly organized by the behavior of the bulk fields near the AdS boundary.   The perturbative derivation that we described earlier straightforwardly accounts for all such corrections.

\subsection{Conductivity}\label{conductivity}
Next, we compute the asymptotics of $\sigma(\Omega)$ at large $\Omega$.   The structure of the computation is very similar to the computation of the asymptotics of $\langle \Oo(\Omega)\Oo(-\Omega)\rangle$.   We compute within linear response the fluctuations of the gauge field $A_x$ and employ the asymptotic expansion (\ref{eq:Ooasym}) to compute $\langle J_x\rangle / \Omega A_x^0$.   Since our background is uncharged,  if we turn on a source for $A_x$, no other bulk fields will become excited (this is a consequence of the fact that $A_x$ is a spin 1 mode under the spatial $\mathrm{SO}(D-1)$ isotropy).  The equations of motion then reduce to \begin{equation}
0 = \frac{1}{\sqrt{g}} \partial_a \left(\sqrt{g} Z(\Phi) F^{ax}\right) = \frac{1}{\sqrt{g}}\partial_r \left(\sqrt{g}Z(\Phi)g^{rr}g^{xx} \partial_r A_x\right) - \Omega^2 Z(\Phi) g^{tt}g^{xx} A_x.  \label{eq:maxwell}
\end{equation}
Upon the variable change (\ref{eq:r2R}), and keeping only finite $\Omega$ corrections from the scalar field $\Phi$ (for the same reasons as we mentioned in the previous section), we find that the leading order corrections to (\ref{eq:maxwell}) at large but finite $\Omega$ can be found by perturbatively solving  

\begin{equation}
\partial_R^2 A_x + \frac{3-D}{R}A_x - A_x \approx - \partial_R\mathfrak{a}(R)\, \partial_R A_x \label{eq:Ax42}
\end{equation}
where \begin{equation}
\mathfrak{a}(R) = \alpha_Z \left[\frac{h}{\Omega^{D-\Delta}} R^{D-\Delta} + \frac{\langle \mathcal{O}\rangle}{(2\Delta-D)\Omega^\Delta}R^\Delta\right].  \label{eq:fraka0}
\end{equation}

We see that this differential equation is extremely similar to the one studied in the previous subsection, upon replacing $\chi$ with $A_x$ and setting (for the remainder of this section)\begin{equation}
b=\frac{D}{2}-1.   \label{eq:b42}
\end{equation}
The zeroth-order solution at $\Omega =\infty$ is given by (\ref{eq:scalar41leadS}), and the Green's function for the left hand side of (\ref{eq:Ax42}) is given by (\ref{eq:green41}).   The only difference is that the source, which we must integrate over to recover the subleading behavior of $A_x$, is a bit more complicated.   Again, we begin by assuming that \begin{equation}
\mathfrak{a}(R) = cR^a.  \label{eq:fraka}
\end{equation}
The coefficient is chosen to simplify the equations in the remainder of the paragraph.  The boundary conditions on $A_x$ are $A_x(0)=1$, $A_x(\infty)=0$.   At first order in $c$, we find \begin{equation}
A_x^{(1)} = \frac{c}{\mathcal{Z}(b)} \int\limits_0^\infty\mathrm{d}R_0\;  G(R;R_0) \left[ R_0^{2b-1}\partial_{R_0}\left(R_0^{1+a-2b}\partial_{R_0} \hat{\mathrm{K}}_b(R_0)\right) - R_0^a \hat{\mathrm{K}}_b(R_0)\right].  \label{eq:42eq2}
\end{equation}
Now, using \begin{equation}
\partial_x^2 \hat{\mathrm{K}}_b(x) = \frac{2b-1}{x}\partial_x \hat{\mathrm{K}}_b(x) + \hat{\mathrm{K}}_b(x), \label{eq:besselhatid}
\end{equation}
we may simplify the integral in square brackets in (\ref{eq:42eq2}) to \begin{equation}
A_x^{(1)} =\frac{c}{\mathcal{Z}(b)} \int\limits_0^\infty\mathrm{d}R_0\;  G(R;R_0) \left[ a R_0^{a-1} \partial_{R_0} \hat{\mathrm{K}}_b(R_0) \right], 
\end{equation}
As in the subsection above after (\ref{eq:scalarint}), let us assume that $a$ is large enough that the following integration by parts manipulations are acceptable:  \begin{align}
\int\limits_0^\infty \mathrm{d}R_0\; \hat{\mathrm{K}}_b R_0^{a-2b} \partial_{R_0} \hat{\mathrm{K}}_b(R_0) &= \int\limits_0^\infty \mathrm{d}R_0\; R_0^{a-2b} \frac{1}{2}\partial_{R_0} \left(\hat{\mathrm{K}}_b(R_0)\right)^2 \notag \\
&\approx \left(b-\frac{a}{2}\right) \int\limits_0^\infty \mathrm{d}R_0 \; R_0^{a-1-2b} \hat{\mathrm{K}}_b(R_0)^2 \notag \\
&=  \left(b-\frac{a}{2}\right) \Psi(a;b).   \label{eq:besselibp}
\end{align}

Hence, we find (for a single power $a$) \begin{align}
A_x^{(1)}(R\rightarrow 0) \approx \frac{c}{\mathcal{Z}(b)}\mathrm{I}_b(R)R^b \times a\left(b-\frac{a}{2}\right)\Psi(a;b).   \label{eq:Ax42single}
\end{align}
Keeping in mind that the true source is (\ref{eq:fraka0}), we conclude that \begin{align}
A_x(R\rightarrow 0) &\approx \frac{\hat{\mathrm{K}}_b(R)}{\mathcal{Z}(b)} - \frac{R^{2b}}{2b} \frac{\alpha_Z h}{\Omega^{D-\Delta}} \frac{D-\Delta}{\mathcal{Z}(b)^2}\left(1-\frac{\Delta}{2}\right)\Psi(D-\Delta;b) \notag \\
&-  \frac{R^{2b}}{2b} \frac{\alpha_Z \langle \mathcal{O}\rangle}{(2\Delta-D)\Omega^{\Delta}} \frac{\Delta}{\mathcal{Z}(b)^2}\left(1-\frac{D-\Delta}{2}\right)\Psi(\Delta;b). \label{eq:Ax42single2}
\end{align} 
Hence,   
\begin{align}
\sigma(\mathrm{i} \Omega) &= \frac{L^{D-3}}{e^2}\Omega^{D-3}\left[\sigma_\infty - \frac{\alpha_Z h}{\Omega^{D-\Delta}} \frac{D-\Delta}{\mathcal{Z}(\frac{D}{2}-1)^2}\left(1-\frac{\Delta}{2}\right)\Psi\left(D-\Delta;\tfrac{D}{2}-1\right) \right. \notag \\
&-\left. \frac{\alpha_Z \langle \mathcal{O}\rangle}{(2\Delta-D)\Omega^{\Delta}} \frac{\Delta}{\mathcal{Z}(\frac{D}{2}-1)^2}\left(1-\frac{D-\Delta}{2}\right)\Psi\left(\Delta;\tfrac{D}{2}-1\right) \right].  \label{eq:sigmaholfinal}
\end{align}
It is simple to compare to the predictions of conformal perturbation theory.   Combining (\ref{eq:ABsigma}), (\ref{eq:COOholography}) and (\ref{eq:AJJOholography}) and simplifying, we indeed obtain (\ref{eq:sigmaholfinal}). 

\subsection{Viscosity}
The computation of the viscosity proceeds similarly to the previous two subsections.   The $\mathrm{SO}(D-1)$ isotropy of the background implies that $g_{xy}$ will not couple to any other modes.  We hence write the perturbation to the metric as \begin{equation}
\delta (\mathrm{d}s^2) = 2 \frac{L^2}{r^2}h_{xy} \mathrm{d}x\mathrm{d}y,  \label{eq:hxydef}
\end{equation}   
and find the differential equation governing $h_{xy}$.   We do this computation in Appendix \ref{app:weyl}, and here quote the leading order answer (at finite $\Omega$):
\begin{align}
&\left(\frac{L}{r}\right)^{D-1}\left[\partial_r^2 h_{xy} - \frac{D-1}{r}\partial_r h_{xy} - r^2\Omega^2 h_{xy}\right]\notag\\
=& \frac{8}{D-1} \partial_r \left[\left(\frac{L}{r}\right)^{D-3}Y(\Phi) 2(D-1)\Omega^2 \partial_r h_{xy}\right]   \notag \\
&- \frac{8}{D-1}\left(\frac{L}{r}\right)^{D-3}Y(\Phi) \left((D-2)\Omega^4 h_{xy} + \Omega^2 \partial_r^2 h_{xy}\right)  \notag \\
&+ \frac{8}{D-1} \partial_r^2 \left[\left(\frac{L}{r}\right)^{D-3}Y(\Phi) \left(\Omega^2 h_{xy} + (D-2) \partial_r^2 h_{xy}\right) \right]. \label{eq:weyleq}
\end{align}
After the variable change (\ref{eq:r2R}), we obtain \begin{align} 
\partial_R^2 h_{xy} &- \frac{D-1}{R}\partial_R h_{xy} - R^2 h_{xy}= \frac{8}{D-1}\left(\frac{L}{R}\right)^{-2}Y(\Phi) \left((D-2) h_{xy} +  \partial_R^2 h_{xy}\right) \notag \\
&- \frac{8}{D-1} \left(\frac{R}{L}\right)^{D-1} \partial_R \left[\left(\frac{L}{R}\right)^{D-3}Y(\Phi) 2(D-1) \partial_R h_{xy}\right]  \notag \\
&+ \frac{8}{D-1}\left(\frac{R}{L}\right)^{D-1}  \partial_R^2 \left[\left(\frac{L}{R}\right)^{D-3}Y(\Phi) \left(h_{xy} + (D-2) \partial_R^2 h_{xy}\right) \right]. \label{eq:weyleq2}
\end{align}
We now proceed similarly to the previous subsection.   When $\Omega \rightarrow \infty$, $Y$ vanishes and (\ref{eq:weyleq2}) is solved by (\ref{eq:scalar41leadS}) with \begin{equation}
b=\frac{D}{2}.
\end{equation}
Since $Y(\Phi)$ will be a sum of two power laws in $R$,  to first order in perturbation theory we can determine $h_{xy}$ by solving (\ref{eq:weyleq2}) after replacing $Y(\Phi)$ with $L^2 \mathfrak{a}(R)$, with $\mathfrak{a}(R)$ as defined in (\ref{eq:fraka}).    At first order in $c$, assuming convergence of integrals, we find \begin{align}
h_{xy}(R\rightarrow 0) &\approx \frac{-8c}{D-1} \mathrm{I}_b(R)R^b \int \mathrm{d}R_0 \hat{\mathrm{K}}_b(R_0) \left[R_0^{a+3-2b}\left((D-2)\hat{\mathrm{K}}_b(R_0) + \partial_{R_0}^2\hat{\mathrm{K}}_b(R)\right) \right. \notag \\
&-\left. 2(D-1) \partial_{R_0} \left(R_0^{a+3-2b}\partial_{R_0}\hat{\mathrm{K}}_b(R)\right) + \partial_{R_0}^2 \right. \notag \\
&+\left. \partial^2_{R_0}\left(R_0^{a+3-2b}\left(\hat{\mathrm{K}}_b(R_0) +(D-2) \partial_{R_0}^2\hat{\mathrm{K}}_b(R)\right)\right) \right].
\end{align}
Repeated application of the identities (\ref{eq:Psidef}), (\ref{eq:scalarint}), (\ref{eq:besselhatid}) and (\ref{eq:besselibp}) leads to \begin{align}
h_{xy}(R\rightarrow 0) &= -8c \mathrm{I}_b(R)R^b \left[ a(a+1)\Psi(a+2;b) + \frac{a}{2}(2b-a)(2b-2)(a+1) \Psi(a;b)\right] \notag \\
&= -8c \mathrm{I}_b(R)R^b \times \frac{a}{4}(2b-a)(a+2)(2b-a-2) \Psi(a;b) 
\end{align} 
Now, we have \begin{equation}
(a,c) = \left(D-\Delta, \alpha_Y h\right)\text{ or} \left(\Delta,\frac{\alpha_Y\langle \mathcal{O}\rangle}{2\Delta-D}\right),
\end{equation}
so we conclude that \begin{align}
\eta &=  \frac{L^{D-1}}{2\kappa^2}\Omega^{D-1}\left[\mathcal{C}_{TT} - \frac{2\alpha_Y h}{\mathcal{Z}(\frac{D}{2})^2} \Delta(D-\Delta)(D-\Delta+2)(\Delta-2) \Psi\left(D-\Delta;\frac{D}{2}\right) \frac{1}{\Omega^{D-\Delta}} \right. \notag \\
&\left. -  \frac{2\alpha_Y \langle \mathcal{O}\rangle }{(2\Delta-D)\mathcal{Z}(\frac{D}{2})^2} \Delta(D-\Delta)(\Delta+2)(D-\Delta-2) \Psi\left(\Delta;\frac{D}{2}\right) \frac{1}{\Omega^\Delta}\right]
\end{align}
Comparing this equation to (\ref{eq:ABvisc}), (\ref{eq:COOholography}) and (\ref{eq:ATTOhol}), we again find that our holographic answer agrees with conformal perturbation theory.

\section{Holography: Full Frequency Response}\labell{sec5}
A natural advantage that holography provides to quantum field theories is the capability of directly exploring the real time response functions at all frequencies. In this section, we investigate the holographic linear response to scalar deformations of the CFT.  Following \cite{Myers:2016wsu}, we account for the thermal expectation values $\langle \mcO \rangle_T \sim T^\Delta$ that generally arise in CFTs. Using this model, we can solve the equations of motion for the dynamical fields at all frequencies and calculate the response functions studied in the previous sections. 
\subsection{A Minimal Model}\labell{sec51}
We will study holographic models of the form (\ref{eq:holomodel}), for simple choices of $W$, $V$, $Z$ and $Y$.    Essentially, we will truncate the asymptotic expansions of these $\Phi$-dependent functions at lowest non-trivial order:    \begin{subequations} \labell{eq:potentials}\begin{align}
V(\Phi) = & -\frac{D(D-1)}{2L^2\kappa^2} + \frac{\Delta(D-\Delta)}{2L^2}\Phi^2\\
Z(\Phi) = & 1 + \alpha_Z L^{\frac{D-1}{2}}\Phi\\
Y(\Phi) = & \alpha_Y L^{\frac{D+3}{2}}\Phi\\
W(\Phi) = & \frac{\Delta_0(D-\Delta_0)}{L^2} + \alpha_W L^{\frac{D-5}{2}} \Phi
\end{align}
\end{subequations} 
Using these simple couplings, we will study the finite temperature response both for $\omega \gg T$ and $\omega \lesssim T$.      

Assuming that $\kappa \rightarrow 0$, we may solve for the background geometry without considering fluctuations of the matter content:  $A_\mu$, $\Phi$ or $\psi$.   We will focus on the planar AdS-Schwarzchild black hole solution\footnote{We consider other background metrics in Section \ref{sec:RN}.}  to the resulting equations of motion, whose real-time metric reads \cite{lucasreview}:
\begin{equation}
\mathrm{d}s^2 = \left(\frac{4\pi T}{D}\right)^2 \frac{L^2}{u^2}(-f(u)\mathrm{d}t^2 + \eta_{ij}\mathrm{d}x^i\mathrm{d}x^j) + \frac{L^2\mathrm{d}u^2}{u^2 f(u)}\labell{eq:metric},
\end{equation}
where $f(u) = 1-u^D$ and $T$ is the Hawking temperature of the black hole, as well as the temperature of the dual field theory.  The coordinate $u$ is dimensionless:   $u=0$ corresponds to the AdS boundary, while $u=1$ corresponds to the black hole horizon.   In the asymptotically AdS regime ($u\rightarrow 0$), the coordinate $u$ is a simple rescaling of the coordinate $r$ in (\ref{eq:adsmetric}):   $u= 4\pi Tr/D$.

We now study the perturbative fluctuations of the matter fields $\Phi$, $A_\mu$ and $\psi$ about this background solution.  Varying \reef{eq:holomodel} with respect to $\Phi$, we obtain:
\begin{equation}
0 = (\nabla^2 - m^2)\Phi + \frac{\alpha_YL^{\frac{D+3}{2}}}{\kappa^2} C_{abcd}C^{abcd}. \labell{eq:scalareom}
\end{equation}
Because the black hole background \reef{eq:metric} is translation invariant in the boundary directions,   $C_{abcd}C^{abcd}$ depends only on $u$,  and thus we may look for a solution of the form $\Phi(u)$.   The resulting ordinary differential equation is 
\begin{equation}
0 = \Phi''(u) + \left(\!\frac{f'(u)}{f(u)} \!-\! \frac{D-1}{u}\!\right)\!\Phi'(u) + \frac{\Delta(D-\Delta)}{u^2 f(u)}\Phi(u) + \frac{D(D-1)^2(D-2)\alpha_Y\, L^{\frac{D-1}{2}} \, u^{2D-2}}{\kappa^2f(u)}\labell{scalar ode}
\end{equation}
This equation can be solved using standard Green's function techniques analogous to the previous section.  We find \begin{equation}
\begin{split}
\Phi(u) & = \hyperf{\tfrac\Delta D}{\tfrac\Delta D}{\tfrac{2\Delta} D}{u^D}\ \left(\Phi_1 - \frac{D(D-1)^2(D-2)\alpha_Y L^{\frac{D-1}{2}}}{(2\Delta-D)\kappa^2} g_\Delta(u)\right)\,u^\Delta\\
+\ \  & \!\!\hyperf{\tfrac{D-\Delta} D}{\tfrac{D-\Delta} D}{2 \tfrac{D-\Delta} D}{u^D} \ \left(\Phi_0 + \frac{D(D-1)^2(D-2)\alpha_Y L^{\frac{D-1}{2}}}{(2\Delta-D)\kappa^2} h_\Delta(u)\right)\,u^{D-\Delta}\,, \labell{eq:solution}
\end{split} 
\end{equation}
where $\Phi_0$ and $\Phi_1$ are integration constants, $\hyperf{z_1}{z_2}{z_3}{z_4}$ denotes the standard hypergeometric function, and  $g_\Delta(u)$ and $h_\Delta(u)$ are dimensionless functions given by\footnote{This representation of the solution is only valid for $\Delta<2D$. In particular, the integral defining $g_\Delta(u)$ in \reef{eq:solute} diverges for $\Delta\ge 2D$. Further, the two independent solutions presented in \reef{eq:solution} are actually identical for  $\Delta=D/2$. Of course, the coefficients of $g_\Delta(u)$ and $h_\Delta(u)$ also diverge for this particular value of $\Delta$. However, we note that the conductivity is still a smooth function of $\Delta$ at this special value and so where results are presented for $\Delta=D/2$, we have added a small positive number to the scaling dimension:  $\Delta = D/2 + \epsilon$ where $\epsilon\sim 10^{-7}$. \labell{footy7}}
\begin{subequations}\labell{eq:solute}\begin{align}
g_\Delta(u) = & \int\limits_0^{u} \mathrm{d}y\,y^{2D-1-\Delta}\  \hyperf{\tfrac{D-\Delta} D}{\tfrac{D-\Delta} D}{2\tfrac{D-\Delta} D}{y^D} \,, \\
h_\Delta(u) = & \int\limits_0^{u} \mathrm{d}y\,y^{D-1+\Delta}\ \hyperf{\tfrac\Delta D}{\tfrac\Delta D}{\tfrac{2\Delta} D}{y^D} \,. 
\end{align}
\end{subequations}
Note that $g_\Delta(0)=h_\Delta(0)=0$.   Hence, at the AdS boundary, the asymptotic behavior of $\Phi(u)$ is
\begin{equation}
\phi(u) = \Phi_0 u^{D-\Delta}\Big(1 + \mathrm{O}(u^D)\Big) + \Phi_1 u^\Delta \Big(1+\mathrm{O}(u^D)\Big)\,. \labell{boots}
\end{equation}
$\Phi_0$ and $\Phi_1$ encode the source $h$ and response $\langle \mathcal{O}\rangle$ respectively, as we discussed previously in (\ref{eq:Ooasym}).     However, since we are using the dimensionless radial coordinate $u$ here, it is important to note that there will be additional powers of $T$ relating the physical source and response with the coefficients $\Phi_{0,1}$:  \begin{equation}
\Phi_1 = \frac{\langle \mcO \rangle }{(2\Delta - D)L^{\frac{D-1}{2}}}\left(\frac{D}{4\pi T}\right)^\Delta\ ;\quad \Phi_0 = \frac{h}{L^{\frac{D-1}{2}}}\left(\frac{D}{4\pi T}\right)^{D-\Delta}\,.
\end{equation}
The integration constant $\Phi_1$ is fixed by demanding regularity at the black hole horizon: 
\begin{equation}
\begin{split}
\Phi_1 = & - \Phi_0 \frac{\Gamma\left(2-\frac{2\Delta}{D}\right)\Gamma\left(\frac{\Delta}{D}\right)^2}{\Gamma\left(1-\frac{\Delta}{D}\right)^2\Gamma\left(\frac{2\Delta}{D}\right)} \\
& + \frac{\alpha_Y L^{\frac{D-1}{2}}D(D-1)^2(D-2)}{\kappa^2(2\Delta - D)} \left( g_\Delta(1) - \frac{\Gamma\left(2-\frac{2\Delta}{D}\right)\Gamma\left(\frac{\Delta}{D}\right)^2}{\Gamma\left(1-\frac{\Delta}{D}\right)^2\Gamma\left(\frac{2\Delta}{D}\right)}h_\Delta(1)\right)\,.  \label{eq:Phi1}
\end{split}
\end{equation}
Note that $g_\Delta(1)$ and $h_\Delta(1)$ are finite and can be determined numerically.   Some sample plots of $\Phi(u)$, upon setting $h=0$, are given in Figure \ref{scalar plots}.
\begin{figure}\centering
	\includegraphics[width = 0.6\textwidth]{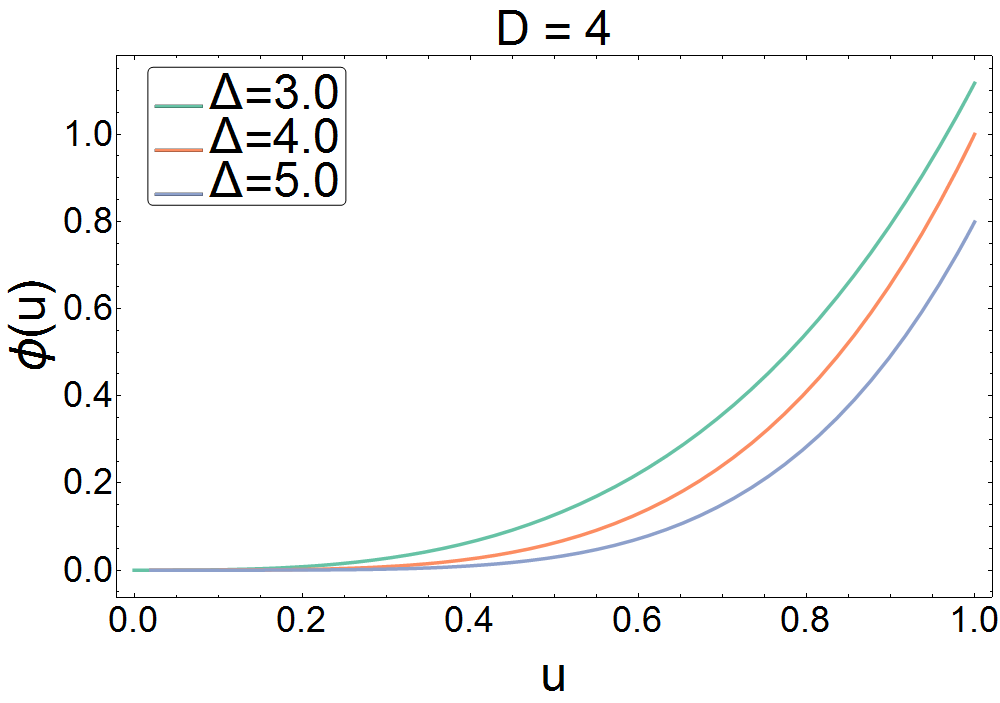}  
	\caption{The scalar field with $D=4$ as a function of $u$ for various scaling dimensions $\Delta$ with $\Phi_1$ fixed to 1. }
	\labell{scalar plots}
\end{figure}

Due to the $C^2$ coupling in this holographic model,  the scalar field $\Phi$ spontaneously acquires an expectation value upon 
subjecting the CFT to a finite temperature $T$:   $\langle \mathcal{O}\rangle_T \sim \Phi_1 T^\Delta$.     At finite detuning $h$,  $\langle \mathcal{O}\rangle$ picks up a simple linear correction in $h$, as can be seen from (\ref{eq:Phi1}).   This is an artifact of the `probe' limit where we have neglected the backreaction of $\Phi$ on the metric.   Despite this limitation,  this holographic system is a useful way of modeling the finite temperature response of a CFT at all frequencies.



\subsection{Scalar Two-Point Functions}\labell{sec:scalartwopoint}
In this section we will construct the full frequency response for the minimal model \reef{eq:holomodel} and show that the analysis done in Section \ref{sec:41} accurately predicts the high-frequency behavior of the scalar two point function $\langle \mcO_0 \mcO_0\rangle$.    
Given the explicit model introduced in Section \ref{sec51}, the equation of motion for $\psi$ is
\begin{equation}
\begin{split}
0 = & \left(\nabla^2 + \frac{\Delta_0(D-\Delta_0)}{L^2} + \alpha_W L^{\frac{D-5}{2}}\Phi\right)\psi \\ 
= & \frac{1}{\sqrt{-g}} \partial_u(\sqrt{-g}g^{uu}\partial_u\psi) - \omega^2 g^{tt}\psi + \frac{\Delta_0(D-\Delta_0)}{L^2} \psi + \alpha_W L^{\frac{D-5}{2}}\Phi \psi\,.
\end{split}
\end{equation}
As before, $\psi$ is only dependent on $u$;  the resulting ordinary differential equation reads
\begin{equation}
0 = \psi'' + \left(\frac{f'}{f} - \frac{D-1}{u}\right)\psi' + \frac{\omega^2D^2\psi}{(4\pi T)^2f^2} + \frac{\Delta_0(D-\Delta_0)}{u^2f}\psi + \frac{\alpha_W L^{\frac{D-1}{2}}\Phi\psi}{u^2f}\,.\labell{eq:scalareom}
\end{equation}
Because this equation is homogeneous, one of the two boundary conditions we impose is `arbitrary': the observable quantity we are extracting is the ratio of $\langle \Oo\rangle/h_0$, as in (\ref{eq:sec32pt}), and this is unaffected under rescaling: $\psi \rightarrow \lambda \psi$.   The boundary conditions to \reef{eq:scalareom} require some care \cite{lucasreview}.   Near the horizon, $\psi(u)$ acquires a log-oscillatory divergence:  this is a consequence of the fact that we require that matter must be `infalling' into the black hole.   In order to regulate the divergence, we will introduce another field $\psi(u) = f(u)^b \Psi(u)$ with $b=-\mathrm{i}\omega/4\pi T$,  and enforce regularity on  $\Psi(u)$ via the following mixed boundary condition:  \begin{equation}
\Psi'(1) = \frac{\Psi(1)}{D(2b+1)}\left(\Delta_0(D-\Delta_0) + \alpha_W L^{\frac{D-1}{2}}\Phi(1)\right).  \label{eq:Psip1}
\end{equation}
Numerically, we solve this equation of motion by `shooting' any solution obeying (\ref{eq:Psip1}) from the black hole horizon to the asymptotic boundary, and obtain $\langle \Oo\Oo\rangle$ from the resulting asymptotic behavior.   We know from (\ref{eq:Ooasym}) that $\psi(u) = \psi_0 u^{D-\Delta} + \psi_1 u^\Delta + \cdots$ as $u\rightarrow 0$.   Hence,  we use regression on points sampled near the boundary to compute $\psi_0$ and $\psi_1$.  Using (\ref{eq:phinormalization}) and (\ref{eq:sec32pt}), which are valid for arbitrary asymptotically AdS geometries, we numerically obtain the scalar two point function, which is plotted in Figure~\ref{scalar plots}. 

\begin{figure}\centering
	\includegraphics[width = 0.45\textwidth]{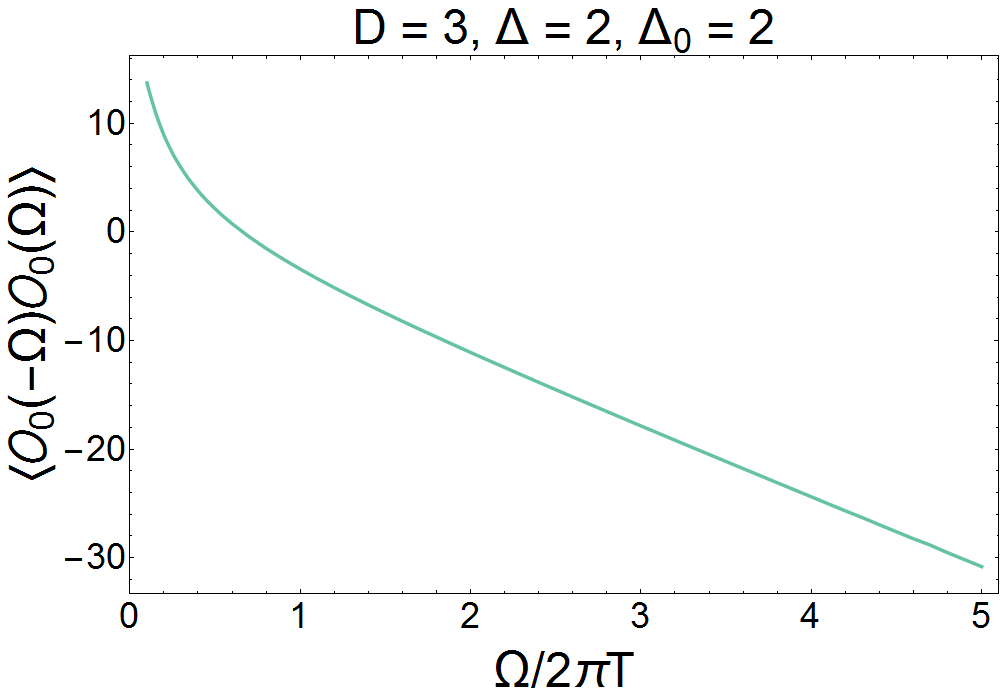}\hspace{0.8cm}
	\includegraphics[width = 0.45\textwidth]{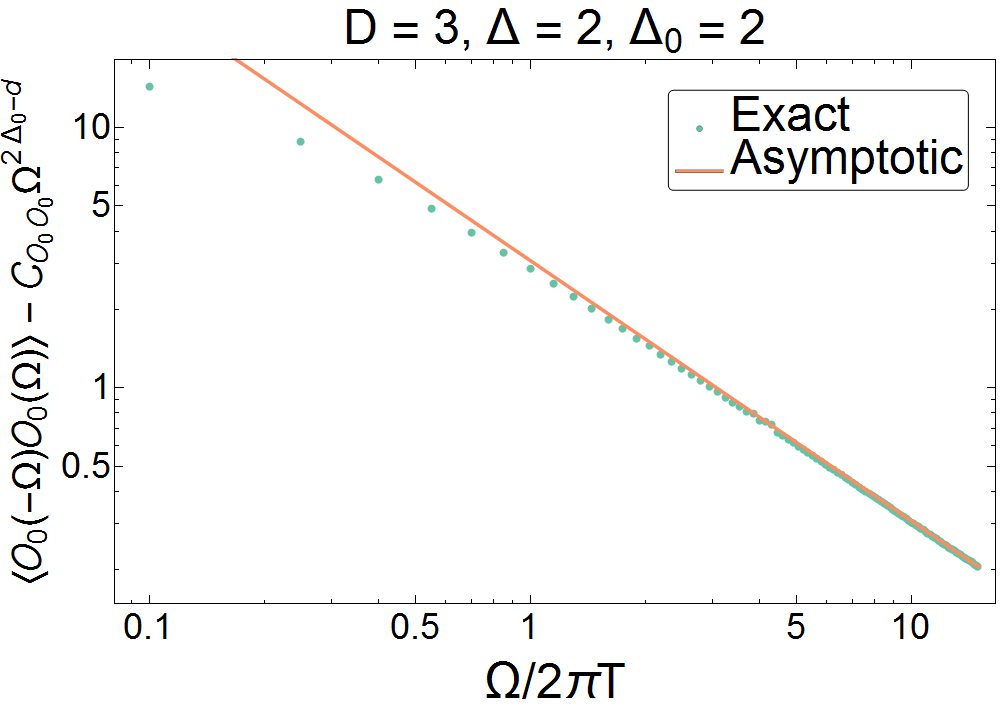} 
	\caption{Scalar two point function $\langle \mcO_0(-\Omega) \mcO_0(\Omega)\rangle$ in $D=3$ at finite temperature, with scaling dimensions $\Delta_0 = 2$ and $\Delta = 2$. We have set $\alpha_Y\alpha_W =1$. 
Left: full frequency response for the scalar two point function. Right: comparison between the analytic prediction for the leading correction to the linear response plotted against the numerically calculated response with the leading term removed 
(plotted in a log-log plot).} 
	\labell{scalar plots}
\end{figure}

\subsection{Conductivity}\labell{sec:conductivity}
In this section we calculate the full frequency-dependent conductivity for the model of Section \ref{sec51}, and confirm the high-frequency analysis done in Section \ref{sec4}.   The computation proceeds similarly as in the previous section. In the $u$ coordinate, the conductivity is given by solving the equation of motion for $A_x$, assuming implicit frequency dependence of the form $\mathrm{e}^{-\mathrm{i}\omega t}$:  \begin{equation}
A''_x(u) + \left(\frac{Z'(u)}{Z(u)} + \frac{f'(u)}{f(u)} - \frac{D-3}{u}\right)A'_x(u) + \left(\frac{D\omega}{4\pi T}\right)^2 \frac{A_x(u)}{f(u)^2} = 0\,, \labell{eq:gauge eom}
\end{equation}
where $Z(u) \equiv Z(\Phi(u))$.   
Just as in the previous subsection, we must solve (\ref{eq:gauge eom}) with an appropriate infalling boundary condition.  As before, we write $A_x(u) = f(u)^bF(u)$;  the regularity condition for $F(u)$ at the horizon is
\begin{equation}
F'(1) = -\frac{bF(1)}{1+2b}\left(2 + \frac{Z'(1)}{Z(1)}\right)\,.
\end{equation}
We employ the same shooting method in order to construct our solution, and hence extract $\sigma(\omega)$.  

We extract $\sigma(\omega)$ through the asymptotic behavior via (\ref{eq:Ooasym}).   Assuming $A_x \sim A_0 + \cdots + A_1 u^{D-2}$, and $D$ odd, the conductivity is given by 
\begin{equation}
\sigma(\omega) = -\frac{L^{D-3}}{e^2} \frac{(D-2)A_1}{\mathrm{i}\omega A_0} \left(\frac{4\pi T}{D}\right)^{D-2}.   \labell{eq:sec5sigma}
\end{equation}
When $D$ is even, the presence of logarithms in the asymptotic expansion of $A_x(u)$ complicates the story. 
Let us simply note the proper prescription for this case.   Given the asymptotic expansion   
\cite{gary} \begin{equation} 
A_x(u) = A_0 \left[1+8b^2u^2 \log (\Lambda u)\right] + A_1 u^2 + \cdots   \label{eq:logAx}
\end{equation}
one finds \begin{equation}
\sigma(\omega) = -\frac{\pi^2 T^2 L}{\mathrm{i}\omega e^2} \left(\frac{2A_1}{A_0} - \frac{\omega^2}{2}\right).
\end{equation}
Note that the logarithm in (\ref{eq:logAx}) is analogous to the logarithm that arose in (\ref{eq:sigmalog}), within conformal field theory.   It will not affect the real part of $\sigma(\omega)$.

Let us note in passing that we may compute the DC conductivity $\sigma(\omega=0)$ analytically via the `membrane paradigm' \cite{Myers:2010pk,Ritz:2008kh}:
\begin{equation}
\sigma(0) = \left(\frac{4\pi L T}{D}\right)^{D-3} \frac{Z(u=1)}{e^2}\,.\labell{eq:DCconductivity}
\end{equation}
As expected, the full numerical solution reproduces this result.

Figures \ref{fig:ACcondd=3} and \ref{fig:ACcondd=4} show $\sigma(\omega)$ in $D=3$ and $D=4$, respectively, for varying coupling constants $\alpha_W\alpha_Z$.   This is analogous to changing the CFT data $\mathcal{C}_{JJ\mathcal{O}}$.   Not surprisingly, we find that the non-trivial structure in the conductivity becomes enhanced as this coupling strength increases.


\begin{figure}\centering
	\includegraphics[width = 0.45\textwidth]{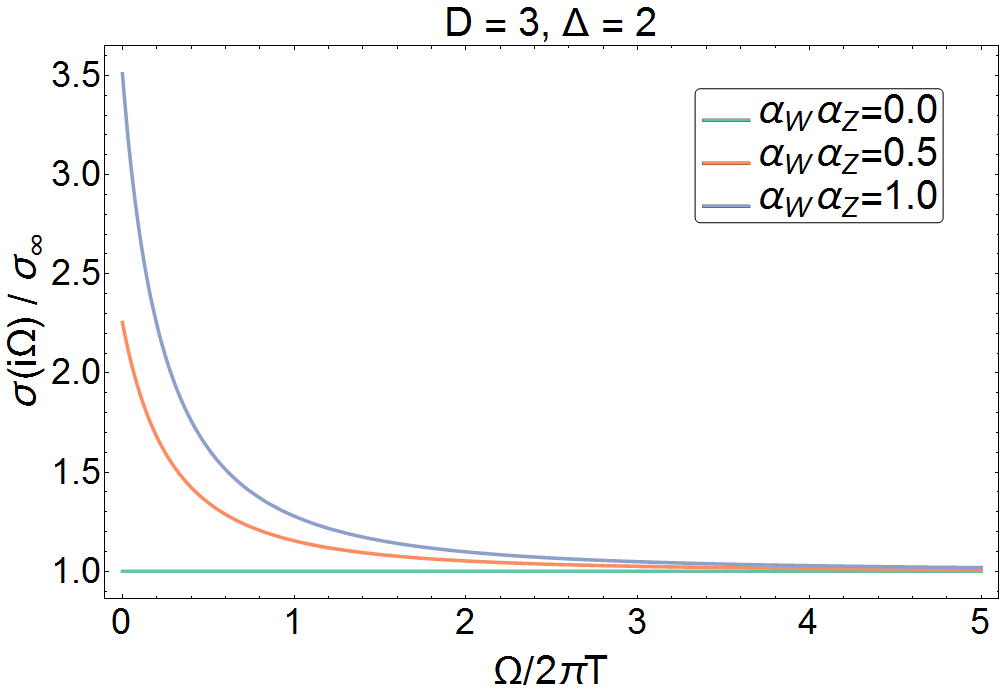}
	\includegraphics[width = 0.45\textwidth]{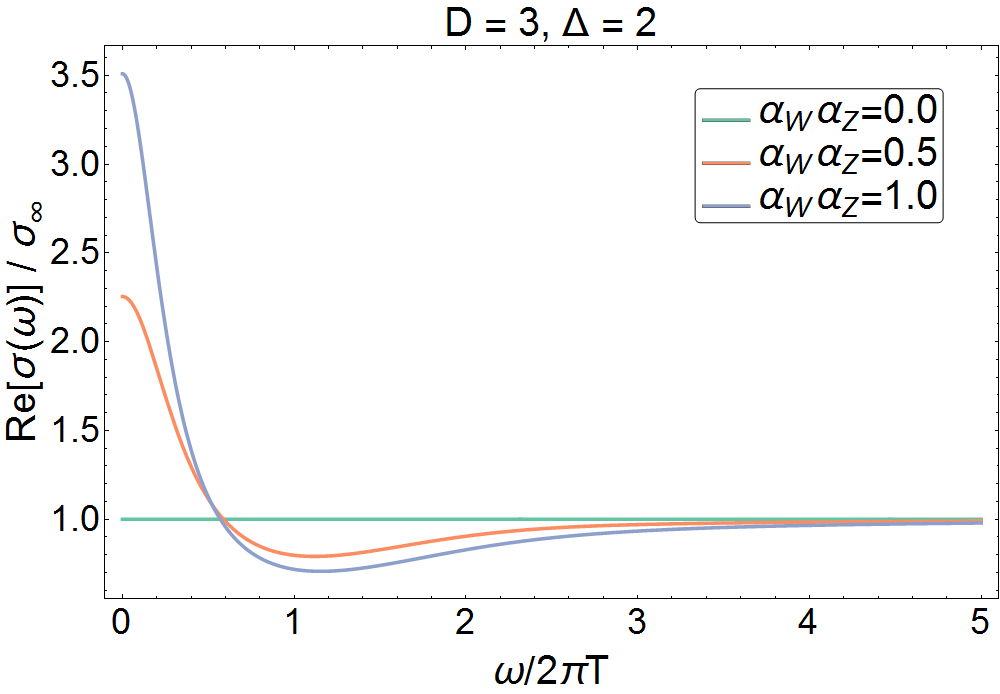}
	\caption{The Euclidean (left) and real (right) AC conductivity for $D=3$ and $\Delta=2$ in the critical theory ($h=0$) for various choices of interaction strength $\alpha_W\alpha_Z$. } \labell{fig:ACcondd=3}	
\end{figure}
\begin{figure}
	\centering
	\begin{subfigure}{0.45\textwidth}{\centering
			\includegraphics[width = \textwidth]{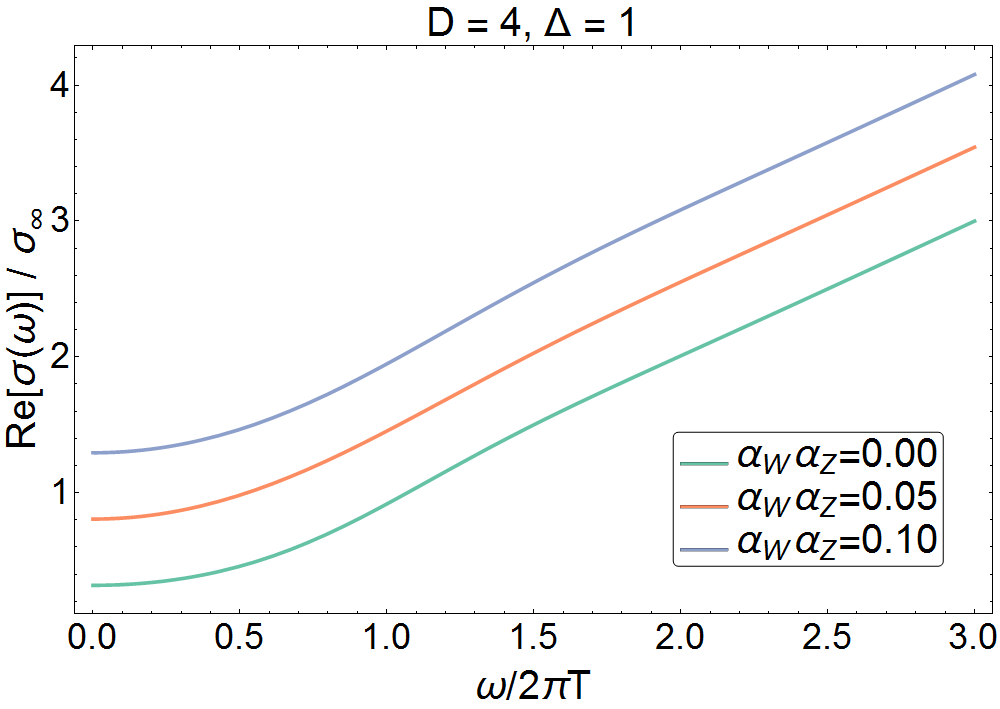}}
	\end{subfigure}
	\begin{subfigure}{0.45\textwidth}{\centering
			\includegraphics[width = \textwidth]{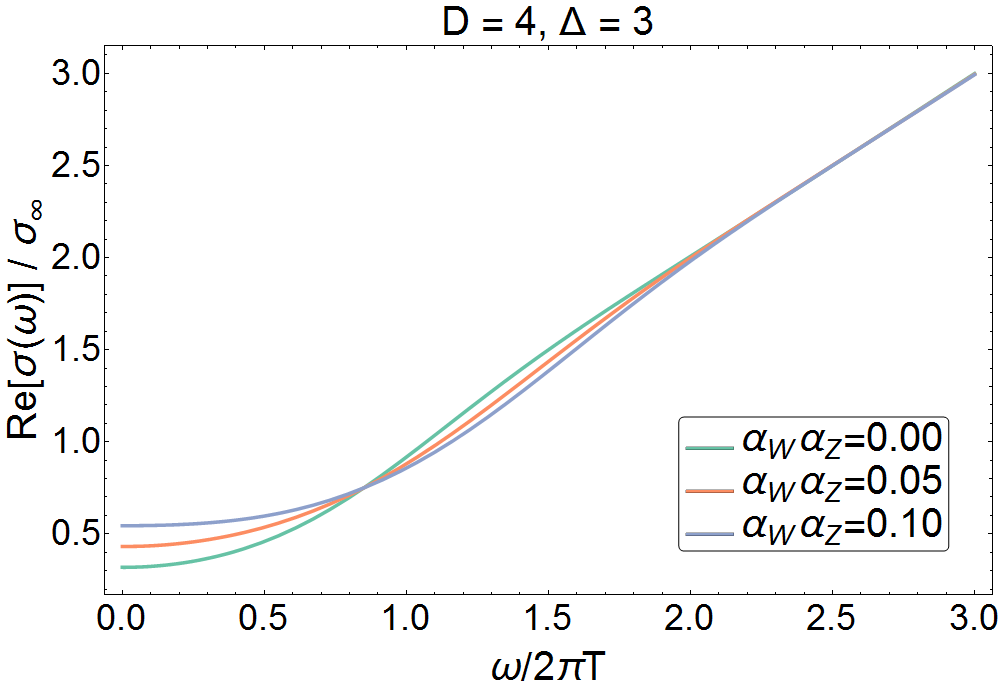}}
	\end{subfigure}
	\caption{The real part of the AC conductivity for $D=4$ and $\Delta=1$ (left) and $\Delta=3$ (right), in the critical theory ($h=0$), for various choices of coupling constants $\alpha_W \alpha_Z$. At large $\omega$,  $\mathrm{Re}(\sigma(\omega)) \approx \sigma_\infty \omega$, with  $\sigma_\infty = \frac{\pi}{2} \tilde{\mathcal{C}}_{JJ}$ given in (\ref{eq:2resigma}). }\labell{fig:ACcondd=4}
\end{figure}

\begin{figure}
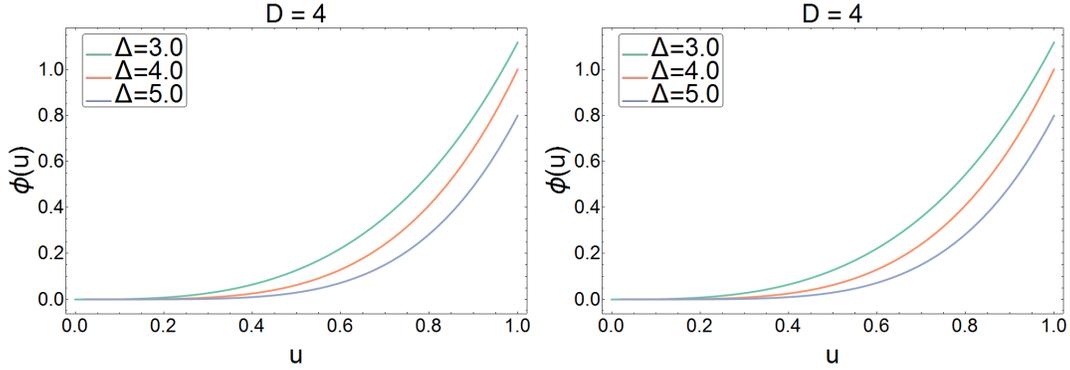
\centering
	\includegraphics[width = 0.45\textwidth]{realseveralDelta.png}
	\includegraphics[width = 0.45\textwidth]{realseveralDelta.png}
	\caption{The real frequency AC conductivities for $D=3$ (left) and $D=4$ (right) for various choices of scaling dimension $\Delta$, with fixed $\alpha_W\alpha_Z = 0.1$.  For $D=3$ the sum rules are observed for $\Delta > 1.0$ and for $D=4$ the sum rules are observed for $\Delta > 2.0$.  Significant enhancement of $\mathrm{Re}(\sigma)$ is observed whenever the sum rules are violated.}\labell{fig:ACcondvariousdelta}
\end{figure}

As we noted after (\ref{eq:mainsum}),  there exist sum rules which tightly constrain the real part of the conductivity whenever $h=0$,  so long as $\Delta > D-2$.   In particular, when such a sum rule is satisfied, it implies that if the conductivity is enhanced at high $\omega$, it must be suppressed at low $\omega$, or vice versa.  Figure \ref{fig:ACcondvariousdelta} shows the dependence of the real part of $\sigma(\omega)$ on the dimension of the operator $\mathcal{O}$ in $D=3$ and $D=4$.   We clearly observe that the conductivity is more affected by relevant operators of smaller dimension.   When the sum rules are violated, we observe dramatic enhancement of the conductivity at all frequencies.  In $D=4$, we note that an operator of dimension $\Delta=1$, leads to a constant shift in the real part of the conductivity at high frequency, which is cleanly observed in the figures \ref{fig:ACcondd=4} and \ref{fig:ACcondvariousdelta}.

\begin{figure}\centering
	\includegraphics[width = 0.55\textwidth]{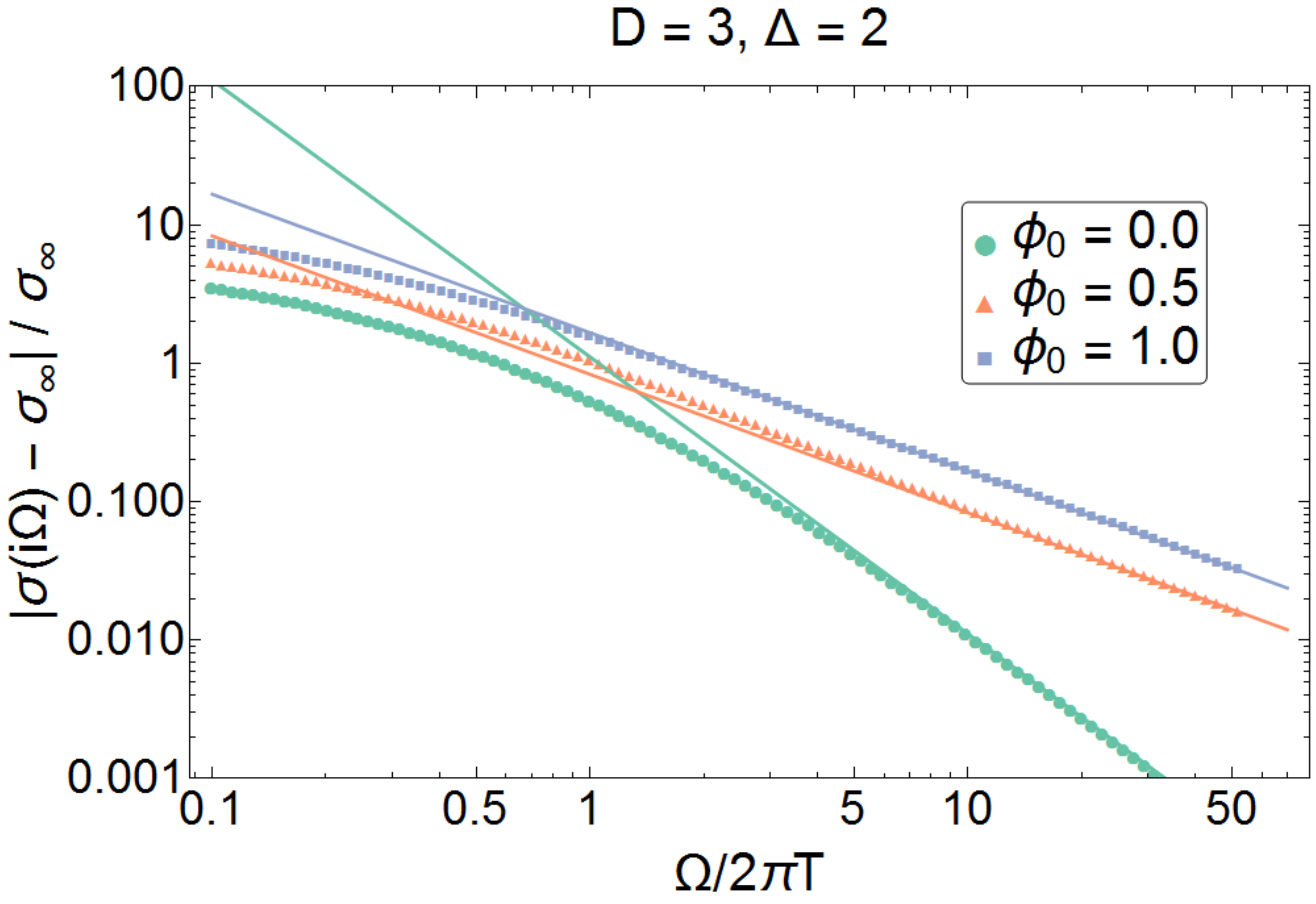}
	\caption{A comparison of our numerical calculation of $\sigma(\mathrm{i}\Omega)$ (shown in dots) to the first subleading contribution in the asymptotic expansion (\ref{eq:mainres}) (shown as thin solid line),  when $\Omega \gg T$.   We take $\Delta=2$ and $D=3$. As predicted in (\ref{eq:mainres}),  when $h=0$ the first subleading power $\Omega^{-\Delta}$ is smaller than when $h\ne 0$.   For $h\ne 0$, the asymptotic corrections are proportional to $h$ and governed by the same power $\Omega^{\Delta-D}$, leading to a slower decay. }\labell{fig:ACcondcompare} 
\end{figure}

Figure \ref{fig:ACcondcompare} compares the analytic predictions for the asymptotic approach of $\sigma(\omega)$ to its CFT value from Section \ref{sec4} to our numerical calculation. We find excellent agreement. We observe that the figure numerically confirms that when the scalar deformation is critical ($\Phi_0 = 0$) the high-frequency conductivity is proportional to $\Omega^{D-3-\Delta}$, while when the scalar is tuned away from the critical point ($\Phi_0 \ne 0$) the high-frequency conductivity is proportional to $\Omega^{\Delta - 3}$.

\subsubsection{Reissner-Nordstr\"om Model} \labell{sec:RN}
So far, the background geometry has been completely thermal.   As a consequence, we have not checked the validity of our asymptotic 
results when $\langle \mathcal{O}\rangle$ depends nonlinearly on non-thermal detuning  parameters.   

To perform such a check, we now consider the real time action 
\begin{equation}
S = \int \mathrm{d}^{D+1}x\sqrt{-g}\left(\frac{R}{2\kappa^2} - Z(\Phi) \frac{F^2}{4e^2} - \frac{1}{2}(\partial \Phi)^2 - V(\Phi)\right),
\end{equation}
with $Z$ and $V$ given in (\ref{eq:potentials}).   Compared to (\ref{eq:holomodel}), we have removed the $\psi$ field as well as the $\Phi C^2$ coupling.
As before, let us imagine that the backreaction of $\Phi$ on the other matter fields can be ignored, but let us now deform the CFT by a finite temperature $T$ and a finite charge density $\rho$.   The dual geometry to such a theory is known analytically for any density \cite{lucasreview}, and is called the AdS-Reissner-Nordstr\"om black hole: \begin{subequations}\label{eq:EM sol}
\begin{align}
\mathrm{d}s^2 = &   \frac{L^2}{r_+^2 u^2} \left(-f(u) \mathrm{d}t^2 + \delta^{ij}\mathrm{d}x_i \mathrm{d}x_j\right) + \frac{L^2 \mathrm{d}u^2}{u^2 f(u)}\\
A = & \sqrt{\frac{D-1}{D-2}}\frac{e\, r_0 \xi}{\kappa L} ( 1 - u^{D-2}) \mathrm{d}t 
\end{align}
\end{subequations}
where \begin{equation}
f(u) = 1-(1+\xi^2)u^D + \xi^2 u^{2(D-1)}  \label{eq:fxi}
\end{equation}
and $r_+$ and $\xi$ are related to the charge density $\rho$ and temperature $T$ of the black hole: \begin{subequations}\begin{align}
T &= \frac{D-(D-2)\xi^2}{4\pi r_+}  , \\
\xi^2 &= \frac{\kappa^2e^2\rho^2r_+^{2D-2}}{(D-1)(D-2)L^{2D-4}} .
\end{align}\end{subequations}
The dimensionless radial coordinate $u$ is once again chosen in such a way that the ``outer" black hole horizon is located at $u = 1$ and the asymptotic AdS boundary is located at $u\to0$. The black hole horizon is extremal when $T=0$:  this occurs at precisely $\xi ^2 = \tfrac{D}{D-2}$, and so we will restrict our attention to $\xi^2 < \tfrac{D}{D-2}$. 

Around this background geometry, we can compute $\Phi(u)$, whose equation of motion is given by 
\begin{equation}
(\nabla - m^2)\Phi - \frac{\alpha_Z L^{\frac{D-1}{2}}}{4 e^2} F_{ab}F^{ab} =(\nabla - m^2)\Phi  +  \frac{\alpha_Z L^{\frac{D-1}{2}}}{2 e^2}  \frac{u^4 A_t'^2}{r_+^2} = 0.
\end{equation}
When $\xi \ne 0$, we will need to solve this equation numerically to determine $\Phi(u)$. 

Due to the presence of the finite charge density, the equations of motion for $A_x$ become more complicated within linear response \cite{lucasreview}.   In particular, they couple to fluctuations in the metric components $g_{tx}$ and $g_{rx}$.   Using standard techniques, we are able to reduce the equations of motion to a `massive' equation of motion for $A_x$:  
\begin{equation}
0 =   A_x'' + \left(\frac{Z'}{Z} + \frac{f'}{f} - \frac{D-3}{u}\right) A_x' + \frac{\omega^2 r_+^2}{ f^2}  A_x  - \frac{2 (D-1)(D-2)u^2\xi^2}{f}  A_x.
\end{equation}
This equation is identical to that in \cite{lucasreview}, up to the dependence on $Z(\Phi)$, and it can be evaluated numerically.  Imposing infalling boundary conditions requires writing $A_x(u) = f(u)^b F(u)$ as before;  the boundary condition at the black hole horizon imposing regularity is  
\begin{equation}
\frac{F'(1)}{F(1)} = \frac{1}{(1+2b)f'(1)}\left(2(D-1)(D-2)\xi^2 - bf'(1)\left(\frac{Z'(1)}{Z(1)} + \frac{f''(1)}{f'(1)}- \frac{D-3}{u}\right)\right)\,.
\end{equation}

\begin{figure}
	\centering
	\includegraphics[width=0.55\textwidth]{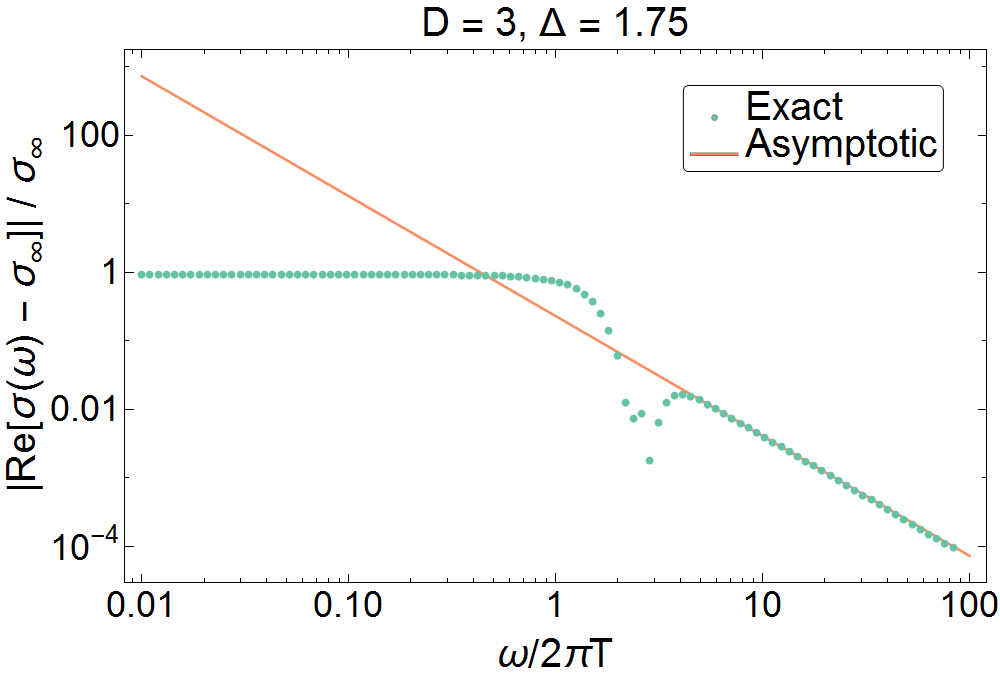}
	\caption{Comparison of $\mathrm{Re}[\sigma(\omega) - \sigma_\infty]$ between our analytic prediction (\ref{eq:mainres}) for the subleading correction to the conductivity, and the full dynamical conductivity
at $\alpha_Y \alpha_Z  = \xi = 1$ in $D\!=\! 3$. For simplicity we have set $h=0$. As in Fig.~\ref{fig:ACcondcompare}, 
we observe excellent agreement when $\omega \gg T$;  at low frequency, the structure of $\sigma(\omega)$ is highly non-trivial in  this background: see Figure \ref{fig:RN_compared}. }    
	\label{fig:RN_perturbed}
\end{figure}

Figure \ref{fig:RN_perturbed} demonstrates agreement between our analytic prediction (\ref{eq:mainres}) for the high frequency corrections to $\sigma(\omega)$, and a full numerical computation.   It is important to keep in mind that in this more generic geometry, the conductivity is not given by the CFT result even when $h=T=0$. 
Indeed, due to the presence of a finite charge density $\rho$, and hence chemical potential $\mu$, 
the conductivity takes the following form:
\begin{equation}
\sigma(\omega) = \sigma_\infty \cdot(\mathrm{i}\omega)^{D-3} \left[1 + \mathcal{A} \frac{h}{(\mathrm{i}\omega)^{D-\Delta}} + \mathcal{B} \frac{\langle \mathcal{O}\rangle}{(\mathrm{i}\omega)^\Delta} - \mathcal{A}^\prime \frac{\mu^2}{\omega^2} + \cdots\right].  \label{eq:53as}
\end{equation}
The coefficients $\mathcal{A}$ and $\mathcal{B}$ are given in (\ref{eq:ABsigma}) and (\ref{eq:sigmaholfinal}). 
The coefficient $\mathcal{A}^\prime$ is related to the four-point correlation function $\langle JJJJ\rangle$ in the CFT, which is non-vanishing in our holographic model due to graviton exchange in the bulk.    
The $\mathcal A'$ term is in fact the counterpart of the higher order correction $(h/\omega^{D-\Delta})^2$, where the role of $\mathcal O$ is played 
by the charge density $J_t$, and the chemical potential $\mu$ plays the role of $h$. 
In (\ref{eq:53as}), we see that if $h=0$, it is necessary to deform this field theory at finite charge density 
by a relevant operator of dimension $\Delta<2$ in order for the expansion (\ref{eq:mainres}) to hold.  

\begin{figure}
	\centering
	\includegraphics[width=0.55\textwidth]{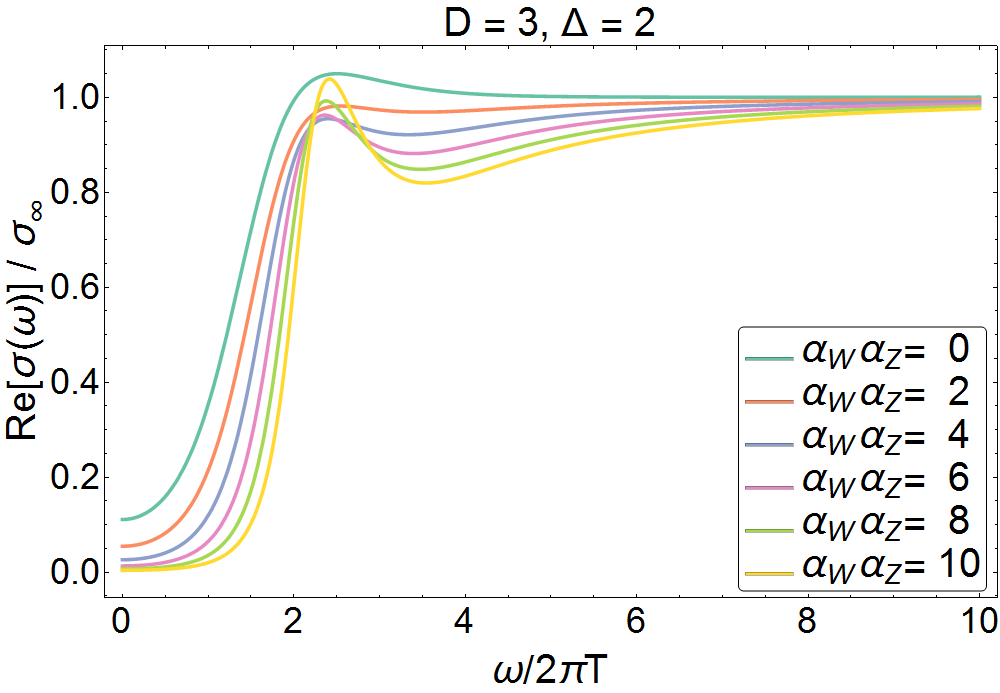}
	\caption{The real part of the conductivity for $D=3$, $\Delta = 2$ and `charge parameter' $\xi = 1$ for various values of $\alpha_Z\alpha_W$.   Coupling to the relevant operator $\mathcal{O}$ leads to pronounced features in $\mathrm{Re}(\sigma)$ at intermediate frequencies, in addition to modifying the high frequency asymptotics as in (\ref{eq:53as}).}
	\label{fig:RN_compared}
\end{figure}

In Figure \ref{fig:RN_compared} we plot $\mathrm{Re}(\sigma(\omega))$ as a function of $\alpha_W\alpha_Z$, corresponding to increasing $\mathcal{C}_{JJ\mathcal{O}}$.    Unlike before, in this field theory at finite charge density 
we observe non-trivial structure even when $\alpha_W\alpha_Z=0$, when the scalar operator $\mathcal{O}$ is decoupled.   This is a consequence of the fact that finite $\mu$ already corresponds to a non-trivial deformation of the CFT.   Upon setting $\alpha_W\alpha_Z>0$, we observe additional structure emerge in Figure \ref{fig:RN_compared}, associated with the coupling to the operator $\mathcal{O}$. 

In Figure \ref{fig:RN_compared},  one may notice the presence of ``missing" spectral weight.   This spectral weight has in fact shifted into a $\delta$-function at $\omega=0$:    `hydrodynamics' constrains the low frequency conductivity to be \cite{lucasreview} \begin{equation}
\sigma(\omega \rightarrow 0) = \frac{\rho^2}{\epsilon+P} \left(\pi \delta(\omega) + \frac{\mathrm{i}}{\omega} \right) + \cdots.
\end{equation}
Here $\epsilon$ is the energy density and $P$ is the pressure.  A uniform and static electric field can impart momentum into a system at finite charge density at a fixed rate via the Lorentz force:  this is the origin of this $\delta$ function.    This $\delta$ function must be taken into account when evaluating sum rules, and upon doing so, the sum rules for the conductivity are restored.

\color{black}
\section{Lifshitz Holography}\label{sec6}
In this section, we will now extend our analysis to holographic quantum critical systems with dynamic critical exponent $z\ne 1$.   This means that there is an emergent scale invariance of the low energy effective theory, but time and space scale separately.  Under a rescaling $\lambda$: \begin{equation}
\mathbf{x} \rightarrow \lambda \mathbf{x}, \;\;\;\; t\rightarrow \lambda^z t.
\end{equation}
Because time and space are no longer related through Lorentz transformations, in this section we will not talk about the spacetime dimension $D$,  but instead the spatial dimension $d=D-1$, as is more conventional in condensed matter.  

As argued in \cite{Lucas:2016fju}, the high frequency behavior of correlation functions in such Lifshitz theories will share many features with CFTs.   For simplicity, we will only talk about the conductivity in this section.  In particular, the asymptotic expansion of the conductivity will be modified from (\ref{eq:mainres}) to \begin{equation}
\sigma(\mathrm i \Omega) = \Omega^{\frac{d-2}{z}} \left[\mathcal{C}_{JJ} + \frac{\mathcal{A} h}{\Omega^{1+\frac{d-\Delta}{z}}} + \frac{\mathcal B\langle \mathcal{O}\rangle}{\Omega^{\frac{\Delta}{z}}}+\cdots\right].  \label{eq:main7}
\end{equation}
We now present a holographic Lifshitz theory where this result can be shown explicitly,  and a relationship between $\mathcal{A}$ and $\mathcal{B}$ can be obtained.

\subsection{Bulk Model}
We first begin by outlining the holographic model which we study.   The bulk action is given by \begin{equation}
S = S_{\mathrm{bg}}[g_{\mu\nu},\Phi] + \int \mathrm{d}^{d+2}x\sqrt{-g} \frac{Z(\Phi)}{4e^2}F_{ab}F^{ab},   \label{eq:sec7action}
\end{equation}
with $S_{\mathrm{bg}}$ the action for the metric and the other bulk fields which support the Lifshitz geometry,  and the remaining term describing a fluctuating gauge field.    We take $Z(\Phi)$ to be given by (\ref{eq:ZPhi}), as before.   Specific forms of $S_{\mathrm{bg}}$ can be found in \cite{Kachru:2008yh}, for example, but they will not be relevant for our purpose.   We will take our background to be uncharged under $F_{ab}$, the bulk field under which we will compute the conductivity.   The quadratic terms in $\Phi$ in $S_{\mathrm{bg}}$ are given by \begin{equation}
S_{\mathrm{bg}} = \int \mathrm{d}^{d+2}x\sqrt{-g} \left(\frac{1}{2}(\partial \Phi)^2 + \frac{\Delta(\Delta-d-z)}{L^2}\Phi^2\right) + \cdots.   \label{eq:sec7actionphi}
\end{equation}
Note that the mass of $\Phi^2$ is distinct from that in (\ref{eq:WPhi}), once $z\ne 1$.

The background metric of the Lifshitz theory is given by \cite{Kachru:2008yh} \begin{equation}
\mathrm{d}s^2 = \frac{L^2}{r^2}\left[\mathrm{d}r^2 - \frac{\mathrm{d}t^2}{r^{2(z-1)}} + \mathrm{d}\mathbf{x}^2\right].   \label{eq:lifmetric}
\end{equation}
The isometries of this metric map on to the Lifshitz symmetries of the dual field theory, and as a consequence the correlation functions of the dual theory are necessarily Lifshitz invariant.   As Lifshitz invariance is a much weaker requirement than conformal invariance \cite{Bekaert:2011qd, Golkar:2014mwa, Goldberger:2014hca}, it is not clear that holographic models can capture the dynamics of generic Lifshitz theories.   Nonetheless, we will be able to demonstrate a generalization of (\ref{eq:mainres}) in at least one class of interacting Lifshitz theories, and although our specific results for $\mathcal{A}$ and $\mathcal{B}$ are not generic, we expect that the qualitative features of this model are more robust.  One can also study more complicated bulk models than (\ref{eq:sec7action}), as the symmetries of Lifshitz theories are far less constraining \cite{Keeler:2015afa}.   

In order  to compute the conductivity, we must  note changes to the holographic dictionary in a Lifshitz background.  Following the discussion of holographic renormalization in \cite{lucasreview}, one finds:  \begin{subequations}\label{eq:lifsasym}\begin{align}
\Phi &= \frac{h}{L^{d/2}} r^{d+z-\Delta} + \frac{\langle \mathcal{O}\rangle}{(2\Delta-d-z)L^{d/2}} r^\Delta + \cdots, \\
A_x &=  A_x^0 + \frac{e^2}{L^{d-2}} \frac{\langle J_x\rangle}{d+z-2} r^{d+z-2} + \cdots. 
\end{align}\end{subequations}

\subsection{Asymptotics of the Conductivity}
The equations of motion for the gauge field in the background (\ref{eq:lifmetric}) are \begin{equation}
0 = \frac{1}{\sqrt{g}} \partial_a \left(Z\sqrt{g} F^{ax}\right) = r^{d+z+1}\partial_r \left(Z(\Phi(r))r^{1-d-z}\partial_r A_x\right) - \Omega^2 Z(\Phi(r)) r^{2z}A_x.   \label{eq:sec7gaugeeom}
\end{equation}
Let us now define the variable \begin{equation}
R \equiv \frac{\Omega}{z}r^z.  \label{eq:6R}
\end{equation}
By the same logic as before, the limit $\Omega \rightarrow \infty$ becomes completely regular, with (\ref{eq:sec7gaugeeom}) becoming \begin{equation}
0 = R^{\frac{d-2}{z}}\partial_R \left(Z\left(\left(\frac{zR}{\Omega}\right)^{1/z}\right) R^{-\frac{d-2}{z}}\partial_R A_x\right) -  Z\left(\left(\frac{zR}{\Omega}\right)^{1/z}\right) A_x
\end{equation}
At $\Omega=\infty$, the solution to this equation with the correct boundary conditions is \begin{equation}
A_x = \frac{\hat{\mathrm{K}}_{\frac{d+z-2}{2z}}\left(R\right)}{\mathcal{Z}\left(\frac{d+z-2}{2z}\right)}. \label{eq:axsec7}
\end{equation}
At finite $\Omega$, we must solve \begin{align}
\partial_R^2 A_x - \frac{d-2}{zR}\partial_R A_x - A_x &= \mathfrak{a}(R)A_x -   R^{\frac{d-2}{z}}\partial_R \left(\mathfrak{a}(R) R^{-\frac{d-2}{z}} \partial_R  A_x\right)
\end{align}
perturbatively in $\mathfrak{a}(R)$.   This is exactly analogous to the solution of (\ref{eq:Ax42}).   Following (\ref{eq:Ax42single}), and using that \begin{equation}
\mathfrak{a}(R) = \alpha_Z h \left(\frac{zR}{\Omega}\right)^{\frac{d+z-\Delta}{z}} +   \frac{\alpha_Z \langle \mathcal{O}\rangle}{2\Delta-d-z} \left(\frac{zR}{\Omega}\right)^{\frac{\Delta}{z}} ,
\end{equation}
we obtain (analogously to (\ref{eq:Ax42single2})): \begin{align}
A_x(R\rightarrow 0) &\approx \frac{\hat{\mathrm{K}}_{\frac{d+z-2}{2z}}\left(R\right)}{\mathcal{Z}\left(\frac{d+z-2}{2z}\right)} + \frac{R^{\frac{d+z-2}{z}}}{\frac{d+z-2}{z}} \times \notag \\
&\left\lbrace \alpha_Z h \left(\frac{z}{\Omega}\right)^{\frac{d+z-\Delta}{z}} \frac{(d+z-\Delta)(\Delta-2)}{2z^2\mathcal{Z}(\frac{d+z-2}{2})^2}  \Psi\left(\frac{d+z-\Delta}{z},\frac{d+z-2}{2z}\right) \right.\notag \\
&\left. +  \frac{\alpha_Z \langle \mathcal{O}\rangle}{2\Delta-d-z} \left(\frac{z}{\Omega}\right)^{\frac{\Delta}{z}} \frac{\Delta(d+z-\Delta-2)}{2z^2\mathcal{Z}(\frac{d+z-2}{2z})^2}  \Psi\left(\frac{\Delta}{z},\frac{d+z-2}{2z}\right)\right\rbrace.
\end{align}
Hence, the conductivity is \begin{align}
\sigma(\mathrm i \Omega) &= \frac{L^{d-2}}{e^2} \Omega^{\frac{d-2}{z}}\left[\mathcal{C}_{JJ}+ \frac{\alpha_Z  z^{\frac{2-z-\Delta}{z}}(d+z-\Delta)(\Delta-2)}{2\mathcal{Z}(\frac{d+z-2}{2z})^2} \Psi\left(\frac{d+z-\Delta}{z},\frac{d+z-2}{2z}\right) \frac{h}{\Omega^{\frac{d+z-\Delta}{z}}}\right. \notag \\
&\left.+ \frac{\alpha_Z z^{\frac{2-d-2z+\Delta}{z}} \Delta(d+z-\Delta-2) }{2(2\Delta-d-z)\mathcal{Z}(\frac{d+z-2}{2z})^2}\Psi\left(\frac{\Delta}{z},\frac{d+z-2}{2z}\right) \frac{\langle \mathcal{O}\rangle}{\Omega^{\frac{\Delta}{z}}} \right].  \label{eq:72sigma}
\end{align}

\subsection{Three-Point Function and (Holographic) Lifshitz Perturbation Theory}
Now, we follow the construction of Section \ref{sec:AJJO} and compare our holographic result to ``Lifshitz perturbation theory".

The properly normalized solution to the scalar equation of motion in a Lifshitz background that follows from (\ref{eq:sec7actionphi}), \begin{equation}
\frac{\Delta(\Delta-d-z)}{L^2}\Phi = \nabla_a \nabla^a \Phi,
\end{equation}
is \begin{equation}
\Phi(p_3,r) = \mathrm{e}^{\mathrm{i}p_3\cdot x} r^{d+z-\Delta}\frac{\hat{\mathrm{K}}_{\frac{2\Delta-d-z}{2z}}(\frac{\omega}{z}r^z)}{\mathcal{Z}(\frac{2\Delta-d-z}{2z})L^{d/2}}  j_{\mathcal{O}}(p_3)  .\label{eq:sec73scalar}
\end{equation}
The cubic contribution to the bulk action becomes \begin{align}
S_{\mathrm{bulk}} &= \int \mathrm{d}^{d+2}x \sqrt{g} \frac{\alpha_Z L^{d/2}\Phi(p_3)}{e^2} g^{ab}g^{xx} \partial_a A_x(p_1)\partial_b A_x(p_2) \notag \\
&= -\int \mathrm{d}^{d+2}x A_x(p_1) \partial_a\left[ \sqrt{g} \frac{\alpha_Z L^{d/2}\Phi(p_3)}{e^2} g^{ab}g^{xx} \partial_b A_x(p_2)\right] \notag \\
&= -\int \mathrm{d}^{d+2}x A_x(p_1)  \sqrt{g} \frac{\alpha_Z L^{d/2}}{e^2} g^{ab}g^{xx} \partial_b A_x(p_2) \partial_a \Phi(p_3) \notag \\
&= \frac{\alpha_Z L^{d/2}}{2e^2} \int \mathrm{d}^{d+2}x \sqrt{g} A_x(p_1)A_x(p_2) \left[g^{xx}\frac{\Delta(\Delta-d-z)}{L^2}\Phi(p_3)  +  \frac{2}{r}g^{xx} g^{rr} \partial_r \Phi(p_3)\right] .
\end{align}
Using (\ref{eq:axsec7}) and (\ref{eq:sec73scalar}), along with (\ref{eq:besselders}) we obtain (suppressing the $\delta$-function for momenta, and the integrals in the boundary directions) \begin{align}
S_{\mathrm{bulk}} &= \frac{\alpha_Z L^{d-2}}{2e^2} \int \frac{\mathrm{d}r}{r^{d+z-1}} \frac{\hat{\mathrm{K}}_{\frac{d+z-2}{2z}}\left(\frac{p_1}{z}r^z\right)}{\mathcal{Z}\left(\frac{d+z-2}{2z}\right)}\frac{\hat{\mathrm{K}}_{\frac{d+z-2}{2z}}\left(\frac{p_2}{z}r^z\right)}{\mathcal{Z}\left(\frac{d+z-2}{2z}\right)}\left[ \Delta(\Delta-d-z) \frac{\hat{\mathrm{K}}_{\frac{2\Delta-d-z}{2z}}(\frac{p_3}{z}r^z)}{\mathcal{Z}(\frac{2\Delta-d-z}{2z})} \right. \notag \\
&\left. + \frac{2\Delta \hat{\mathrm{K}}_{\frac{2\Delta-d-z}{2z}}(\frac{p_3}{z}r^z)}{\mathcal{Z}(\frac{2\Delta-d-z}{2z})} - \frac{2z \hat{\mathrm{K}}_{\frac{2\Delta-d+z}{2z}}(\frac{p_3}{z}r^z)}{\mathcal{Z}(\frac{2\Delta-d-z}{2z})}   \right]r^{d+z-\Delta} 
\end{align}
Changing variables to $\rho = r^z/z$, we obtain \begin{align}
S_{\mathrm{bulk}} &= \frac{\alpha_Z L^{d-2}}{2e^2} \int \frac{\mathrm{d}\rho (z\rho)^{\frac{2-z-\Delta}{z}} }{\mathcal{Z}(\frac{d+z-2}{2z})^2 \mathcal{Z}(\frac{2\Delta-d-z}{2z})} \left[\Delta(\Delta+2-d-z)\hat{\mathrm{K}}_{\frac{2\Delta-d-z}{2z}}(p_3\rho)\right. \notag \\
&\left. - 2z\hat{\mathrm{K}}_{\frac{2\Delta-d+z}{2z}}(p_3\rho)\right] \hat{\mathrm{K}}_{\frac{d+z-2}{2z}}(p_1\rho)\hat{\mathrm{K}}_{\frac{d+z-2}{2z}}(p_2\rho).
\end{align}
Analogously to (\ref{eq:42Axfin}), we obtain 
\begin{align}
&\langle J_x(\Omega_1)J_x(\Omega_2)\mathcal{O}(\Omega_3) \rangle_0 \notag\\
=& \frac{\alpha_Z L^{d-2} z^{\frac{2-z-\Delta}{z}}}{e^2 \mathcal{Z}(\frac{d+z-2}{2z})^2 \mathcal{Z}(\frac{2\Delta-d-z}{2z})}\left\lbrace z I\left(\frac{d+z}{2z},\frac{d+z-2}{2z},\frac{d+z-2}{2z},\frac{2\Delta-d+z}{2z}\right) \right. \notag \\
&\left. - \frac{\Delta}{2}(\Delta+2-d-z)I\left(\frac{d-z}{2z},\frac{d+z-2}{2z},\frac{d+z-2}{2z},\frac{2\Delta-d-z}{2z}\right)\right\rbrace.  \label{eq:JJOz}
\end{align}

We now attempt to follow the arguments of conformal perturbation theory.   We expect to find that the conductivity takes an analogous form to (\ref{eq:mainres}): \begin{equation}
\sigma(\mathrm i  \Omega) = \Omega^{\frac{d-2}{z}}\left[\sigma_\infty + \frac{\mathcal{A}h}{\Omega^{\frac{d+z-\Delta}{z}}} + \frac{\mathcal{B}\langle \mathcal{O}\rangle}{\Omega^{\frac{d+z-\Delta}{z}}}  \right],
\end{equation}
and now use the form of the three-point correlator (\ref{eq:JJOz}) to fix $\mathcal{A}$ and $\mathcal{B}$.   We take $p_{1,2,3}$ to be given by (\ref{eq:p1p2p3}),  and Taylor expand in $p/\Omega$.   We find a regular term as $p\rightarrow 0$, given by \begin{align}
\langle J_x(\Omega)&J_x(-\Omega)\mathcal{O}(0) \rangle_0 =  \frac{\alpha_Z L^{d-2} z^{\frac{2-z-\Delta}{z}}\Omega^{\frac{\Delta-z-2}{z}}}{e^2 \mathcal{Z}(\frac{d+z-2}{2z})^2 \mathcal{Z}(\frac{2\Delta-d-z}{2z})} \left\lbrace z \mathcal{Z}\left(\frac{2\Delta-d+z}{2z}\right) \right. \notag \\
&\left. -\frac{\Delta}{2}(\Delta+2-d-z)\mathcal{Z}\left(\frac{2\Delta-d-z}{2z}\right)  \right\rbrace \Psi \left(\frac{d+z-\Delta}{z}, \frac{d+z-2}{2z}\right) \notag \\
&= -\frac{\alpha_Z L^{d-2} z^{\frac{2-z-\Delta}{z}}\Omega^{\frac{\Delta-z-2}{z}}}{e^2 \mathcal{Z}(\frac{d+z-2}{2z})^2 } \left(1-\frac{\Delta}{2}\right)(d+z-\Delta)\Psi \left(\frac{d+z-\Delta}{z}, \frac{d+z-2}{2z}\right).
\end{align}
Similarly, we find a non-analytic contribution in $p$:  
\begin{align}
\langle J_x(\Omega)&J_x(-\Omega)\mathcal{O}(p) \rangle_0 = \cdots  -  \frac{\alpha_Z L^{d-2} z^{\frac{2-z-\Delta}{z}}\Omega^{\frac{\Delta-z-2}{z}}}{e^2 \mathcal{Z}(\frac{d+z-2}{2z})^2 \mathcal{Z}(\frac{2\Delta-d-z}{2z})} \times \notag \\
& \frac{\Delta(\Delta+2-d-z)}{2} \mathcal{Z}\left(\frac{d+z-2\Delta}{2z}\right) \Psi \left(\frac{\Delta}{z}, \frac{d+z-2}{2z}\right) \left(\frac{p}{\Omega}\right)^{\frac{2\Delta-d-z}{z}}.
\end{align}
Following the logic of conformal perturbation theory, we fix \begin{subequations}\label{eq:sec73AB}\begin{align}
\mathcal{A} &= -\frac{\alpha_Z L^{d-2} z^{\frac{2-z-\Delta}{z}}}{e^2 \mathcal{Z}(\frac{d+z-2}{2z})^2 } \left(1-\tfrac{\Delta}{2}\right)(d+z-\Delta)\Psi \left(\frac{d+z-\Delta}{z}, \frac{d+z-2}{2z}\right), \\
\mathcal{B} &= -\frac{\alpha_Z L^{d-2} z^{\frac{2-z-\Delta}{z}}}{\mathcal{C}_{\mathcal{OO}} e^2 \mathcal{Z}(\frac{d+z-2}{2z})^2  \mathcal{Z}(\frac{2\Delta-d-z}{2z})} \notag \\
&\;\;\;\;\;\;\;\;\;\;\;\;\; \times  \frac{\Delta(\Delta+2-d-z)}{2} \mathcal{Z}\left(\frac{d+z-2\Delta}{2z}\right) \Psi \left(\frac{\Delta}{z}, \frac{d+z-2}{2z}\right).
\end{align}
\end{subequations}

The computation of $\mathcal{C}_{\mathcal{OO}}$ in a holographic Lifshitz theory proceeds along the lines of Section \ref{app:COO}.   Employing (\ref{eq:lifsasym}) and (\ref{eq:sec73scalar}) we find \begin{equation}
\mathcal{C}_{\mathcal{OO}} = \frac{2\Delta-d-z}{z^{\frac{2\Delta-d-z}{z}}} \frac{\mathcal{Z}(\frac{d+z-2\Delta}{2z})}{\mathcal{Z}(\frac{2\Delta-d-z}{2z})}.
\end{equation}
Plugging this result into (\ref{eq:sec73AB}), we see complete agreement with (\ref{eq:72sigma}).

\section{Sum Rules}\label{sec7}
One of the main applications of the high frequency asymptotics that we have been discussing 
is the derivation of sum rules for dynamical response functions. The most familiar sum rules are associated with the conductivity $\sigma(\omega)$.    Such conductivity sum rules in near-critical systems were shown recently in \cite{Lucas:2016fju}, and we review the arguments here, and generalize them to other response functions.   Our first goal is to understand when the following identity holds: 
\begin{equation}
\int\limits_0^\infty \mathrm{d}\omega \; \mathrm{Re}\left(\sigma(\omega;T,h) - \mathcal{C}_{JJ}\cdot 
(\mathrm{i}\omega)^{\frac{d-2}{z}}\right) =0,
\label{eq:conductivitysum}
\end{equation}
where $d$ is the number of spatial dimensions. This is a highly non-trivial constraint on the conductivity,
which is a scaling function of 2 variables: $\omega/T$ and $h/T$. 
At $z=1$, (\ref{eq:conductivitysum}) was derived under certain conditions for CFTs, which we review 
below \cite{katz, Lucas:2016fju,sum-rules,ws}.   
We will also analyze its validity for Lifshitz theories $z\neq 1$. 

The asymptotic expansion (\ref{eq:main7}) allows us to understand precisely when (\ref{eq:conductivitysum}) will hold.   We need the first subleading term in an asymptotic expansion of $\sigma(\omega)$ to decay faster than $\omega^{-1}$ as $\omega \rightarrow \infty$.   If this is the case, then we may evaluate the integral of (\ref{eq:conductivitysum}) via contour integration, choosing a semicircular contour in the upper half plane and using analyticity of $\sigma(\omega)$ to show that no poles or branch cuts may be enclosed by the contour (see e.g. \cite{Son09, katz, Lucas:2016fju}).    From (\ref{eq:main7}), we conclude that the sum rule is valid when 
\begin{equation}
d+z-2<\Delta<2,   \label{eq:J8}
\end{equation}
a result first shown in \cite{Lucas:2016fju}.   The first inequality (the lower bound on $\Delta$) comes from demanding that the $\mathcal{A}$ term decay faster than $\omega^{-1}$;  the second inequality (the upper bound on $\Delta$) 
comes from demanding the same of the $\mathcal{B}$ term.   Hence, exactly at the critical point ($h=0$), we find the somewhat weaker bound that $\Delta>d-2+z$, which agrees with the result for $d=2$ CFTs \cite{katz}, i.e.\ $\Delta>1$. 

For general $d$ and $z$, the analogue of (\ref{eq:conductivitysum}) for the scalar 2-point function $\langle \Oo\Oo\rangle$ is \begin{equation}
\int\limits_0^\infty \mathrm{d}\omega \; \mathrm{Im}\left(\frac{\langle \Oo(\omega)\Oo(-\omega)\rangle}{\omega} - \mathcal{C}_{\Oo\Oo} \frac{(\mathrm{i}\omega)^{\frac{2\Do-d-z}{z}}}{\omega}\right) = 0.  \label{eq:Oosum}
\end{equation}
where $\mathrm{Im}(\langle \Oo(\omega)\Oo(-\omega)\rangle)$ is the usual spectral function. 
The generalization of (\ref{eq:mainres}) to $z\ne 1$ is \begin{equation}
\langle \Oo(\omega)\Oo(-\omega)\rangle = (\mathrm{i}\omega)^{\frac{2\Do-d-z}{z}} \left[\mathcal{C}_{\Oo\Oo}  + \frac{\mathcal{A} h}{(\mathrm{i}\omega)^{1+\frac{d-\Delta}{z}}} + \frac{\mathcal B\langle \mathcal{O}\rangle}{(\mathrm{i}\omega)^{\frac{\Delta}{z}}}+\cdots\right].   \label{eq:Ooz}
\end{equation}
In this paper, we have demonstrated this asymptotic expansion carefully for CFTs (with $z=1$), but we expect that it also holds true for a large
class of systems with $z\ne 1$ (including the Lifshitz holographic models described above). If $2\Do>d+z$, then the subtraction in (\ref{eq:Oosum}) is not badly behaved near $\omega=0$,  where we expect that at any finite $T$ or $h$,  any divergences in $\langle \Oo\Oo\rangle$ will be resolved.    From (\ref{eq:Ooz}), we conclude that (\ref{eq:Oosum}) holds so long as \begin{equation} 
2\Do - d -z< \Delta < 2(d+z-\Do),   \label{eq:Do8}
\end{equation}
using similar logic to the previous paragraph.   As with the conductivity, the upper bound on $\Delta$ is only needed if $h\ne 0$.   We note that (\ref{eq:J8}) is a special case of (\ref{eq:Do8}), when $\Do = d+z-1$, consistent with the fact that in a Lifshitz theory the operator dimension of $J_x$ is $d+z-1$ (measured in units of inverse length). 

We now turn to sum rules for the shear viscosity. These were studied in certain critical 
phases (CFTs without a relevant singlet $\mathcal O$) in \cite{Son09,caron09,sum-rules}.  
Here we generalize to quantum critical points with general $z$, and also to theories detuned by a relevant operator $\mathcal O$. 
We expect the asymptotic form of the viscosity to be \begin{equation} 
\eta(\omega) = (\mathrm{i}\omega)^{\frac{d}{z}} \left[\mathcal{C}_{TT}  + \frac{\mathcal{A} h}{(\mathrm{i}\omega)^{1+\frac{d-\Delta}{z}}} 
+ \frac{\mathcal B\langle \mathcal{O}\rangle}{(\mathrm{i}\omega)^{\frac{\Delta}{z}}}+  \frac{\mathcal{B}_T P}{(\mathrm{i}\omega)^{1+\frac{d}{z}}}+ \cdots\right].   \label{eq:BTexpansion}
\end{equation}
for generic $d$ and $z$.    As we will shortly see, it is important to keep track of a term proportional to the pressure $P=\langle T_{xx} \rangle$ in the asymptotic expansion.  Such a term can arise due to the stress tensor appearing in the OPE of $T_{xy}T_{xy}$; for CFTs, we have already discussed the presence of the stress tensor in the OPE in (\ref{eq:TmunuOPE}).   We postulate that this term will generically occur, as all Lifshitz theories will have a conserved stress tensor of dimension $[T_{xx}]=d+z$, and we confirm that it arises in holographic theories in Appendix \ref{app:visc}.    A formula for $\mathcal{B}_T\ne 0$ has been derived for the conductivity in \cite{katz} and the viscosity in \cite{David2016}, when $z=1$; it is related to the coefficients of $\langle TTT\rangle_0$ in the CFT \cite{willprl}. We may employ (\ref{eq:Do8}) to find  
 the analogous sum rule 
\begin{equation} \label{eq:eta-sr1}
  \int\limits_0^\infty \mathrm{d}\omega \; \mathrm{Re}\left(\eta(\omega) - \mathcal{C}_{TT}(\mathrm{i}\omega)^{\frac{d}{z}}\right) 
  = - \frac{\pi}{2} \mathcal{B}_T P.
\end{equation}
The factor of $\pi/2$ can be derived by contour integration (e.g.~\cite{Son09,Lucas:2016fju});  it is related to the fact that $\eta(\omega) = \cdots + \mathcal{B}_T P/(\mathrm{i}\omega) + \cdots$ has a $1/\omega$ term. Eq.~(\ref{eq:eta-sr1}) is satisfied whenever   
\begin{equation}
\Delta > d+z,
\end{equation}
exactly at the critical point, $h=0$,  but can never be satisfied when $h\ne 0$ (as the analogous constraint on $\Delta$ becomes $\Delta<0$, which is inconsistent with unitarity). 
%

At finite temperature but zero detuning, $h=0$, one can construct a sum rule that  
holds in the important case where $\mathcal O$ is relevant, $\Delta<d+z$,
\begin{align}
  \int\limits_0^\infty \mathrm{d}\omega \; \mathrm{Re}\left(\eta(\omega) - \mathcal{C}_{TT}(\mathrm{i}\omega)^{\frac{d}{z}}
- \mathcal{B} \langle\mathcal O\rangle_T (\mathrm{i}\omega)^{\frac{d-\Delta}{z}}  \right) = - \frac{\pi}{2} \mathcal{B}_T P.
\end{align}
This sum rule was introduced for CFTs ($z=1$) in \cite{willprl}. It is more complicated than the conductivity sum rule (\ref{eq:conductivitysum})  
because one needs to evaluate the 1-point function of $\langle \mathcal O\rangle_T$ at finite temperature.  

Finally, we mention another type of sum rule, called the ``dual sum rule'', for the \emph{inverse} conductivity $1/\sigma$:  
  \begin{align}
    \int_0^\infty \mathrm{d}\omega \left[\mathrm{Re}\left( \frac{1}{\sigma(\omega)} \right)-\frac{1}{\mathcal{C}_{JJ} 
(\mathrm{i}\omega)^{\frac{d-2}{z}}}\right] =0  
  \end{align}  
which was initially found for CFTs in $d=2$ \cite{ws,katz}. In that context, it acquires a physical interpretation 
by virtue of particle-vortex or S-duality \cite{Witten03,Myers:2010pk,ws}. 
Here we generalize it to $z\neq 1$ and $d\neq 2$. In order
for the sum rule to hold, we need $2+z-d<\Delta < 2d-2$ which follows from the $\mathcal A,\mathcal B$ terms in the 
large $\omega$ expansion. In addition, we need $z>d-2$ 
in order for the subtracted term $1/\omega^{(d-2)/z}$ to be integrable at small frequencies.  Interestingly, in the case of $d=2$ ($D=2+1$) CFTs 
these constraints reduce to the ones for the direct sum rule for $\sigma$, (\ref{eq:conductivitysum}).   

\section{Conclusion} \label{sec:conclusion}
In this paper, we have demonstrated the universal nature of the high frequency response of conformal quantum critical systems (\ref{eq:mainres}), both at finite temperature and when deformed by relevant operators.    Holographic methods have demonstrated that, for certain large-$N$ matrix theories, these results remain correct even when the ground state is far from a CFT.    As a consequence, we find non-perturbative sum rules which place constraints on the low frequency conductivity, regardless of the ultimate fate of the ground state.  Our results are directly testable in quantum Monte Carlo simulations,  and  in experiments in cold atomic gases, where proposals for measuring the optical conductivity have been made \cite{tokuno}.

One interesting generalization of (\ref{eq:mainres}) will occur when the critical theory is deformed not by a spatially homogeneous coupling $h$,  but by a random inhomogeneous coupling $h(\mathbf{x})$.   We expect this randomness to modify the form of the asymptotic expansion (\ref{eq:mainres}); such an
analysis could be performed in large-$N$ vector models or in holography. This generalization would be relevant for recent numerical simulations that yielded the dynamical conductivity across the disorder-tuned 
superconductor-insulator transition \cite{Swanson2013}.      

It is known that in vector large-$N$ models,  interacting Goldstone bosons in a superfluid ground state lead to logarithmic corrections to (\ref{eq:mainres}) \cite{Lucas:2016fju}.    The likely reason why this effect is missing in holography is that in these vector large-$N$ models, there are $\mathcal{O}(N)$ Goldstone bosons,   whereas holographic superfluids have only one Goldstone boson.   We anticipate that the study of quantum corrections in the bulk, along the lines of \cite{Anninos:2010sq},  may reveal the breakdown of (\ref{eq:mainres}) in a holographic superfluid as well.   It would be interesting to understand whether the logarithmic corrections to (\ref{eq:mainres}) are controlled universally by thermodynamic properties of the superfluid:  the calculation of \cite{Lucas:2016fju} suggests that the coefficient of this logarithm is proportional to the superfluid density.

We have also been able to extend the holographic results for $z=1$ QCPs to $z\ne 1$.   The flavor of the expansion is extremely similar for these theories in holography,  and again we expect that key aspects of this expansion remain true for other Lifshitz QCPs.   It would be interesting to find non-holographic models where this expansion can be reliably computed.

\section*{Acknowledgements}
We are grateful to Rob Myers for collaboration in the early stages of this work, and Tom Faulkner for discussions. 
We also thank Snir Gazit and Daniel Podolsky for their 
collaboration on a closely related project. AL was supported by the Gordon and Betty Moore Foundation's EPiQS Initiative through Grant GBMF4302. 
 TS was supported by funding from an NSERC Discovery grant. 
WWK was supported by a postdoctoral fellowship and a Discovery Grant from NSERC, by a Canada Research Chair, and by MURI grant W911NF-14-1-0003 from ARO. WWK further acknowledges the hospitality of the Aspen Center for Physics, where part of this work was done, and which is supported by the National Science Foundation through the grant PHY-1066293.  
Research at Perimeter Institute is supported by the Government of Canada through the Department of Innovation, Science and Economic Development and by the Province of Ontario through the Ministry of Research \& Innovation. 

\appendix  

\section{Two-Point Functions when $\Delta-D/2$ is an Integer}\label{app:integer}
In this appendix, we consider the asymptotics of the correlation function $\langle \Oo(-\Omega)\Oo(\Omega)\rangle$ when $\Delta-D/2$ is a non-negative integer.     
\subsection{$\Delta=D/2$}
The most interesting case is $\Delta=D/2$.   The first thing to note is that the momentum space two-point function of $\mathcal{O}$ has logarithms:  \begin{equation}
\langle \mathcal{O}(p) \mathcal{O}(-p)\rangle_0  = \mathcal{C}_{\mathcal{OO}} \log \frac{p}{\mu}. \label{eq:COOlog} 
\end{equation}
These logarithms appear for all $\Delta-D/2=n=1,2,\ldots$ as well, multiplied by an extra factor of $p^{2n}$.   There are two ways to understand the presence of these logarithms.   Firstly, such a logarithm is necessary for the two-point function in position space to be non-local.   Secondly, the emergence of the scale $\mu$ breaks conformal invariance, and is related to conformal anomalies of the dual field theory \cite{Skenderis:2002wp}.   We will see below how $\mu$ appears in the asymptotics of two-point functions away from criticality.
\subsubsection{Conformal Perturbation Theory}
As in the main text, we start by revisiting conformal perturbation theory. The object that we must study is 
\begin{equation}
\langle \Oo(\Omega)\Oo(-\Omega-p)\mathcal{O}(p)\rangle_0 \approx A_{\Oo\Oo\mathcal{O}} \int\limits_0^\infty \mathrm{d}x \; x^{\frac{D}{2}-1} \Omega^{2\Do-D} \mathrm{K}_{\Do-\frac{D}{2}}(\Omega x)^2 \mathrm{K}_0(px) + \mathrm{O}\!\left(\frac{p^2}{\Omega^2}\right).
\end{equation}
Since the integral is dominated by regions where $\Omega x \sim 1$, we can approximate this integral by \begin{align}
\langle \Oo(\Omega)\Oo(-\Omega-p)\mathcal{O}(p)\rangle_0 &\approx A_{\Oo\Oo\mathcal{O}} \Omega^{2\Do-\frac{3D}{2}} \int\limits_0^\infty \mathrm{d}y \; y^{\frac{D}{2}-1} \mathrm{K}_{\Do-\frac{D}{2}}(y)^2 \times \notag \\
&\;\;\;\;\;\;\;\;\; \left(-\log y -  \log \frac{p}{\Omega} + \log 2 - \gamma_{\textsc{e}}\right) \notag \\
&= A_{\Oo\Oo\mathcal{O}} \Omega^{2\Do-\frac{3D}{2}} \left[ \tilde C+ \Psi\!\left(\tfrac{D}{2},\Do-\tfrac{D}{2}\right) \log \frac{\Omega}{p}\right] \notag \\
&= A_{\Oo\Oo\mathcal{O}} \Omega^{2\Do-\frac{3D}{2}} \Psi\!\left(\tfrac{D}{2},\Do-\tfrac{D}{2}\right) \left[  \log \frac{\Omega}{\hat C\mu} - \log \frac{p}{\mu}\right]  \label{eq:Chat}
\end{align}
with $\tilde C$ and $\hat C$ complicated constants, and $\mu$ an arbitrary scale. The presence of $\mu$ in the answer indicates the running of the couplings, as previously noted. 

Our main result (\ref{eq:mainres}) becomes modified to \begin{equation}
\langle \Oo(-\Omega)\Oo(\Omega)\rangle = \Omega^{2\Do-D}\left(\mathcal{C}_{\Oo\Oo} + \Omega^{-\frac{D}{2}}\left(\mathcal{A}h\log\frac{\Omega}{\mu} + \mathcal{B}\langle\mathcal{O}\rangle\right) + \cdots\right).  \label{eq:mainresA}
\end{equation}
Following the discussion in the main text we find:  \begin{subequations}\label{eq:ABA}\begin{align}
\mathcal{A} &= A_{\Oo\Oo\mathcal{O}} \Psi\left(\tfrac{D}{2},\Do-\tfrac{D}{2}\right), \\
\mathcal{B} &= - \frac{1}{\mathcal{C}_{\mathcal{OO}}}A_{\Oo\Oo\mathcal{O}} \Psi\left(\tfrac{D}{2},\Do-\tfrac{D}{2}\right) = -\frac{\mathcal{A}}{\mathcal{C}_{\mathcal{OO}}}.
\end{align}\end{subequations} 

\subsubsection{Holography}
We now turn to the holographic computation.   The first thing we must revisit is the holographic renormalization prescription for $\Phi$.   Following \cite{Casini:2016rwj}, we obtain \begin{equation}
\Phi(r\rightarrow 0) = \frac{r^{\frac{D}{2}}}{L^{\frac{D-1}{2}}}\left[h \log \frac{\mathrm{e}^{\gamma_{\textsc{e}}}\mu r}{2} - 2 \langle \mathcal{O}\rangle\right] + \cdots,  \label{eq:41D21}
\end{equation}
with $\mu$ an RG-induced scale and $\gamma_{\textsc{e}}$ the Euler-Mascheroni constant.   The specific definition of $\mu$ above will prove useful. 
We compute $\mathcal{C}_{\mathcal{OO}}$ as in Section \ref{app:COO}.   Looking for solutions to the bulk wave equation of the form $\Phi(r)\mathrm{e}^{-\mathrm{i}\Omega t}$, we find \begin{equation}
\Phi(r) = r^{\frac{D}{2}} \mathrm{K}_0(\Omega r) = r^{\frac{D}{2}} \left[ - \log(\Omega r) + \log 2 - \gamma_{\textsc{e}}+\cdots\right] . \label{eq:41D22}
\end{equation}
Comparing (\ref{eq:41D21}) and (\ref{eq:41D22}) to (\ref{eq:COOlog}) directly gives \begin{equation}
\mathcal{C}_{\mathcal{OO}} =- \frac{1}{2}.  \label{eq:COOA}
\end{equation}
The computation of $A_{\Oo\Oo\mathcal{O}}$ then follows very similarly to Section \ref{sec:AOOO}; we find\footnote{There is a minus sign difference from (\ref{eq:AOOOholography}), due to the asymptotic behavior of $\mathrm{K}_0(x) \sim -\log x$, instead of $+\log x$.} \begin{equation}
A_{\Oo\Oo\mathcal{O}} = \frac{\alpha_W}{\mathcal{Z}(\Do-\frac{D}{2})^2}.  \label{eq:AOOOA}
\end{equation}

The computation of the two-point function $\langle \Oo(\Omega)\Oo(-\Omega)\rangle$ then follows very closely the computation of Section \ref{sec:41}.    Employing the notation introduced there, we find that we must solve a differential equation analogous to (\ref{eq:scalar413}): \begin{equation}
\partial_R^2\chi + \frac{1-2b}{R}\partial_R\chi - \chi \approx \frac{\alpha_W}{R^2}\left(h\log \frac{\mathrm{e}^{\gamma_{\textsc{e}}}\mu R}{2\Omega} - 2\langle \mathcal{O}\rangle \right)\frac{R^{\frac{D}{2}}}{\Omega^{\frac{D}{2}}}
\end{equation}
The computation proceeds in an analogous fashion, and we find that the leading correction to $\chi$ is, similarly to (\ref{eq:scalarint}): \begin{align}
\chi^1(R\rightarrow 0) &\approx  -\mathrm{I}_b(R)R^b \int\limits_0^\infty \mathrm{d}R_0 \; R_0 \frac{\hat{\mathrm{K}}_b(R_0)}{\mathcal{Z}(b)} \frac{\alpha_W}{\Omega^{\frac{D}{2}}}R_0^{\frac{D}{2}-2}\left(h\log \frac{\mathrm{e}^{\gamma_{\textsc{e}}}\mu R_0}{2\Omega} - 2\langle \mathcal{O}\rangle \right) \notag \\
&= \mathrm{I}_b(R)R^b \frac{\alpha_W}{\mathcal{Z}(b)\Omega^{\frac{D}{2}}} \left(h\log\frac{\Omega}{\hat C \mu} + 2\langle \mathcal{O}\rangle\right).
\end{align}
Hence, we obtain  \begin{align}
\langle \Oo(\Omega)\Oo(-\Omega)\rangle &= \Omega^{2\Do-D}\left[\mathcal{C}_{\Oo\Oo} + \frac{\alpha_W}{\mathcal{Z}(\Do-\frac{D}{2})^2 \Omega^{\frac{D}{2}}} \left(h\log\frac{\Omega}{\hat C \mu} + 2\langle \mathcal{O}\rangle\right) + \cdots \right]. \label{eq:appAfinal}
\end{align}
Remarkably, the exotic factor of $\hat C$ defined in (\ref{eq:Chat}) shows up in the logarithm in (\ref{eq:appAfinal}).  As we already defined the scale $\mu$ in (\ref{eq:41D21}) so that the two-point function $\langle \mathcal{O}(\Omega)\mathcal{O}(-\Omega)\rangle_0$ 
took exactly the form (\ref{eq:COOlog}), the appearance of the same $\hat C$ in the logarithms of both (\ref{eq:mainresA}) and (\ref{eq:appAfinal}) is not trivial.  Noticing this, and comparing (\ref{eq:ABA}), (\ref{eq:COOA}) and (\ref{eq:AOOOA}), we see that (\ref{eq:appAfinal}) reduces to (\ref{eq:mainresA}).    Hence, the precise agreement of holography to conformal perturbation theory extends to the special operator dimension $\Delta=D/2$. 

Let us note the important feature that if we re-define $\mu$ from our definition in (\ref{eq:41D21}), this necessitates a re-definition of $\langle \mathcal{O}\rangle$.   Thus, $\langle \mathcal{O}\rangle$ hence depends on the choice of $\mu$.   It is crucial that $\langle \mathcal{O}\rangle$ and $\mu$ enter the two-point function in such a way that it is invariant under a simultaneous shift of $\mu$ and $\langle \mathcal{O}\rangle$.  Indeed, comparing (\ref{eq:41D21}) and (\ref{eq:appAfinal}) shows that this is the case.

\subsection{$\Delta>D/2$}
The case $\Delta>D/2$ is much more similar to the case studied in the main text, and so we give a qualitative treatment, beginning with holography.   Let us define \begin{equation}
\Delta = \frac{D}{2}+n.
\end{equation}
  The asymptotic behavior of $\Phi$ is \cite{Skenderis:2002wp}\begin{equation}
\Phi(r\rightarrow0) = \frac{r^{\frac{D}{2}}}{L^{\frac{D-1}{2}}}\left[\frac{h}{r^n}\left(1+\cdots + \tilde c_{2n} r^{2n} \log(\mu r) \right)+ c_{2n} r^n \langle \mathcal{O}\rangle \right].
\end{equation}
The coefficients $c_{2n}$ and $\tilde c_{2n}$ are dimensionless numbers;  their precise values are not important for this discussion.   The crucial thing we notice is that $h$ is the leading order term, and has no logarithmic corrections.    So while there might be further logarithmic corrections (at higher orders in $h$), the qualitative form of (\ref{eq:scalar413}), and hence (\ref{eq:mainres}), remains correct. 

A similar result can be found using conformal perturbation theory.   The leading order correction to the CFT two-point function is proportional to $h/\Omega^{D-\Delta}$, with no logarithmic corrections. 

\section{Evaluating $C_{abcd}C^{abcd}$}\label{app:weyl}
In this appendix we derive the equations of motion for $h_{xy}(t,r)$.   As in the computations of $\langle \Oo \Oo\rangle$ and $\sigma(\Omega)$, we may (to leading order) set the background metric $g$ to that of pure AdS. For the Weyl correction to the action, it is easiest to first evaluate the action in terms of $h_{xy}$.   
  As the Weyl tensor vanishes for pure AdS space, we need only compute it to first order in $h$.  Up to the action of symmetries, the non-vanishing components of $C_{abcd}$ at this order in $h$ are: \begin{subequations}\begin{align}
C_{rxty} &= C_{rytx} = -\frac{L^2}{2r^2}  \partial_r \partial_t h_{xy}, \\
C_{rxry} &= -\frac{L^2}{2r^2} \left[\frac{D-2}{D-1} \partial_r^2 h_{xy} - \frac{1}{D-1} \partial_t^2 h_{xy}\right], \\
C_{txty} &= \frac{L^2}{2r^2} \left[\frac{1}{D-1}\partial_r^2 h_{xy} - \frac{D-2}{D-1} \partial_t^2 h_{xy} + \right], \\
C_{zxzy} &= \frac{L^2}{2r^2} \frac{\partial_r^2 h_{xy} + \partial_t^2 h_{xy}}{D-1}.
\end{align}\end{subequations}
In the above equations, $z$ stands for a generic spatial direction which is not $x$ or $y$.   Hence, we find \begin{align}
\sqrt{g}Y(\Phi)C_{abcd}C^{abcd} &= \frac{2}{D-1} \left(\frac{L}{r}\right)^{D-3} Y(\Phi) \left[2(D-1)(\partial_r \partial_t h_{xy})^2 - 2 \partial_t^2 h_{xy}\partial_r^2 h_{xy} \right. \notag \\
&\left.+ (D-2)(\partial_r^2 h_{xy})^2 + (D-2)(\partial_t^2 h_{xy})^2\right] \label{eq:C2}
\end{align}
The relevant contributions to the equations of motion are \begin{equation}
0 = \frac{\delta S}{\delta h_{xy}} = \frac{r^2}{L^2}\left[R_{xy} + \frac{D}{r^2}h_{xy} \right] +  \frac{\delta}{\delta h_{xy}} \int \mathrm{d}r \sqrt{g}Y(\Phi)C_{abcd}C^{abcd} 
\end{equation}
Using (\ref{eq:C2}) and \begin{equation}
R_{xy} + \frac{D}{r^2}h_{xy} = \frac{1}{2} \left[-\partial_r^2 h_{xy} + \frac{D-1}{r}\partial_r h_{xy} + r^2\Omega^2 h_{xy}\right]
\end{equation}
we obtain (\ref{eq:weyleq}).

\section{Computing $A_{TT\mathcal{O}}$}\label{app:ATTO}
In this appendix, we assume that all momenta $p_{1,2,3}$ point in the $t$ direction.   Starting with (\ref{eq:C2}), and slightly rearranging, the bulk action which we must evaluate to determine $A_{TT\mathcal{O}}$ is given by \begin{align}
S_{\mathrm{bulk}} = \int \mathrm{d}^{D+1}x &\sqrt{g} g^{-1}g^{-1} \frac{4\alpha_Z L^{\frac{D+3}{2}}}{\kappa^2} \Phi(p_3) \left[\partial_\mu \partial_\nu h_{xy}(p_1)\partial_\mu \partial_\nu h_{xy}(p_2) \right. \notag \\
&\left.- \frac{\partial_\mu \partial_\mu h_{xy}(p_1)\partial_\nu \partial_\nu h_{xy}(p_2)}{D-1} \right].  \label{eq:Sbulkapp}
\end{align}
In this appendix, we will denote with $g^{-1} = r^2/L^2$ a generic diagonal element of the AdS metric (\ref{eq:adsmetric});  $\sqrt{g} = (L/r)^{D+1}$ is unchanged.   We will also use an Einstein summation convention on lowered Greek indices, with no factors of the AdS metric  included.

We now perform a series of integration by parts manipulations, with the end goal to remove all derivatives from $h_{xy}$.   We begin by analyzing the second term in more detail.   Integration by parts a few times leads to \begin{align}
\int \mathrm{d}r &\sqrt{g}g^{-1}g^{-1}\Phi(p_3) \partial_\mu \partial_\mu h_{xy}(p_1)\partial_\nu \partial_\nu h_{xy}(p_2) \notag\\
= \int \mathrm{d}r\Big[& \sqrt{g}g^{-1}g^{-1}\Phi(p_3) \partial_\nu \partial_\mu h_{xy}(p_1)\partial_\mu \partial_\nu h_{xy}(p_2) \notag \\
& + \partial_\mu \partial_\nu \left(\sqrt{g}g^{-1}g^{-1}\Phi(p_3)\right) \partial_\mu h_{xy}(p_1)\partial_\nu h_{xy}(p_2) \notag \\
&- \partial_\mu \partial_\mu \left(\sqrt{g}g^{-1}g^{-1}\Phi(p_3)\right) \partial_\nu h_{xy}(p_1)\partial_\nu h_{xy}(p_2)\Big].  \label{eq:BIBP1}
\end{align}
After more integration by parts, the second term on the right hand side above can be replaced with
\begin{align}
\int \mathrm{d}r  \partial_\mu &\partial_\nu \left(\sqrt{g}g^{-1}g^{-1}\Phi(p_3)\right) \partial_\mu h_{xy}(p_1)\partial_\nu h_{xy}(p_2) \notag \\
= \int \mathrm{d}r \Big[&\frac{1}{2}\partial_\mu \partial_\mu \left(\sqrt{g}g^{-1}g^{-1}\Phi(p_3)\right) \partial_\nu h_{xy}(p_1)\partial_\nu h_{xy}(p_2) \notag \\
& + \sqrt{g}g^{-1}g^{-1} \Phi(p_3) \Big( \partial_\nu h_{xy}(p_1) \partial_\nu \partial_\mu \partial_\mu h_{xy}(p_2) \notag \\ 
& + \partial_\nu \partial_\nu h_{xy}(p_1) \partial_\mu \partial_\mu h_{xy}(p_2)\Big) \Big].  \label{eq:BIBP2}
\end{align}
We are finally ready to employ the $\Omega=\infty$ equations of motion, \begin{equation}
\partial_r^2 - \frac{D-1}{r}\partial_r h_{xy} - r^2\Omega^2 h_{xy} = 0,  \label{eq:hxyeom}
\end{equation}
 to simplify the second term above:
 \begin{align}
\int &\mathrm{d}r \sqrt{g}g^{-1}g^{-1} \Phi(p_3)\partial_\nu h_{xy}(p_1) \partial_\nu \partial_\mu \partial_\mu h_{xy}(p_2) \\
= \int &\mathrm{d}r \sqrt{g}g^{-1}g^{-1} \Phi(p_3) \partial_\nu h_{xy}(p_1) \partial_\nu \left(\frac{D-1}{r}\partial_r h_{xy}(p_2)\right) \notag \\
= \int &\mathrm{d}r \sqrt{g}g^{-1}g^{-1}\Phi(p_3) \left[\frac{D-1}{2r} \partial_r \left(\partial_\nu h_{xy}(p_1) \partial_\nu h_{xy}(p_2)\right) - \frac{\partial_\mu \partial_\mu h_{xy}(p_1)\partial_\nu \partial_\nu h_{xy}(p_2)}{D-1} \right].   \label{eq:BIBP3}
\end{align}
In the second line above, we have re-used the equation of motion (\ref{eq:hxyeom}) to replace $\partial_r h_{xy}\partial_r h_{xy}$.   Combining (\ref{eq:BIBP1}), (\ref{eq:BIBP2}) and (\ref{eq:BIBP3}), we find \begin{align}
\tilde S &\equiv \int \mathrm{d}r   \sqrt{g}g^{-1}g^{-1}\Phi(p_3)\left[  \partial_\nu \partial_\mu h_{xy}(p_1)\partial_\mu \partial_\nu h_{xy}(p_2) - \frac{\partial_\mu \partial_\mu h_{xy}(p_1)\partial_\nu \partial_\nu h_{xy}(p_2)}{D-1} \right] = \notag \\
& \frac{1}{2}  \int \mathrm{d}r \partial_\nu h_{xy}(p_1) \partial_\nu h_{xy}(p_2) \left[\partial_\mu \partial_\mu \left(\sqrt{g}g^{-1}g^{-1} \Phi(p_3)\right) + \partial_r \left(\frac{D-1}{r} \sqrt{g}g^{-1}g^{-1}\Phi(p_3)\right)\right].
\end{align} 
From (\ref{eq:Sbulkapp}), $\tilde S$ is proportional to $S_{\mathrm{bulk}}$.    Straightforward integration by parts, along with the equation of motion (\ref{eq:hxyeom}), gives us \begin{equation}
 \int \mathrm{d}r \sqrt{g} g^{-1} \partial_\nu h_{xy}(p_1) \partial_\nu h_{xy}(p_2)   \mathcal{Y} = \int \frac{\mathrm{d}r}{2} h_{xy}(p_1)h_{xy}(p_2) \partial_\mu \left(\sqrt{g}g^{-1}\partial_\mu \mathcal{Y}\right).
\end{equation} 
Upon using the equations of motion for $\Phi$, we find \begin{align}
\partial_\mu \partial_\mu \left(\sqrt{g}g^{-1}g^{-1} \Phi(p_3)\right) &+ \partial_r \left(\frac{D-1}{r} \sqrt{g}g^{-1}g^{-1}\Phi(p_3)\right) = \sqrt{g}g^{-1} \mathcal{Y}
\end{align}
where \begin{align}
\mathcal{Y} &\equiv \frac{\Delta(\Delta-D)-2(D-2)}{2L^2}\Phi(p_3) + \frac{2r}{L^2}\partial_r \Phi(p_3) \notag \\
&= \frac{(\Delta+2)(\Delta+2-D)}{2L^2}\Phi(p_3) + \frac{2r^{D-\Delta}}{L^{\frac{d}{2}}\mathcal{Z}(\Delta-\frac{D}{2})} \hat{\mathrm{K}}_{\Delta-\frac{D}{2}+1}(p_3r)
\end{align}
Upon using the identities (\ref{eq:besselders}) we find \begin{align}
\partial_\mu &\left(\sqrt{g}g^{-1}\partial_\mu \mathcal{Y}\right) = \frac{\Delta(\Delta-D)(\Delta+2)(\Delta+2-D)}{2L^4}\sqrt{g}\Phi(p_3) \notag \\
&+ \frac{2(\Delta+2)(D-\Delta-2)}{L^4}\sqrt{g} \frac{r^{D-\Delta}\hat{\mathrm{K}}_{\Delta-\frac{D}{2}+1}(p_3r)}{L^{d/2}\mathcal{Z}(\Delta-\frac{D}{2})}+ \frac{4}{L^4}\sqrt{g} \frac{r^{D-\Delta}\hat{\mathrm{K}}_{\Delta-\frac{D}{2}+2}(p_3r)}{L^{d/2}\mathcal{Z}(\Delta-\frac{D}{2})}. \label{eq:B9}
\end{align}
Combining (\ref{eq:Sbulkapp}) and (\ref{eq:B9}) with \begin{equation}
\langle T_{xy}(p_1)T_{xy}(p_2)\mathcal{O}(p_3) \rangle_0 = -\frac{\delta^3 S_{\mathrm{bulk}}}{\delta h_{xy}(p_1)\delta h_{xy}(p_2)\delta \Phi (p_3)}
\end{equation}
we recover (\ref{eq:TTOformal}) with $A_{TT\mathcal{O}}$ given by (\ref{eq:ATTOhol}).

\section{Generalization of Holographic Asymptotic Integrals}\label{app:analcont}
In this appendix, we discuss some of the possibilities which we neglected in the main text in our derivation of (\ref{eq:sigmaholfinal}).   The notation of this appendix follows the discussion around (\ref{eq:scalarint}).
\subsection{$a+2<2b$}
As $a>0$ ($0<\Do<D$), we may safely assume that $b>0$.   What happens in this case is that the integral in (\ref{eq:scalarint}) is divergent.   As a consequence, we must explicitly write \begin{align}
\chi = c R^b \mathrm{K}_b(R) \int\limits_0^R \mathrm{d}R_0 \; R_0^{1+a}\mathrm{K}_b(R_0)\mathrm{I}_b(R_0) + cR^b \mathrm{I}_b(R)\int\limits_R^\infty \mathrm{d}R_0 \; R_0^{1+a}\mathrm{K}_b(R_0)^2.  \label{eq:B1}
\end{align}
We now analyze the integrands as $R,R_0 \rightarrow 0$.   Denoting $\mathfrak{p}(x)$ as an arbitrary regular Taylor series in $x$ (no divergences or non-analytic contributions), the first integral in (\ref{eq:B1}) is 
\begin{align}
R^b\mathrm{K}_b(R) \int\limits_0^R \mathrm{d}R_0 \; R_0^{1+a}\mathrm{K}_b(R_0)\mathrm{I}_b(R_0) &= (\mathfrak{p}(R)\! +\! R^{2b}\mathfrak{p}(R))\! \int\limits_0^R \mathrm{d}R_0\, R_0^{1+a}(\mathfrak{p}(R_0)\! +\! R_0^{2b}\mathfrak{p}(R_0)) \notag \\
&= \mathfrak{p}(R) R^{2+a} + \mathfrak{p}(R) R^{2+a+2b} + \mathfrak{p}(R) R^{2+a+4b}.
\end{align}
Hence, this integral cannot contribute to the $\mathrm{O}(R^{2b})$ term.   The second integral is divergent as $R\rightarrow 0$ and can be analytically evaluated to: \begin{equation}
\int\limits_R^\infty \mathrm{d}R_0 \; R_0^{1+a}\mathrm{K}_b(R_0)^2 = \Psi(a+2;b) + R^{2+a-2b}\mathfrak{p}(R) + R^{2+a}\mathfrak{p}(R) + R^{2+a+2b}\mathfrak{p}(R),
\end{equation}
with $\mathfrak{p}(R)$s above given by specific hypergeometric functions.   Hence, we see that the only term which will scale as $R^{2b}$ is proportional to $\Psi(a+2;b)$.   This justifies the ``analytic continuation" of $a+2-2b$ through 0 using holography.
\subsection{$b<0$}
When $b<0$, the dominant term in $\psi(R)$ as $R\rightarrow 0$ is the response, and the subleading term is the source.  Unitarity in a conformal field theory requires that for any non-trivial scalar operator \cite{Rychkov:2016iqz} \begin{equation}
\Delta, \Do > \frac{D}{2}-1,
\end{equation}
which implies that $b>-1$.  Now, when $b<0$,  $x^b\mathrm{K}_b(x) \sim x^{2b}$ as $x\rightarrow 0$.   (\ref{eq:scalarint}) converges so long as $a+1+2b>-1$, which becomes equivalent to $a>0$.  Since we have assumed that the operator $\mathcal{O}$ is relevant, this is indeed the case.  Thus, the computation of the response presented in the main text remains correct.

\section{Viscosity in a Holographic Lifshitz Model}\label{app:visc}
In this appendix we argue that the $\mathcal{B}_T$ term in (\ref{eq:BTexpansion}) arises in a generic holographic model.    For simplicity, we focus on a background geometry described by the ``Lifshitz black brane" (in Euclidean time):   \begin{equation}
\mathrm{d}s^2 = \frac{L^2}{r^2}\left[\frac{\mathrm{d}r^2}{f(r)} + \frac{f(r)}{r^{2z-2}}\mathrm{d}t_{\textsc{e}}^2 + \mathrm{d}\mathbf{x}^2\right], \;\;\; f(r) = 1-\left(\frac{r}{r_+}\right)^{d+z}.
\end{equation}
The constant $r_+$ is related to the temperature,  and in particular the thermodynamic pressure obeys \begin{equation}
P \propto r_+^{-d-z}. \label{eq:appEP}
\end{equation}
More generally, we expect that the pressure is related to the $\mathrm{O}(r^{d+z})$ contributions to $f(r)$, though we note that the holographic renormalization for $T_{\mu\nu}$ can require some care, especially in Lifshitz theories \cite{Chemissany:2014xsa}.   With this particular choice for $f(r)$, we find that for a rather general family of $\mathbf{x}$-independent matter backgrounds that support a Lifshitz geometry, the equations of motion for the metric perturbation $h_{xy}$, defined in (\ref{eq:hxydef}), are independent of the details of the bulk matter: \begin{equation}
0 = r^{d+z+1}f\partial_r\left(\frac{f}{r^{d+z-1}}\partial_r h_{xy}\right) -  r^{2z}\Omega^2h_{xy} .
\end{equation}
In terms of the coordinate $R$ defined in (\ref{eq:6R}) we find \begin{equation}
0 = R^{\frac{d}{z}}f\left(\left(\frac{zR}{\Omega}\right)^{\frac{1}{z}}\right)\partial_R\left(\frac{1}{R^{\frac{d}{z}}}f\left(\left(\frac{zR}{\Omega}\right)^{\frac{1}{z}}\right)\partial_R h_{xy}\right) - h_{xy}.
\end{equation}
When $\Omega=\infty$, we can approximate $f=1$;  the resulting equation is regular in $R$ and we may solve it with suitable boundary conditions with a modified Bessel function, as before.    Hence about $\Omega=\infty$ we look for a solution of the form $h_{xy}= h^{(0)}_{xy} +h^{(1)}_{xy} $  where \begin{equation}
R^{\frac{d}{z}}\partial_R\left(R^{-\frac{d}{z}}\partial_Rh^{(1)}_{xy}\right)-h^{(1)}_{xy}= \frac{1}{r_+^{d+z}}\left(\frac{zR}{\Omega}\right)^{1+\frac{d}{z}} h^{(0)}_{xy} + R^{\frac{d}{z}}\partial_R\left(\frac{1}{r_+^{d+z} R^{\frac{d}{z}}}\left(\frac{zR}{\Omega}\right)^{1+\frac{d}{z}} \partial_R h^{(0)}_{xy}\right).
\end{equation}
We conclude by linearity and (\ref{eq:appEP}) that $h^{(1)}_{xy} \propto P\Omega^{-1-\frac{d}{z}}$.  As $h^{(1)}_{xy}$ is sourced by a complicated function, we do not expect the $\mathrm{O}\left(R^{1+\frac{d}{z}}\right)$ coefficient of $h^{(1)}_{xy}$, responsible for the corrections to $\eta(\mathrm{i}\Omega)$, will be non-zero.    Hence,  the coefficient $\mathcal{B}_T$ in (\ref{eq:BTexpansion}) must be generically non-zero.

\bibliography{references}

\end{document}